\documentstyle[preprint,aps]{revtex}
\tighten
\draft
\widetext
\input{epsf.sty}
\preprint{CLNS 00/1711}
\font\blackboard=msbm10 % scaled \magstep1
\font\blackboards=msbm7 \font\blackboardss=msbm5
\newfam\black \textfont\black=\blackboard
\scriptfont\black=\blackboards \scriptscriptfont\black=\blackboardss
\def\Bbb#1{{\fam\black\relax#1}}

\def\sigz{\sigma_0}
\def\sigL{\sigma_L}

\def\Le{\Lambda_{eff}}
\def\Li{\Lambda_i}
\def\kL{k_0}
\def\kM{k_1}
\def\kR{k_2}
\def\yL{y_0}
\def\yR{y_2}
\def\yM{y_1}
\def\li{L_i}
\def\ki{k_i}
\def\yi{y_i}
\def\qi{q_i}

\def\Di{\Delta l_i}
\def\lm{L_{i-1}}

\def\kp{k_{i+1}}
\def\yp{y_{i+1}}

\def\be{\begin{equation}}
\def\ee{\end{equation}}
\def\baray{\begin{eqnarray}}
\def\earay{\end{eqnarray}}

\def\Y{$\surd$}
\def\N{$\times$}
\def\P{?}

\newcounter{multibraneseps}
\newcounter{bulktypeseps}
\newcounter{2braneNewtonconsteps}
\newcounter{Hubblegraphportraiteps}
\newcounter{figRSeps}
\newcounter{bulkbehavioureps}
\newcounter{hyperboloideps}
\newcounter{kphaseeps}

\begin{document}

\title{A Brane World Perspective on the Cosmological Constant 
and the Hierarchy Problems}
\author{\'Eanna Flanagan\footnote{flanagan@spacenet.tn.cornell.edu}, 
Nicholas Jones\footnote{nicholas-jones@cornell.edu}, 
Horace Stoica\footnote{fhs3@mail.lns.cornell.edu},
S.-H. Henry Tye\footnote{tye@mail.lns.cornell.edu} and
Ira Wasserman\footnote{ira@spacenet.tn.cornell.edu}}

\address{Laboratory for Nuclear Studies and Center for Radiophysics
and Space Research\\
Cornell University \\
Ithaca, NY 14853}

\medskip
\date{\today}
\maketitle

\begin{abstract}

We elaborate on the recently proposed static brane world scenario, 
where the effective 4-D cosmological constant is
exponentially small when parallel $3$-branes are far apart.
We extend this result to a compactified model with
two positive tension branes.
Besides an exponentially small effective 4-D cosmological 
constant, this model incorporates a Randall-Sundrum-like solution to 
the hierarchy problem. Furthermore, the exponential factors for the 
hierarchy problem and the cosmological constant problem obey
an inequality that is satisfied in nature.
This inequality implies that the cosmological constant problem can be 
explained if the hierarchy problem is understood.
The basic idea generalizes to the multibrane world scenario.
We discuss models with piecewise adjustable 
bulk cosmological constants (to be determined by the $5$-dimensional
Einstein equation), a key element of the scenario. 
We also discuss
the global structure of this scenario and clarify the physical 
properties of the particle (Rindler) horizons that are present.
Finally, we derive a 4-D effective theory in which all observers on 
all branes not separated by particle horizons measure the same 
Newton's constant and 4-D cosmological constant.

\end{abstract}

\section{Introduction}

Recent observational data\cite{rev} indicate that our universe has
a positive cosmological constant which, compared to the 
Planck or the electroweak scale, is many 
orders of magnitude smaller than expected 
within the context of ordinary gravity and quantum 
field theory. This is the well-known cosmological constant problem 
\cite{weinberg,reviews}. Recently, two of us proposed a static brane world 
solution to this problem in which 
the cosmological constant becomes exponentially small compared to 
all other scales in the model\cite{ira}. Here, by ``static'', we mean 
the situation where the branes are stationary relative to each other
\footnote{More precisely, the bulk spacetime possesses a timelike
Killing vector field which is not hypersurface orthogonal, so that the
spacetime is stationary but not static, and the branes are invariant
under the diffeomorphism generated by this vector field.}.
We still allow for expansion of the Universe in the other three spatial
directions, but confine ourselves to situations in which the Hubble
expansion rate $H$ is time-independent. In more complete Big Bang
cosmological models, this would correspond to late times, when the
cosmological constant becomes dominant.

It is generally believed that fine-tuning is necessary for a very small
cosmological constant in $4$-dimensional theories\cite{weinberg}.
This leads one to search for a naturally small cosmological constant
in higher dimensional theories. However, for a usual compactification
of a higher dimensional theory to an effective $4$-dimensional theory,
one ends up with a normal $4$-dimensional theory, and the fine-tuning
problem generically reappears. This is the case for
usual Kaluza-Klein (KK) compactification, and for the generic
compactifications with large extra dimensions\cite{ADD}.
The Randall-Sundrum (RS) model\cite{RS1,RS2} provides a hope of avoiding 
this pathology.  The RS model considers
$3$-branes inside $5$-dimensional spacetime, with Anti-deSitter (AdS) spaces
in the bulk. Normal finite gravitational interaction is reproduced
even if the $5$th spatial dimension is not compactified.
For a single $3$-brane (or a stack of $3$-branes) with brane
tension (or vacuum energy density) $\sigz$, the effective 4-D
cosmological constant as seen by observers on the brane is taken to be zero,
that is\cite{RS1,oldcosmo}
\be
\label{RStuning}
\Le \propto \sigz - \sqrt{6\Lambda/\kappa^2} = 0
\ee
where $\kappa^2$ is the 5-dimensional gravitational coupling,
and $-\Lambda$ is the negative bulk cosmological constant
\footnote{Throughout this paper, the cosmological constants
will have units of energy density in the appropriate dimension}.
We see that the brane tension, which is at least partly determined by
the dynamics of the standard model, is not required to vanish.
The cancellation between the brane tension and the bulk cosmological
constant is imposed in any brane world scenario. However, the non-linear
relation ({\em i.e.}, $\sigz$ versus $\sqrt \Lambda$),
as well as the lack of the need to compactify the
$5$th dimension, follow from the non-factorizable metric in the RS model.
Since the $5$th dimension may stay uncompactified, any 4-D description
will be inadequate. If one now insists on a 4-D description for
observers on the brane, one ends up with a novel 4-D description via 
the AdS/CFT correspondence\cite{AdS,gubser}, which offers hope to void
the argument for the need of fine-tuning to obtain a very small 4-D
cosmological constant. 
This has motivated a number of attempts to find a solution to the 
cosmological constant problem along this direction\cite{selftune}.
Here we shall elaborate on the approach of Ref\cite{ira}.
In the two brane Randall-Sundrum model\cite{RS1}, 
the cancellation in (\ref{RStuning})
is a fine-tuning. In addition, the two branes 
are required to have equal and opposite tensions. These two 
fine-tunings are removed in Ref\cite{ira}. 

The two brane model in Ref\cite{ira} does not solve the hierarchy problem,
which is presumably solved in the RS model. However, the 
two brane model of Ref\cite{ira} is easy to 
generalize to a multibrane model in which both the cosmological constant
problem and the hierarchy problem can be explained simultaneously. 
The generalization to $N$ 3-branes is straightforward because 
an observer on any brane can only see properties around his/her own 
neighborhood. This allows us to analyze the various possibilities generally. 

The simplest realization that solves both fine-tuning 
problems is a two brane compactified model. 
The model has an exponentially small $\Le$, as in Ref\cite{ira}, and
implements the Randall-Sundrum-like solution to
the hierarchy problem, but with a positive tension for the visible brane.
Consider two parallel $3$-branes with tensions $\sigma_0>\sigma_1>0$ 
sitting in a compactified $5$th dimension, with radius $L_2/{2 \pi}$. 
The $\sigma_0$ (Planck, hidden) brane sits at $L_0=0$ and the 
$\sigma_1$ (TeV, visible) brane sits at 
$L_1$. Since $L_2$ is identified with $L_0$, this means the branes 
are separated by $L_1$ on one side and by $L_2-L_1$ on the other side. 
Without loss of generality, let $L_2-L_1>L_1$. We find that, for large 
$L_1$, the warp factor for the hierarchy problem is
\be
{m_{Higgs}^2\over m_{Planck}^2} \simeq{A(L_1)\over A(0)} 
\simeq e^{-\kappa^2(\sigma_0-\sigma_1)L_1/3}
\ee 
and the effective 4-D cosmological constant is
\be
\Le \simeq \frac{2\sigma_0(\sigma_0 + \sigma_1)}{(\sigma_0-\sigma_1)}
e^{-\kappa^2[(\sigma_0+\sigma_1)(L_2-L_1)+(\sigma_0-\sigma_1)L_1]/6}.
\ee
Since $\sigma_0>\sigma_1>0$,
and $L_2-L_1>L_1$, 
we see that
\be
\label{important}
\ln\biggl(\frac{m^4_{Planck}}{\Le}\biggr) > \ln\biggl(\frac{m^2_{Planck}}{m^2_{Higgs}}\biggr)
\ee
as is the case in nature, where $\ln({m^4_{Planck}}/{\Le}) \sim 2.3\times 122$
and $\ln({m^2_{Planck}}/{m^2_{Higgs}}) \sim  2.3 \times 32$. 
To get a feeling for the magnitudes of the various quantities, 
we may choose $L_2-L_1 \sim 2 L_1$, and $\sigma_0 \sim 2 \sigma_1$.
If the brane tensions are around the Planck scale, the exponents that
describe nature follow if $L_1 \sim 10 m^{-1}_{Planck}$. 
In this model, the branes are stationary (that is, they satisfy 
the brane equations of motion) and both $L_1$ and $L_2$ 
are stable. But we still need a dynamical reason why 
$L_1$ (or $L_2$) is large. However, if we understand the hierarchy problem
via some means ({\em e.g.}, comparing the renormalization group flows of 
couplings (marginal operators) versus the mass terms (relevant operators) 
in 4-D quantum field theory), this brane world
model provides an explanation of the cosmological constant problem.   
As we shall see, the above inequality (\ref{important}) is robust in a 
generic multibrane world where the hierarchy problem is solved. 

The exponential behaviors are related to the warp factor in the 
Randall-Sundrum scenario, but with two key differences:
\newcounter{Lcount}
\begin{list}{(\arabic{Lcount})}
  {\usecounter{Lcount}}
  \item The RS model demands $\Le$ to be exactly zero.  This requires 
    a fine-tuning. Here we find that this fine-tuning is not necessary 
    for an exponentially small $\Le$, as long as the branes are  
    relatively far apart.
  \item The RS model also requires the fine-tuning in (\ref{RStuning}).
    Not surprisingly, although $\kappa^2 \sigz^2 - 6\Lambda$ in the above 
    model\cite{ira} is not zero, it turns out to be exponentially small. 
    {\em A priori}, this is still a fine-tuning. 
    To avoid this fine-tuning, the bulk $\Lambda$ is treated as an
    ``integration constant'' that is determined by the 5-D Einstein equation. 
    There are a number of well-known realizations where the cosmological 
    constant is an integration 
    constant\cite{Einstein,Ng,Zee,unruh,HT,aurilia,fst1,fst2,fst3}.
    Since they are crucial to the above model,
    we shall review some of them, and then generalize them so that the 
    bulk cosmological constant is 
    piecewise constant, and each piece is an independent integration constant 
    to be determined by the 5-D Einstein equation. We note that this 
    idea can just as easily be incorporated into the RS model.
\end{list}

The idea of treating the cosmological constant as an ``integration
constant'' has a long history\cite{weinberg}. Classically,
it is not determined and so the cosmological constant problem is not
solved. In the brane world, it is the bulk cosmological constant that
is treated as an ``integration constant''. In the single brane model,
this bulk cosmological constant and so the effective 4-D cosmological 
constant are undetermined, so it takes a fine-tuning to suppress
the 4-D cosmological constant. When the metric is non-factorizable,
the introduction of another brane
provides additional constraints which fix the value(s) of the bulk
cosmological constant and hence the effective 4-D cosmological constant
as seen by observers on the brane. In this sense, the multibrane world 
scenario is key to the solution.
In Ref\cite{ira}, the bulk cosmological constant is treated as an
integration constant in a generic model-independent way.
Here, more details are worked out for the introduction of $5$-form field 
strengths and its generalization. 
In specific models, we have additional equations, namely,
an equation of motion for each brane. {\em A priori}, these equations are 
non-trivial for branes that are charged under the $4$-form potential.
We show that the static solution in Ref\cite{ira} automatically
satisfies these brane equations of motion.

If the number of parallel $3$-branes is not too large, sitting in the
uncompactified (or compactified with large radius) $5$th dimension, 
then we have a low density of $3$-branes
and their mean separation will be large in general.
In the compactified version, observers on each brane will see an 
exponentially small effective 4-D cosmological constant. In the 
uncompactified case, all but one brane will have this property.
In this sense, the smallness of $\Le$ is quite generic.

In many cases, the metric between branes vanishes somewhere. 
These metric zeros are particle horizons which we show are coordinate
singularities analogous to those in Rindler space.  We eliminate those
singularities, and deduce the global structure of the maximal extensions of
the bulk spacetimes. We show that bulk regions separated by
particle horizons often lie in separate connected components of the
maximally extended solution, and therefore cannot physically influence
one another

In the multibrane model, an issue arises concerning the proper method
for determining the 4-D Newton's constant $G_N$. The two standard 
methods to determine $G_N$ are either to solve for 
the trapped gravity mode and Green's function, or to calculate the 
Hubble constant and determine the coefficient of the matter density 
term\cite{oldcosmo}.  
Naively, one would expect the normalization of the trapped gravity 
mode, and hence its wavefunction at a particular brane, to
depend on the total number and placement of 
all of the branes, 
and so should the value of $G_N$ found from the trapped mode. By contrast,
the value deduced from the cosmological expansion rate should be
determined locally, by the brane tension and the bulk cosmological constants
on either side of the brane, and, one would think, should not depend on 
how many other branes there are, or where they are located.  
The resolution is that the normalization of the trapped gravity mode (or any 
other mode) should be computed only over the region between particle 
horizons.  
This also implies that the integration over the $5$th dimension in the 
5-D action to obtain the low-energy effective 4-D action\cite{RS1} 
should be carried out over the same region.  This means that we cannot 
compare the mass scales between branes that are separated by particle 
horizons, which are normalized independently.  Any solution to the mass 
hierarchy problem can be addressed only between branes that are not 
separated by particle horizons.

Now we can see the generic origin of the inequality (\ref{important}).
To solve the hierarchy problem, both the Planck and the visible branes 
must be inside the same particle horizons. The cosmological constant
is exponentially small roughly as a function of the distance of the
Planck brane from the particle horizon while the hierarchy scale (warp)
factor is exponentially small roughly as a function of the distance of the
Planck brane from the visible brane. Since the visible brane must be
between the Planck brane and the particle horizon, the inequality
(\ref{important}) follows.

Using the effective 4-D action approach, we show that both $G_N$ and the
effective cosmological constant $\Le$ are the same for all observers 
on all branes within a pair of particle horizons, irrespective of the 
type of brane (positive or negative tension, Planck or visible) on 
which they live.  
Generically, the visible brane will have different
bulk cosmological constants on its two sides.  The correction to the
Newton's law is calculated in this case; it has the same dependence on
particle separation as in the symmetric case\cite{RS2}, but with a modified
coefficient.

In general, $\Le (L_i)$ is expected to 
be more complicated than given above, because we expect additional 
non-gravitational brane-brane interactions at small separations.  For 
large separations, it is reasonable to assume that the inter-brane 
dynamics is dominated by pure gravity as described here.  In a more 
realistic situation, the matter density on the visible brane (and dark 
matter on the other brane) should be included, and the branes 
should be allowed to move.  We shall not consider these 
effects in this paper. However, we do note that if the kinds of 
explanations offered here are indeed behind the smallness of the
cosmological constant and Higgs mass relative to the Planck scale,
branes must have moved very little since the epoch of cosmological
nucleosynthesis. Otherwise, the Higgs mass scale might have changed
enormously, which would be inconsistent with the success of Big Bang
cosmology in explaining the light element abundances.

The paper is organized as follows. 
In \S\ref{sec:integconst}, we review models in 
which the bulk cosmological constant emerges as an ``integration 
constant'' and discuss the generalization where it becomes piecewise 
constant and piecewise adjustable. 
In \S\ref{sec:multibrane}, we present the multibrane solution of the
model and review the 4-D effective action.
\S\ref{sec:twobrane} reviews and elaborates on the two brane 
model of Ref\cite{ira}. \S\ref{sec:ccandhier}
considers a two brane compactified 
model where the Randall-Sundrum solution to the hierarchy problem is 
incorporated together with an exponentially small effective 4-D
cosmological constant.
\S\ref{multibraneoverview} gives a general analysis of the multibrane world.
\S\ref{sec:globalstructure} 
discusses the global structure of the metric. In \S\ref{sec:gmh},
we discuss the implications of the particle horizons on the determination
of $G_N$ and the mass hierarchy issue. \S\ref{sec:corrnewton}
considers the correction
to Newton's gravity law in the case where the branes are sitting between 
two different AdS spaces. \S\ref{sec:discussion} contains some overall 
discussions and \S\ref{sec:summary} gives a brief summary. 
Some of the details are relegated to 
the appendices.  Appendix~\ref{app:eqnofmotion} solves the brane equation 
of motion for a stationary brane.  Appendix~\ref{app:twobranedetails} derives 
some details of the two brane compactified model.  Appendix~\ref{app:coords} 
gives the details of the coordinate transformation used in the analysis of 
the global structure.  For the visible brane in Ref\cite{RS1} or the 
two brane compactified model, a naive determination of Newton's constant
using the Hubble constant approach sometimes yields a wrong result.
In Appendix~\ref{app:GNandLE}, we comment on the 
ambiguity/problem with this naive approach. Appendix~\ref{app:newtonsforcelaw} 
gives a Feynman diagram analysis of Newton's force law on the relation between 
different metrics.

\section{Bulk Cosmological Constants as Integration Constants}
\label{sec:integconst}

One of the key points underlying the two brane model\cite{ira} is that 
the bulk cosmological constant is not an input parameter of the model, but an 
``integration constant'' to be determined  
by the 5-D Einstein equation for the bulk 
together with the boundary conditions at the branes. 

Suppose the 5-D theory discussed above arises from the compactification
of a higher dimensional theory. Then
the bulk cosmological constant, the brane tensions, as well as
the brane charges depend on the various compactification radii.
If we do not fix the compactification volume but instead make the
compactification radii dynamical, this will provide a way to
adjust, among other quantities, the bulk cosmological constant.
This means that the bulk cosmological constant becomes dynamical. 
We will treat possible dynamical models for the brane world elsewhere.

Here we shall review two approaches in which this ``integration constant'' 
appears:
\setcounter{Lcount}{0}
\begin{list}{(\roman{Lcount})}
  {\usecounter{Lcount}}
  \item unimodular gravity\cite{Einstein,Ng,Zee}, which is suitable for the 
    two brane model;
  \item $5$-form field strength model\cite{aurilia,fst1,fst2,fst3}, which 
    is suitable for the multibrane model, either compactified or 
    uncompactified, but not suitable for the orbifold version.
\end{list}
We then present a generalized model. Special cases of this generalized 
model reduce to the $5$-form field strength model and the fully 
covariantized variation of the unimodular gravity\cite{HT}. To adapt 
it for the multibrane orbifold model, we introduce brane couplings so that
the bulk cosmological constant becomes piecewise constant.
We note that $5$-form field strength also appears in gauged supergravity 
and/or superstring realizations of the RS scenario\cite{stringMfive}.

In 4-D gravity, when the cosmological constant is an integration constant,
it is typically left undetermined.  In the multibrane model, the bulk
5-D cosmological constant is piecewise constant, determined by the 5-D 
Einstein equation, including the jump conditions at the branes and the 
brane equations of motion.  As we shall see, the number of such adjustable 
bulk cosmological constants should be equal to the number of bulks between 
branes,  so, for $N$ branes in either the uncompactified case or the 
$S^1/\Bbb{Z}_2$ orbifold case, we require $N-1$ adjustable bulk 
cosmological constants.  In the compactified version, we need $N$ 
adjustable bulk cosmological constants.

\subsection{Unimodular Gravity}
\label{sec:unimod}

It was first pointed out by Einstein\cite{Einstein} that the 
cosmological constant could arise as an integration constant if the
original equations of general relativity were replaced by their trace-free
forms.  Einstein\cite{Einstein} had in mind traceless source terms due to
electromagnetic fields alone, but the same traceless field equations are
also obtained in ``unimodular gravity''\cite{Ng,unruh}, where the conformal
part of the metric is constrained and is not allowed to vary.
The fields in this theory consist of a metric $g_{ab}$, a fixed,
background, non-dynamical volume form ${\hat \epsilon}_{a_1a_2 \ldots
a_d}$, and other matter fields.  
The action of the theory is the  
standard Einstein action that depends only on the metric and the
matter fields.  However, only metrics $g_{ab}$ whose volume forms
coincide with the background volume form ${\hat \epsilon}$ are allowed.  
(An equivalent description of the restriction on the space of metrics is
that only metrics whose determinants in a particular fixed coordinate
system are $-1$ are allowed \cite{Ng}).
Unimodular gravity is consistent with
a massless spin-$2$ graviton propagating with $((d-1)(d-2)-2)/2$ 
polarizations in the linearized version of gravity in 
$d$-dimensions\cite{Ng}. 
Varying the action over the allowed class of metrics
yields in $d$ dimensions only the traceless part of Einstein's equation,

\be
\label{Traceless_Einstein_Eq}
R^{ab}-\frac{1}{d}g^{ab}R=\kappa^2\left(T^{ab}-\frac{1}{d}g^{ab}T\right)
\ee
where $T=T^a_a$.

The standard argument of invariance of the action under arbitrary
linearized coordinate transformations still applies and shows that
energy-momentum tensor satisfies the usual conservation law,
\be
\label{T_conservation}
\nabla_a \, T^{ab}=0.
\ee
and of course the Bianchi identities,
\be
\label{Bianchi}
\nabla_b \left(R^{ab}-\frac{1}{2}g^{ab}R\right)=0,
\ee
still hold. 
Combining Eqs.\ (\ref{Traceless_Einstein_Eq}) -- (\ref{Bianchi}) gives
\be
\nabla_a \left[ (d/2-1) R + \kappa^2 T \right] = 0.
\label{tracelessconst}
\ee
Equation (\ref{tracelessconst})
implies that $(d/2-1)R+\kappa^2T$ 
is a constant, which we will denote by $\kappa^2 \Lambda d$. 
As was noted by Einstein\cite{Einstein}, 
when the relation $(d/2-1)R + \kappa^2 T = \kappa^2 \Lambda d$ is inserted in
Eq.~(\ref{Traceless_Einstein_Eq}), the result is the full Einstein
equation with cosmological constant $\Lambda$: 
\be
G^{ab} + \kappa^2 \Lambda g^{ab} = \kappa^2 T^{ab}.
\ee
In this model, the cosmological constant $\Lambda$ is
merely an integration constant, and
has nothing to do with any input parameter in the action, the 
microphysics, or the (quantum) vacuum fluctuations. 
(Note that we may absorb any cosmological constant term inside 
$T^{ab}$ into $\Lambda$.)
In a set-up where $\Lambda$ is not determined, 
it becomes a free parameter. 
In standard 4-D gravity, this altered status of 
the cosmological constant does not help 
to explain its smallness.
However, in the two brane world scenario
in 5-D, $\Lambda$ is no longer a free parameter. Rather, it is the 
bulk cosmological constant determined by the 5-D Einstein equation 
and the boundary (jump) conditions at the branes\cite{ira}

The presence of a non-dynamical background field in unimodular gravity
is of course an unattractive feature of the theory.  However, it is
possible to find other theories with no background fields which,
classically and on-shell, are equivalent to unimodular gravity \cite{Zee}.
For example, one can treat $\Lambda$ as a dynamical field and 
introduce an appropriate Lagrangian multiplier (namely, a vector field)
to render it spacetime-independent\cite{HT}.  This theory is a special
case of our generalized model of Sec.\ \ref{sec:genmodel} below.

\subsection{$5$-Form Field Strength}
\label{sec:fiveform}

Another way to get a piecewise constant bulk 
cosmological constant is to introduce $(d-1)$-form fields 
\cite{aurilia,fst1,fst2,fst3} in d-dimensional spacetime.  This scenario 
is natural in compactified versions in string/M theory, in which 
$D$-branes are charged.  This permits multibrane scenarios with
the cosmological constant taking different constant values in different 
regions of the spacetime.  The discontinuities are due to charged 
$(d-2)$-branes, which act as sources for the $(d-1)$-form field.
A worldvolume action for these branes should, therefore, contain a 
Wess-Zumino term: the integral of the pullback of the $(d-1)$-form potential.
In the multibrane world, the action $S$ consists of a bulk action 
containing gravity,with metric $ {g}_{ {a} {b}}$, coupled to a $(d-1)$-form potential
$ {A}_{(d-1)}$, with field strength ${F}_{(d)}=d{A}_{(d-1)}$, and a set of standard 
worldvolume actions of
$({d}-2)$-branes containing the above-mentioned WZ terms with
dynamical coordinate fields ${X}_{n}^{{a}}$ 
This means the location of the $n$th brane in the bulk is given by
the embedding functions $x^a = X^a_n(\xi_n^\mu)$. 
In addition, the $n$th brane possesses a brane metric
$\gamma_n^{\mu\nu}$ which is a function of the brane 
coordinates $\xi_n^\mu$.(for $d=5$, $a=0,1,2,3,5$ and $\mu=0,1,2,3$).
The action for such a system is\cite{polch,fst3}
\baray
\label{Action}
&& S=\int d^dx\sqrt{\vert g\vert}\left[\frac{1}{2\kappa^2}R-\Lambda-\frac{1}{2\cdot d!}F^2_{\left(d\right)}\right] \nonumber \\
&& +\sum_n\left\{-\frac12\sigma_n\int d^{\left(d-1\right)}\xi_n\sqrt{\vert\gamma_n\vert}\left[\gamma_n^{\mu\nu}
\partial_{\mu}X^a\partial_{\nu}X^b g_{ab}\left(X_n\right)-C\right]\right. \nonumber \\
&&-\left.\frac{e_n}{\left(d-1\right) !}\int d^{\left(d-1\right)}\xi_nA_{a_1\cdots a_{d-1}}\left(X_n\right)
\partial_{\mu_1}X^{a_1}\cdots\partial_{\mu_{d-1}}X^{a_{d-1}}\epsilon^{\mu_1\cdots\mu_{d-1}} \right\}
\earay
where $\sigma_n$ and $e_n$ are the tension and the charge of the $n$th brane.
The dimensionless constant $C$ must have the value $p-1$ in order to
recover the correct equations of motion (see below).
The tensor density $\epsilon^{\mu_1 \ldots \mu_{d-1}}$ is totally 
antisymmetric with $\epsilon^{01 \ldots (d-2)}=1$. 
Varying the action $S$ with respect to the metric, we have
\baray
\label{eq:Einstein1}
&& G^{ab}=\kappa^2\left\{-\Lambda g^{ab}+\frac1{\left(d-1\right)!}\left(F^{ac_1\cdots c_{d-2}}F^{b}_{c_1\cdots c_{d-2}}
-\frac1{2d}g^{ab}F_{(d)}^2\right)\right. \nonumber \\
&& -\left.\sum_n\frac{\sigma_n}{\sqrt{\vert g\vert}}\int d^{\left(d-1\right)}\xi_n\sqrt{\vert\gamma_n\vert}
\left[\gamma_n^{\mu\nu}\partial_{\mu}X^a\partial_{\nu}X^b\delta^{\left(d\right)}\left(x-X_n\right)\right]\right\}.
\earay
Varying the action $S$ with respect to ${A_{(d-1)}}$, the brane metric 
$\gamma_{\mu \nu}$ and the coordinates $X^a$, we have, respectively,
\baray
\label{eq:potential}
&& \sqrt{\vert g\vert}\nabla_aF^{a,b_1\cdots b_{d-1}}-\sum_n e_n \int d^{\left(d-1\right)}\xi_n\partial_{\mu_1}X^{b_1}\cdots
\partial_{\mu_{d-1}}X^{b_{d-1}}\epsilon^{\mu_1\cdots\mu_{d-1}}\delta^{\left(d\right)}\left(x-X_n\right)=0
\earay
\baray
\label{eq:wmetric}
&& \gamma_{n\mu\nu}=\partial_{\mu}X^a_n\partial_{\nu}X^b_ng_{ab}\left(X_n\right)
\earay
and the brane equation of motion for the $n$th brane  
\baray
\label{eq:wscalars}
&& \sigma_n \sqrt{\vert\gamma_n\vert}\left[\nabla_{\mu}\nabla^{\mu}X_n^a+\Gamma^a_{bc}\partial_{\mu}X^b_n\partial_{\nu}X^c_n
\gamma_n^{\mu\nu}\right]+\nonumber \\
&& \frac{e_n}{\left(d-1\right)!}F^a_{b_1\cdots b_{d-1}}\partial_{\mu_1}X_n^{b_1}\cdots
\partial_{\mu_{d-1}}X_n^{b_{d-1}}\epsilon^{\mu_1\cdots\mu_{d-1}}=0.
\earay
Eq.~(\ref{eq:wmetric}) simply states that the worldvolume metrics are
those induced on the worldvolumes by the embedding coordinates
$ {X}_{n}^{{a}}$. In equation Eq.~(\ref{eq:wmetric}) there is a 
coefficient $C/\left(p-1\right)$, for a $p$-brane ($p=3$ in our case), 
which we have set equal to 1, {\em i.e.}, $\gamma_{\mu\nu}$ is the 
induced metric on the brane. 
Had we written the brane tension term in the Nambu-Goto action form, the 
same equations would have resulted.  In the bulk
\be
\partial_{a}\left(\sqrt{|g|} {F}^{a {c}_{1}\cdots {c}_{d-1}}\right)=0
\ee
which has solution
\be 
{F}^{{a}_{1}\cdots {a}_{d}}=\frac{e}{\sqrt{|g|}}\epsilon^{
{a}_{1}\cdots {a}_{d}},
\ee
where $e$ is a constant that can vary from one interbrane region to
another, and the tensor density $\epsilon^{{a}_{1}\cdots {a}_{d}}$ is 
totally antisymmetric with $\epsilon^{01 \ldots (d-1)}=1$.
Inserting this solution in Eq.~(\ref{eq:Einstein1}) 
we see that the only contribution of the field strength is to the bulk 
cosmological constant $\Lambda$.
In this paper we are intersted in the parallel stationary 3-branes case,
so we may choose the Gaussian normal coordinate:
$ds^2=g_{\mu\nu}\left(x,y\right)dx^{\mu}dx^{\nu}+dy^2$.
The brane equation of motion (\ref{eq:wscalars}) for a stationary brane 
is discussed in Appendix~\ref{app:eqnofmotion}.
Using worldvolume reparametrization
invariance we can set $d-1$ coordinates (static gauge) to the values
$ {X}^{a}_{n} = \delta^{a}_{\mu}\xi_{n}^{\mu}$. 
The remaining coordinate, namely $X^d\equiv y$, is along the $d$th direction,
\be
\label{eq:ansatzwscalar}
 {X}^{d}_{n}\equiv L_{n}\, ,
\ee
where the $L_{n}$'s are constants. We can
perform the volume integrals in
Eqs.~(\ref{eq:Einstein1},\ref{eq:potential}) leaving only 1-dimensional
delta functions $\delta(y-L_{n})$.  Also, this implies for the
worldvolume metrics that: $\gamma_{n\mu\nu}=g_{\mu\nu}\left(L_n\right)$.
This can also be seen by solving the equation 
Eq.~(\ref{eq:wmetric}) in the Gaussian normal system of coordinates 
in the bulk. Since the branes are parallel, the geodesics orthogonal to one brane
will be orthogonal to all the branes. We obtain:
\be
 \gamma_{n\mu\nu}=\partial_{\mu}\xi_n^{a}\partial_{\nu}\xi_n^{b}g_{ab}
\left(X_n\right)=g_{\mu\nu}\left(L_n\right)
\ee
where we have taken $g_{5\lambda}=0$ for Gaussian normal coordinates.
It follows that $\sqrt{\vert\gamma\vert}=\sqrt{\vert g\vert}$, and 
(\ref{eq:potential}) becomes
\be
\sqrt{\vert g\vert}\nabla_aF^{a,b_1\cdots b_{d-1}}-
\sum_n e_n\epsilon^{b_1\cdots b_{d-1}}\delta\left(y-L_n\right)=0.
\ee
The determinant of the metric is a continuous function everywhere in 
the space, so as we go across the $n$th brane with charge $e_n$,
the field strength jumps by $e_n$, from $e_{(n)}$ to $e_{\left(n+1\right)}=
e_{(n)}+e_n$.
In the multibrane world, the resulting cosmological constant in the 
bulk between the $(n-1)$th and the $n$th branes is given by
\be
        -\Lambda_n = \Lambda + \frac12 e_{(n)}^2=
\Lambda + \frac12\left(e_{(0)}+ e_0+e_1+...+e_{n-1}\right)^2,
\ee
where $e_{(0)}$ is a constant background field strength.
Now the Einstein equation (\ref{eq:Einstein1}) becomes
\baray
\label{EinsteinTensor}
  G^{ab}=\kappa^2\left[\sum\limits_{n=0}^N 
    \Lambda_n\Theta(y-L_{n-1})\Theta(L_n-y) g^{ab}
    -\sum\limits_{n=0}^{N-1}\frac{\sigma_n}2\delta^a_{\mu}
    \delta^b_{\nu}\gamma^{\mu\nu}
    \delta\left(y-L_n\right)\right]
\earay 
where $\Lambda_n>0$ for AdS space, with $-L_{-1}=L_{N}=\infty$ in the 
uncompactified case.  The multibrane world with an uncompactified $5$th
dimension is shown in Figure~1. 

\newcounter{figcount}
\setcounter{figcount}{1}
\setcounter{multibraneseps}{\value{figcount}}
\begin{center}
  \epsfbox{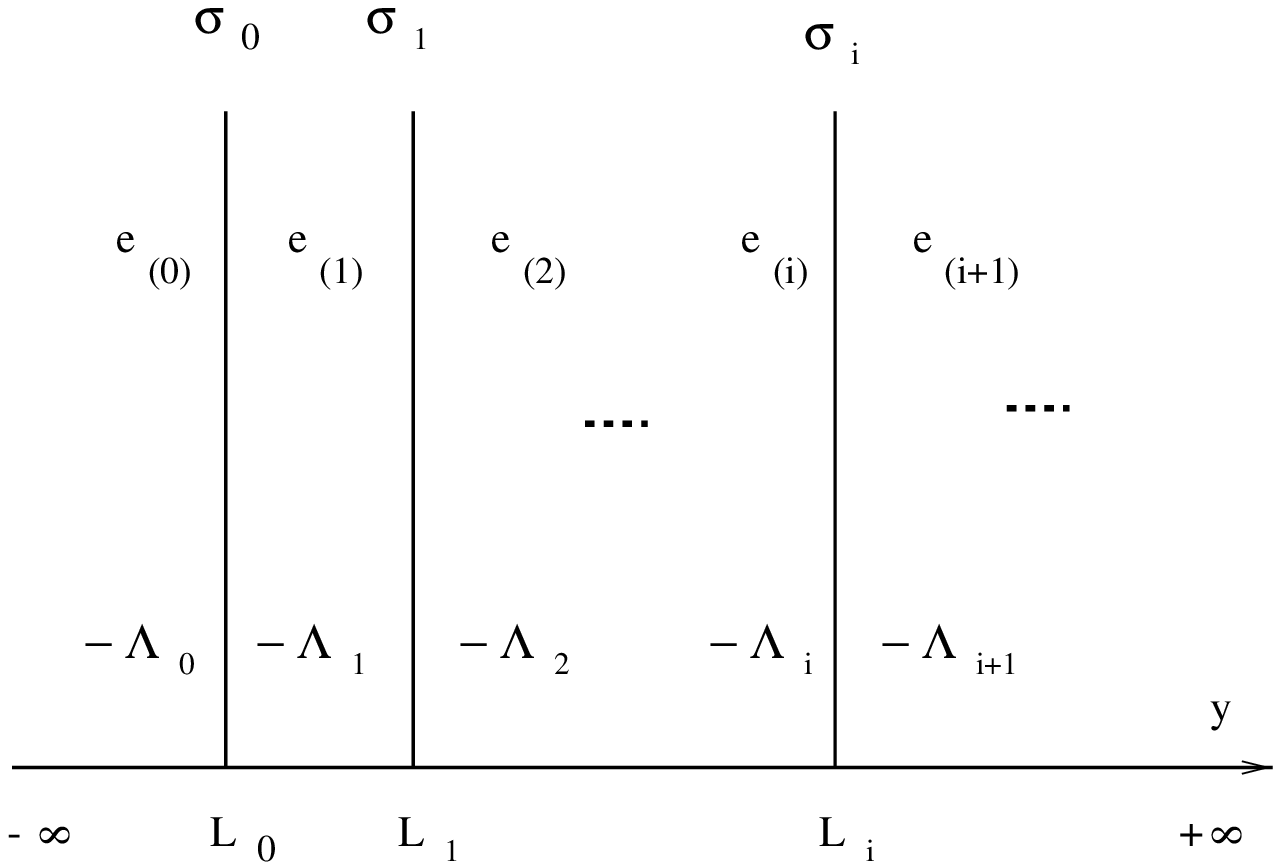}
  \parbox{12cm}{\vspace{1cm}
    FIG. \thefigcount \hspace{2pt} Schematic multibrane setup.}
\end{center}

Although the charges $e_n$ of the branes are taken to be fixed, 
the background field strength $e_{(0)}$ is an integration constant
(to be determined by the Einstein equation.)
To have AdS spaces in the bulk, we require $\Lambda$ to be negative 
enough so that all $\Lambda_n>0$.
Since the $(d-1)$-form field has no dynamical degrees of freedom,
it is consistent to flip the $F^2$ term in the action $S$ so that the 
field strength contributes only negatively to the bulk cosmological constant, 
avoiding the need to introduce a negative $\Lambda$.
However, if the $(d-1)$-form field arises as a component of an 
antisymmetric field in higher dimensional space, then 
positivity of its higher dimensional kinetic energy density 
does not allow this freedom.

Suppose the $d$th dimension ({\em i.e.}, the $y$ direction) is compactified
in the $N$ $3$-brane scenario.
Boundary condition requires the sum of the brane charges to be zero, 
that is, $e_0+e_1+e_2+...+e_N=0$, independent of the value of $e_{(0)}$. 
If the model is a $S^1/\Bbb{Z}_2$ orbifold, the symmetry constraint requires 
that $e_{(0)}=0$. 
In this case, the bulk $\Lambda_n$ are interrelated, so we have 
to find another way to allow $\Lambda_n$ to be adjustable.

In the two brane world, $e_{(0)}$ is determined by the Einstein equation.
In the multibrane world, we may need more freedom than one integration 
constant.
To achieve this, we can consider a scenario where there is more than 
one $(d-1)$-form field.  Considering the case $d=5$, the introduction of 
more than one $4$-form field is quite natural in string/M theory. To be 
specific, consider M theory which has $2$-branes and $5$-branes, which 
are electrically and magnetically charged, respectively,
under the $3$-form field 
$A_{(3)}$. 
The dual of this $3$-form field is a $6$-form field $A_{(6)}$.  If 6 of 
the 11 dimensions of M theory are compactified toroidally, 
then of the $6$-form field $A_{(6)}$ there are, among other fields,
15 $4$-form fields $A^J_{(4)}$ ($J=1,2,...,15$) in the uncompactified 
5-D spacetime. The $3$-branes in this remaining 5-D spacetime are 
$5$-branes of the 11-D theory with two of their spatial dimensions 
compactified.  There are 15 types of such $3$-branes, each charged under 
one $4$-form field $A^J_{(4)}$.  In more realistic compactifications, 
some of these $4$-form fields may be projected out, although generically, 
we expect a number to remain.

Let each $3$-brane be a stack composed of these compactified $5$-branes.
Suppose we have $M$ $4$-form fields $A^J_{(4)}$ and their corresponding 
field strength $F^J_{(5)}$, $J=1,2,.....,M$, under which the $n$th $3$-brane 
has charge $e^J_n$, $J=1,2,.....,M$.  Then the resulting cosmological 
constant in the bulk between the $(n-1)$th and the $n$th branes is given by
\be
-\Lambda_n = \Lambda + \frac12\sum_J (e^J_{(n)})^2
=\Lambda+\frac12\sum_J\left(e^J_{(0)}+e^J_0+e^J_1+...+e^J_{n-1}\right)^2.
\label{manycharges}
\ee
Again we require $\Lambda_n>0$ for AdS spaces between branes.
Now there are $M$ background field strengths $e^J_{(0)}$, $J=1,2,.....,M$, 
that are integration constants to be determined.
Depending on the charges of the brane, this will allow some number 
of bulk cosmological constants to be adjusted to satisfy the Einstein 
equation.

\subsection{A Generalized Model}
\label{sec:genmodel}

We are interested in models where the bulk cosmological constant is
piecewise constant, with each piece an integration constant that will
be determined by the Einstein equation.
The cosmological constant in the unimodular model is the same everywhere.
This scenario is suitable for the two brane model, but is inadequate for
the multibrane world when there are more than two branes.
The model with $M$ types of $5$-form field strengths is suitable for
the $(M+1)$-brane model, in the uncompactified case, or the $M$-brane
model, in the compactified case. However, it is not suitable for the
$S^1/\Bbb{Z}_2$ orbifold model, since the background $5$-form field strength
must be set to zero due to the $\Bbb{Z}_2$ symmetry. As a consequence,
the bulk cosmological constants are fixed,
so it is useful to consider scenarios where the bulk cosmological constant
can be piecewise adjustable. Here we present such a model, which is a
generalization of the above models.

Let us first consider a model without branes, which includes a scalar field 
$\phi$ and its effective potential, $\Lambda(\phi)$:
\be
\label{initial_action}
S=\int d^{\left(d\right)}x\sqrt{\vert g\vert}\left\{\frac1{2\kappa^2}R
-\Lambda\left(\phi\right)-T^a\partial_a\phi \right\}.
\ee
The independent variables of the model are $g_{ab}(x),\;T^a(x)$ and 
$\phi(x)$;  the Einstein equation is
\be
\label{metric_var}
G_{ab} = \kappa^2 \left[-\Lambda\left(\phi\right)
    - T^c\partial_c\phi\right] g_{ab}.
\ee
The variations of $S$ with respect to $T^a$ and $\phi$ give the following
equations:
\be
\label{T_var}
\frac{\delta S}{\delta T^a}=0 
\Longrightarrow \sqrt{\vert g\vert}\partial_a\phi =0
\ee
\be
\label{phi_var}
\frac{\delta S}{\delta \phi}=0 \Longrightarrow \partial_a\left(\sqrt{\vert 
g\vert}T^a\right)=\sqrt{\vert g\vert}\frac{\partial \Lambda}{\partial \phi}.
\ee
Eq.~(\ref{T_var}) implies that $\phi\left( x\right)=$ constant, so
that (\ref{metric_var}) becomes
\be
G_{ab}=-\Lambda\left(\phi=constant\right)g_{ab}
\ee
where $\Lambda(\phi)$ is an integration constant. 
Let us now consider special choices of $\Lambda(\phi)$. For example, if
\be
\Lambda\left(\phi\right)=\frac12\phi^{2},
\ee
the action $S$ reduces to the $d$-form 
field strength case we have already discussed\cite{aurilia}.
To see this, 
define $T^a$ as the dual of a $\left(d-1\right)$ form $A_{\left(d-1\right)}$:
\be
T^a=\frac{\epsilon^{ab_1\cdots b_{d-1}}}{(d-1)!\sqrt{\vert g\vert}}
A_{b_1\cdots b_{d-1}}.
\ee
Then Eq.~(\ref{phi_var}) becomes
\be
 \sqrt{\vert g\vert} \phi= \frac{\epsilon^{ab_1\cdots b_{d-1}}}{(d-1)!} 
\partial_a\ A_{b_1\cdots b_{d-1}}= \frac{\epsilon^{a_1
\cdots a_d}}{d!} F_{a_1\cdots a_d},
\ee
and substituting this back into the action (or equivalently, perform the 
functional integration over $\phi$ in the generating functional) 
and using  $(\epsilon^{a_1\cdots a_d} F_{a_1\cdots a_d})^2 
=-(\vert g\vert)d!F_{a_1\cdots a_d}F^{a_1\cdots a_d}$, gives
\be
S=\int d^{\left(d\right)}x\sqrt{\vert g\vert}\left\{\frac1{2\kappa^2}R-
\frac1{2 \cdot d!}F_{a_1\cdots a_d}F^{a_1\cdots a_d}\right\}.
\ee
Alternatively, if we choose
\be
\Lambda\left(\phi\right)=\phi
\ee
we recover the model of Ref\cite{HT}, which may
be considered as a covariantized version of the unimodular gravity.
The value of $\phi$ can be arbitrary in the orbifold model.
We may also combine the two approaches to obtain an action $S$
\be
S=\int d^{\left(d\right)}x\sqrt{\vert g\vert}\left\{\frac1{2\kappa^2}R-
\frac1{2 \cdot d!}F_{a_1\cdots a_d}F^{a_1\cdots a_d}
-\phi - T^a\partial_a\phi \right\}.
\label{mixedS}
\ee
We can easily further generalize this formalism to $M$ fields $T^a_J$
and $\phi_J$, $J=1,2\cdots, M$ and couple them to the
$\left(d-2\right)$-branes.
\baray
&& S=\int d^{\left(d\right)}x\sqrt{\vert g\vert}\left\{\frac1{2\kappa^2}R-
\sum_{J=1}^M \phi_J + T^a_J\partial_a\phi_J\right\}+\nonumber \\
&& \sum_i-\frac{\sigma_i}{2}\int d^{\left(d-1\right)}\xi\sqrt{\vert\gamma_n
\vert}\left[\gamma_n^{\mu\nu}\partial_{\mu}X^a\partial_{\nu}X^b g_{ab}
\left(X_n\right)-C\right]+\nonumber \\
&& \sum_{i,J}\mu_i^J\int d^{\left(d-1\right)}\xi \, \sqrt{|g|} \,
T^a_J\epsilon_{ab_1\cdots   
b_{d-1}}\partial_{\mu_1}X^{b_1}\cdots
\partial_{\mu_{d-1}}X^{b_{d-1}}\epsilon^{\mu_1\cdots\mu_{d-1}}.
\label{HTcouple}
\earay
If the $n$th $3$-brane has charges $\mu^J_n$, $J=1,2,.....,M$, then the 
resulting cosmological constant in the bulk between the $(n-1)$th and the 
$n$th branes is given by
\be
-\Lambda_n =  \sum_J \phi_n^J = 
\sum_J\left(\phi^J + \mu^J_0 +...+ \mu^J_{n-1}\right).
\label{manyTcharges}
\ee
Again we require $\Lambda_n>0$ for AdS spaces between branes.
There are $M$ background constants $\phi^J$, $J=1,2,.....,M$,
that are integration constants to be determined.  In this case, only the sum 
of $\phi$'s is determined by the Einstein equation,
so only one bulk cosmological constant can be adjusted to satisfy the Einstein
equation.
Here, $\Lambda(\phi)=\phi$ can be either positive or negative.
We can combine one $\phi$ field with $M$ $4$-form potentials.
We shall comment on the dynamics of this system when applied to
the static multibrane solution.

\section{Solution to Einstein Equation and General Formalism}
\label{sec:multibrane}

We consider a multibrane world scenario.  The objective is to provide a
model which concurrently solves both traditional fine tunings, the 
hierarchy and the cosmological constant problems. We first give the 
general multibrane solution and the 4-D effective action formalism.

\subsection{Multibrane World Solution}

Let us first recall the set-up of the static multibrane world.  
Consider $N$ parallel $3$-branes, located at $y = \li , i = 0,...N-1$.
The brane at $y=\li$ has brane tension $\sigma_i$. Unless stated 
otherwise, 
the brane tensions 
are taken to be positive, except for branes at orbifold fixed points, 
for which we allow the possibility of negative brane tensions.
In the uncompactified case, there are $N+1$ AdS bulk spaces, with 
5-D energy-momentum 
tensor $T_{ab} = \Lambda_i g_{ab}$ in the bulk between
the $(i-1)$th and the $i$th branes.
We start with the metric
\be\label{metric}
ds^2=dy^2+A(y)[-dt^2+\exp(2Ht)\delta_{ij}dx^idx^j].
\ee
Note that a rescaling of $(t, x^i)$ rescales both $A(y)$ and $H$, but leaves
$H^2/A(y)$ invariant, so we have the freedom to fix the overall 
normalization of $A(y)$.
The $G_{05}$ component of the Einstein's equation
$G_{ab}=\kappa ^2T_{ab}=8{\pi }G_{5} T_{ab}$ (see for example Eq.(\ref{EinsteinTensor})) 
is satisfied trivially, while the $G_{00}$ and the 
$G_{55}$ components give, respectively,
\baray
{A^{\prime\prime}\over A}&=&{2H^2\over A}+{2\kappa^2\over 3}
\left[\sum\limits_{i=0}^{N}\Li\Theta(y-\lm)\Theta(\li-y)
-\sum\limits_{i=0}^{N-1}\sigma_i\delta(y-\li)\right]\nonumber\\
\biggl({A^\prime\over A}\biggr)^2&=&{4H^2\over A}
+{2\kappa^2\over 3}\left[\sum\limits_{i=0}^{N}
    \Li\Theta(y-\lm)\Theta(\li-y)\right]
\label{Einsteinbig}
\earay
(where $-L_{-1} = L_{N} = \infty$ in the uncompactified case). The $G_{ij}$ 
components do not yield any additional equations. It is convenient to define 
$q_i\equiv\kappa^2\sigma_i/3$ and $\ki\equiv\sqrt{\kappa^2\Lambda_i/6}$,
where both have mass dimension 1.

Integrating the $G_{00}$ component of Einstein's equation across 
the $i$~th brane yields the Israel jump condition\cite{Israel}:
\baray
\lim_{y\to\li+}\biggl({A^\prime\over A}\biggr)
-\lim_{y\to\li-}\biggl({A^\prime\over A}\biggr)
&=&-{2\kappa^2\sigma_i\over 3}
\equiv -2q_i.
\label{Israel}
\earay
In general, there is also an equation of motion for each brane; these 
are discussed for the $5$-form field strength case in 
Appendix~\ref{app:eqnofmotion}.  For stationary branes only the $5$th 
component of the equation of motion for the embedding coordinates is 
non-trivial.  Since the derivatives of the metric at the 
brane are discontinuous, we have to average over the two sides of the brane.
In the above metric ansatz, the equation for the $i$th brane reduces to
\be
\sigma_i \left[\lim_{y\to\li+}\biggl({A^\prime\over A}\biggr)
+\lim_{y\to\li-}\biggl({A^\prime\over A}\biggr)\right]
=\Lambda_{i+1} - \Lambda_i.
\label{branemot}
\ee
(see Appendix~\ref{app:eqnofmotion} for details).
In the $5$-form field strength model with $M$ $5$-form field 
strengths, Eq.~(\ref{manycharges}) gives
\be
 - \frac12 \sum_J e^J_i \left(e^J_{(i)} +e^J_{(i+1)}\right)
=\Lambda_{i+1} - \Lambda_i.
\label{someq}
\ee
This means that, in the stationary situation, the average of 
$A^\prime/A$ of each brane is simply the difference of the bulk 
cosmological constants on its two sides.  
In the symmetric case, 
{\em i.e.}, $\Lambda_{i+1} = \Lambda_i$, 
the limiting value of $A^\prime/A$ from the two sides are equal 
in magnitude but have opposite signs, as expected.
Eq.~(\ref{branemot}) suggests that equations analogous to Eq.~
(\ref{someq}) ought to hold for other models in which the
bulk cosmological constant emerges as an integration constant.
This is indeed the case for the generalized model (\ref{HTcouple}),
where
\be
\sigma_i \left[\lim_{y\to\li+}\biggl({A^\prime\over A}\biggr)
+\lim_{y\to\li-}\biggl({A^\prime\over A}\biggr)\right]
=\sum_J \mu^J_i
=\Lambda_{i+1} - \Lambda_i.
\ee
This completes the set-up of the problem.
Not surprisingly, solutions of the Einstein equation, Eq.~(\ref{Einsteinbig}),
for stationary branes automatically satisfy the brane equations of motion,
Eq.~(\ref{branemot}), when the jump conditions Eq.~(\ref{Israel})
are imposed.
The approach we shall take is to first solve the Einstein equation
(\ref{Einsteinbig}) in the bulk, then impose the jump condition 
(\ref{Israel}), and finally show that the brane Eq.~(\ref{branemot}) 
of motion (for each brane) is automatically satisfied.

Similar systems have been studied in Ref\cite{kaloper}.
Defining $\ki\equiv\sqrt{\kappa^2\Lambda_i/6}$, the solution to 
Einstein's equation (\ref{Einsteinbig}) in the bulks is
\be\label{scale}
A(y)={H^2\sinh^2[\ki(y-\yi)]\over\ki^2}\qquad(\lm<y<\li)
\ee
(as before $-L_{-1}=L_N=\infty$ in the uncompactified case).
Continuity of the 
metric at each of the branes imposes the constraints,
\be\label{boundary}
{\sinh^2[\ki(\li-\yi)]\over\ki^2}={\sinh^2[\kp(\li-\yp)]\over\kp^2},
\ee
and the jump condition Eq.~(\ref{Israel}) gives
\be\label{jumpn}
\qi = {\ki\cosh[\ki(\li-\yi)]\over\sinh[\ki(\li-\yi)]}
-{\kp\cosh[\kp(\li-\yp)]\over\sinh[\kp(\li-\yp)]}.
\ee
It is easy to check, using both the continuity and the jump equations, 
that the brane equation of motion (\ref{branemot}) is automatically 
satisfied for each brane.
The 4-D cosmological constant as seen by observers on the $i$th brane is 
\be
\label{hubbleconstdef}
  H^2(L_i)\equiv{H^2\over A(L_i)}={\ki^2\over\sinh^2[\ki(\li-\yi)]},
\ee
where $A(L_i)$ is the value of the warp factor at the $i$th brane {\em i.e.}
$A(L_i)=A(y=L_i)$.
Hence there is an inherent sign ambiguity in the definition of $H(L_i)$:
\be\label{hubble}
  H(L_i) = \pm{\ki\over\sinh[\ki(\li-\yi)]}.
\ee
Since $\ki>0$ by definition and we assume $H(L_i)>0$, the sign we must choose
is determined by the sign of $\sinh[\ki(\li-\yi)]$, or, equivalently,
the sign $s_i^-$ of $A^\prime$ at $y\to\li^-$. Inverting this 
expression, we can express the separation between two consecutive branes,
$\Di\equiv\ki(\li-\lm)$, in terms of the Hubble constants on those branes:
\be\label{eq:separations}
  \Di = -s_{i-1}^+\sinh^{-1}\left({\ki\over H(L_{i-1})}\right)
    +s_i^-\sinh^{-1}\left({\ki\over H(L_i)}\right)
\ee
where $s_{i-1}^+$ is the sign of $A^\prime$ at $y\to L_{i-1}^+$.

\addtocounter{figcount}{1}
\setcounter{bulktypeseps}{\value{figcount}}
\begin{center}
  \epsfbox{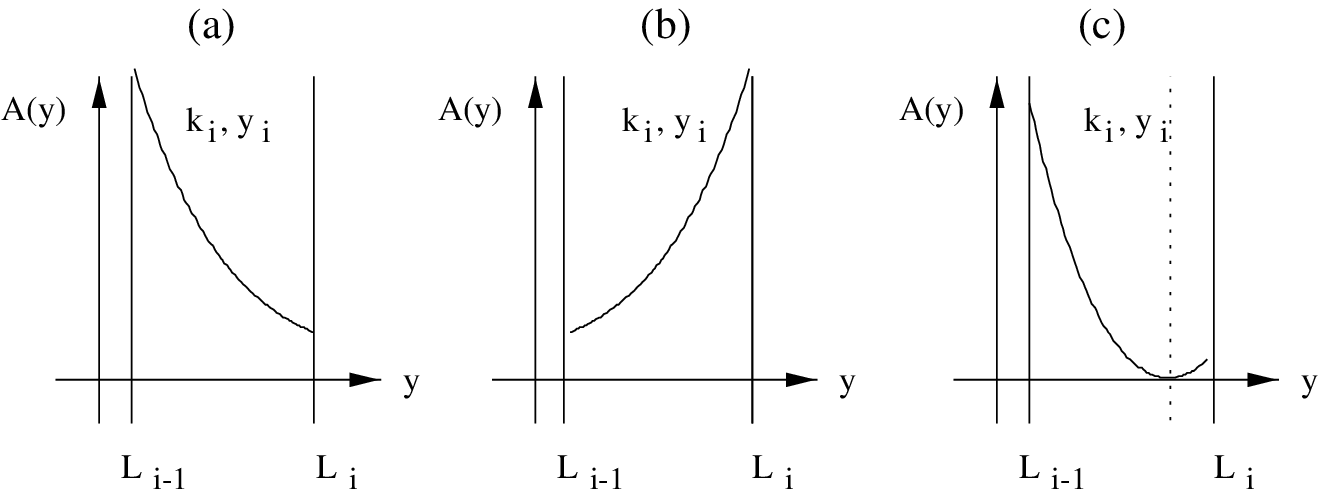}
  \parbox{12cm}{\vspace{1cm}
    FIG. \thefigcount \hspace{2pt} Schematic diagram of the possible 
    types of behavior of the metric coefficient $A(y)$ in the bulk 
    between the branes. \vspace{12pt}}
\end{center}

The metric coefficient $A(y)$ can have three different types of behaviors
in the bulk between the branes, as illustrated in Figure~2.
We see from Eq.~(\ref{scale}) that $A(y)$ may decrease monotonically, 
increase monotonically, or must vanish at an intermediate
$y$, which corresponds to a coordinate singularity of the metric, Eq.~
(\ref{metric}), a particle horizon.  Such horizons are discussed further 
in Sec.\ref{sec:globalstructure}.

Because the definition of $\Di$ incorporates both the brane
separation and $\ki$, these equations can be thought of as determining 
$(\li-\lm)$ from $H(L_i)$ and $H(L_{i-1})$ for given $\ki$, or as 
determining $\ki$ from $H(L_i)$ and $H(L_{i+1})$, with the brane 
separations governed by some higher energy dynamics.
In terms of the bulk cosmological constants on either side of the $i$th brane,
\be
\label{khubble}
  H^2(L_i) = {[\qi^2-(\kp+\ki)^2][\qi^2-(\kp-\ki)^2]\over4\qi^2},
\ee
in agreement with Ref.\cite{horace}.
Note that Eq.~(\ref{khubble}) appears to give the expansion rate on the
$i$th brane in terms of local quantities -- the bulk cosmological constants on either
side of the brane, and its tension -- but in the context of a full multibrane
spacetime, these quantities may depend on the positions of the other branes,
and the locations of particle horizons, which are not necessarily nearby.
The form of the metric in Eq.~(\ref{metric}) assumes a constant $H$, but 
when matter density $\rho$ on the brane is included, one can easily
solve for $H^2$ on a particular brane without solving for the  
metric in all regions of spacetime\cite{oldcosmo,cosmo1,cosmo,horace}.
The result is simple: one simply replaces $q_i \to q_i + \kappa^2 \rho/3 $. 
This result provides a determination of the 4-D Newton's constant $G_N$
by requiring $H^2$ to have the standard form\cite{oldcosmo}
\be
H^2 \approx 8 \pi G_N (\Le + \rho + ...)/3.
\label{LeGN}
\ee
As we shall see later, this way of determining $G_N$ is valid only when
$A(y)$ is peaked at the particular brane where the expansion rate is
evaluated.

\subsection{The 4-D Effective Action}
\label{sec:effaction}

Consider the 5-D action ${\bf S}^{(5)}$ for the 5-D gravity plus 
scalar fields confined on branes: 
\be
{\bf S}^{\left(5\right)}=\int dyd^4x \sqrt{-g}\left[\frac{R^{\left(5\right)}}
{2\kappa^2} + \Lambda + \sum_i \delta (y-L_i)\sqrt{g^{55}}
(-\sigma_i + {\bf L}_i)\right]+ 
\int_{\Sigma}\frac{K\sqrt{-\gamma}}{\kappa^2}d^4x
\ee
where the bulk cosmological constant $\Lambda$ ($\Lambda>0$ for AdS)
is piecewise constant, $\sigma_i$ is the $i$th brane tension, 
the surface term is the Hartle-Hawking term, needed for theories with 
boundaries, and
\be 
{\bf L}_i = \frac12 g^{\mu\nu}\phi_{i,\mu} \phi_{i,\nu} -
\frac12M_{Hi}^2\phi_i^2-\lambda_i\phi_i^4-\frac{\phi_i^6}{M_{ci}^2}
+\cdots
\ee
where the scalar field $\phi_i$ is confined to the brane at $L_i$.
Recall that $8\pi G_5 = \kappa^2 = 8\pi/M^3$.  
Generically, the scalar field masses $M_{Hi}$ 
(which are real after spontaneous symmetry breaking) and the cut-off 
masses $M_{ci}$ are expected to be
comparable to $M$ in order of magnitude. The characteristic scales 
brane tensions and the bulk cosmological constant are also 
expected to be set by $M$, {\em i.e.},
$\sigma_i\sim M^4$ while $\Lambda\sim M^5$.
The couplings $\lambda_i$ are of order unity.

Following Ref\cite{RS1}, we can integrate out the $y$ direction to 
obtain an effective 4-D theory. 
As we shall discuss in more detail, the zeros of the metric $A(y)$
are particle horizons. That is, for an observer on a brane, 
it takes infinite time for a light-like signal to travel from the brane to 
the particle horizon.
This implies that the integration over $y$ is only over 
the region between particle horizons 
that includes the position $L$ of the visible brane. 
We can decompose the $\mu \nu$ components of the 5-D Ricci tensor
$R_{\mu \nu}^{\left(5\right)}$ into the 4-D Ricci tensor
$R_{\mu \nu}^{\left(4\right)}(\hat\gamma_{\mu \nu})$ and the
extrinsic curvature $K_{\mu\nu}$ (and $K=g^{\mu \nu} K_{\mu \nu}$):
\be
R_{\mu \nu}^{\left(5\right)}= R_{\mu \nu}^{\left(4\right)}-
g^{55}\partial_yK_{\mu\nu}-g^{55}K_{\mu\nu}K+2g^{55}K_{\mu}^{\lambda}
K_{\lambda \nu}.
\ee
where we recall the 5-D metric
\be
ds^2=A\left(y\right)\hat \gamma_{\mu\nu}\left(x\right)dx^{\mu}dx^{\nu}+dy^2
\ee
so that the pullback on the $L_i$ brane $\gamma_{\mu\nu} = g_{\mu\nu} = 
A(y=L_i)\hat\gamma _{\mu\nu} (x)$, and
$\sqrt{-g}=A^2\left(y\right)\sqrt{-\hat\gamma}$.
Using the fact that $K_{\mu\nu}=g_{\mu\nu}A^{\prime}/{2A}$
(where the prime indicates derivative with respect to $y$) 
and $K=2A^{\prime}/A$, this gives
\baray
&& \sqrt{-g}R^{\left(5\right)}=\sqrt{-\hat \gamma}\left[A(y) 
R^{\left(4\right)}- g^{\mu\nu}\partial_yK_{\mu\nu}-K^2+
2K^{\lambda\rho}K_{\lambda \rho}
+R_{55}\right]=\nonumber\\
&& \sqrt{-\hat \gamma} \left[A(y) R^{\left(4\right)}-(A(y)^{\prime})^2 
-4A(y)^{\prime\prime}A(y)\right].
\earay

We substitute $A(y)$ and its
derivatives into the 5-D action ${\bf S}^{\left(5\right)}$ and integrate
over $y$ to obtain the 4-D low energy effective theory.
This ``integrating out'' of the $5$th dimension 
in ${\bf S}^{(5)}$ yields the effective 4-D action 
${\bf S}^{\left(4\right)}$:
\baray
&& {\bf S}^{\left(4\right)}=\int d^4x \sqrt{-\hat\gamma}
 R^{\left(4\right)}/{2\kappa_N^2} - \int d^4x \sqrt{-\hat\gamma} \Le
+ \nonumber \\
&& \sum_i \int d^4xA^2\left(L_i\right)\sqrt{-\hat\gamma}
\left[\frac1{2A\left(L_i\right)}\hat\gamma^{\mu\nu}\phi_{i,\mu} \phi_{i,\nu}-
\frac12M_H^2\phi_i^2-\lambda_i\phi_i^4-
\frac{\phi_i^6}{M_{ci}^2} +\cdots\right].
\earay
We see that the 4-D gravitational coupling is given by
\be
\label{GNdefine}
 \frac{1}{2\kappa_N^2} = \frac{1}{2\kappa^2} \int A(y) dy.
\ee
Suppose that $A(y)$ is peaked at the $i$th brane, namely the Planck brane,
it is convenient to fix the normalization $A(L_i)=1$;
then, to a good approximation, we have
\be
\label{cosmoGN01}
G_N \simeq  G_5\frac{2k_ik_{i+1}}{k_i + k_{i+1}}
\ee
for the region between particle horizons that includes the
$i$th brane, and $8\pi G_5=\kappa^2$. 
Collecting all the contributions to the effective cosmological constant
$\Le$, we have
\baray
\label{dcosmoconst}
&&  \Le = \int [-\Lambda A(y)^2+((A(y)^{\prime})^2 
+4A(y)^{\prime\prime}A(y))/{2\kappa^2}]dy   \nonumber \\
&&      - 2A(y)^{\prime}A(y)/{\kappa^2}|_{boundaries}  
+ \sum_i  A^2(L_i) \sigma_i \nonumber \\
&& = - \int \Lambda A(y)^2 dy
- \frac{3}{2\kappa^2} \int (A'(y))^2 dy +\sum_i  A^2(L_i) \sigma_i  
\earay
where $A^{\prime\prime}$ is singular at the branes. An integration 
by parts on the $A^{\prime\prime}A$ term removes 
the Hartle-Hawking boundary term, as expected.
The integration over $y$ is between particle horizons 
and $\Lambda$ is piecewise constant. Besides the contributions from the 
brane tensions and the bulk $\Lambda$, we see that there is a contribution 
from the 5-D Ricci scalar $R^{\left(5\right)}$ to $\Le$.
We can now re-express $k_i$ 
(or $\Lambda$) and $y_i$ in $\Le$ in terms of the brane 
tensions $q_i$ (or $\sigma_i$) and the brane separations $L_i$.
{\em A priori}, we expect $\Le$ to be of order Planck scale, but for 
large brane separations, we shall see that the boundary conditions at 
the branes fix the piecewise constant bulk $\Lambda$ so that an almost exact 
cancellation among the terms in Eq.~(\ref{dcosmoconst}) renders 
$\Le$ exponentially small.

We now redefine the field $\hat \phi_i=\sqrt{A\left(L_i\right)}\phi_i$  
to absorb the factor $A\left(L_i\right)$ in the kinetic term 
of the Lagrangian ${\bf L}_i$, so the effective 4-D low energy 
action becomes
\baray
\label{eff4da}
&& {\bf S}^{\left(4\right)}=\int d^4x \sqrt{-\hat\gamma}
\left[\frac{R^{\left(4\right)}}{2\kappa_N^2}-\Le+{\bf L}_{CFT}
+\right. \nonumber \\
&& \left.\sum_i\left(\frac12\hat\gamma^{\mu\nu}
\hat\phi_{i,\mu}\hat\phi_{i,\nu}-\frac{m_{Hi}^2\hat\phi_i^2}2
-\lambda_i \hat\phi_i^4-\frac{\hat \phi_i^6}{m_{ci}^2} +
\cdots\right)\right]
\earay
where $m_{Hi}^2=A(L_i)M_{Hi}^2$ and $m_{ci}^2=A(L_i) M_{ci}^2$.
It is important to note that the Newton's constant 
and the effective cosmological constant $\Le$ are the same everywhere, 
so observers on the Planck (hidden) brane see the same $G_N$ and $\Le$
as observers on the TeV (visible) brane, even if the visible brane 
tension is negative, as is the case in the Randall-Sundrum model.
To be more realistic, ${\bf L}_{visible}$ may be replaced by the 
standard model Lagrangian density.
In ${\bf S}^{\left(4\right)}$, we have also included a conformal 
field theory term ${\bf L}_{CFT}$. Using the AdS/CFT 
correspondence\cite{AdS}, the effect of the Kaluza-Klein (KK) gravity modes 
may be incorporated into a conformal field theory on the brane.
This should be the case when the model covers the region between 
particle horizons,
where the gravity KK modes have a continuous mass spectrum. 
In the orbifold model, where the gravity KK modes 
may be discrete, in which case ${\bf L}_{CFT}$ should be replaced 
by another appropriate strongly interacting field theory.

Note that the warp factor $A(y)$ does not appear in the 4-D effective action.
Although we have chosen the warp factor at the 
Planck brane to be one, we could have chosen the warp factor at the 
visible brane to be one instead. The physics depends only on the ratio of
warp factors on the two branes, which is unaltered by this rescaling. However,
if we choose $A(L_{visible})=1$, then the electroweak scale should be 
taken to be the fundamental scale, while the Planck mass is a derived quantity.

\section{Two Brane World}
\label{sec:twobrane}

To render the general analysis more transparent, consider the simplest
case, a two brane model. 
Setting $L_0 = 0$, $L_1 = L$, the solution for the bulks 
outside the branes is
\baray
A(y)&=&{H^2\sinh^2[\kL(y-\yL)]\over\kL^2}\qquad(y<0)\nonumber\\
A(y)&=&{H^2\sinh^2[\kM(y-\yM)]\over\kM^2}\qquad(0<y<L)\nonumber\\
A(y)&=&{H^2\sinh^2[\kR(y-\yR)]\over\kR^2}\qquad(y>L),
\label{solution}
\earay
where $\yL$, $\yM$ and $\yR$ are constants, and give the locations of
the zeros of $A(y)$ when they are included in the domain of the spacetime
in the fifth dimension.

Continuity of the metric at the branes implies that
\baray
{\sinh^2(\kL\yL)\over\kL^2}&=&{\sinh^2(\kM\yM)\over\kM^2}\nonumber\\
{\sinh^2[\kM(L-\yM)]\over\kM^2}&=&{\sinh^2[\kR(L-\yR)]\over\kR^2}.
\label{match}
\earay

\addtocounter{figcount}{1}
\setcounter{2braneNewtonconsteps}{\value{figcount}}
\begin{center}
  \epsfbox{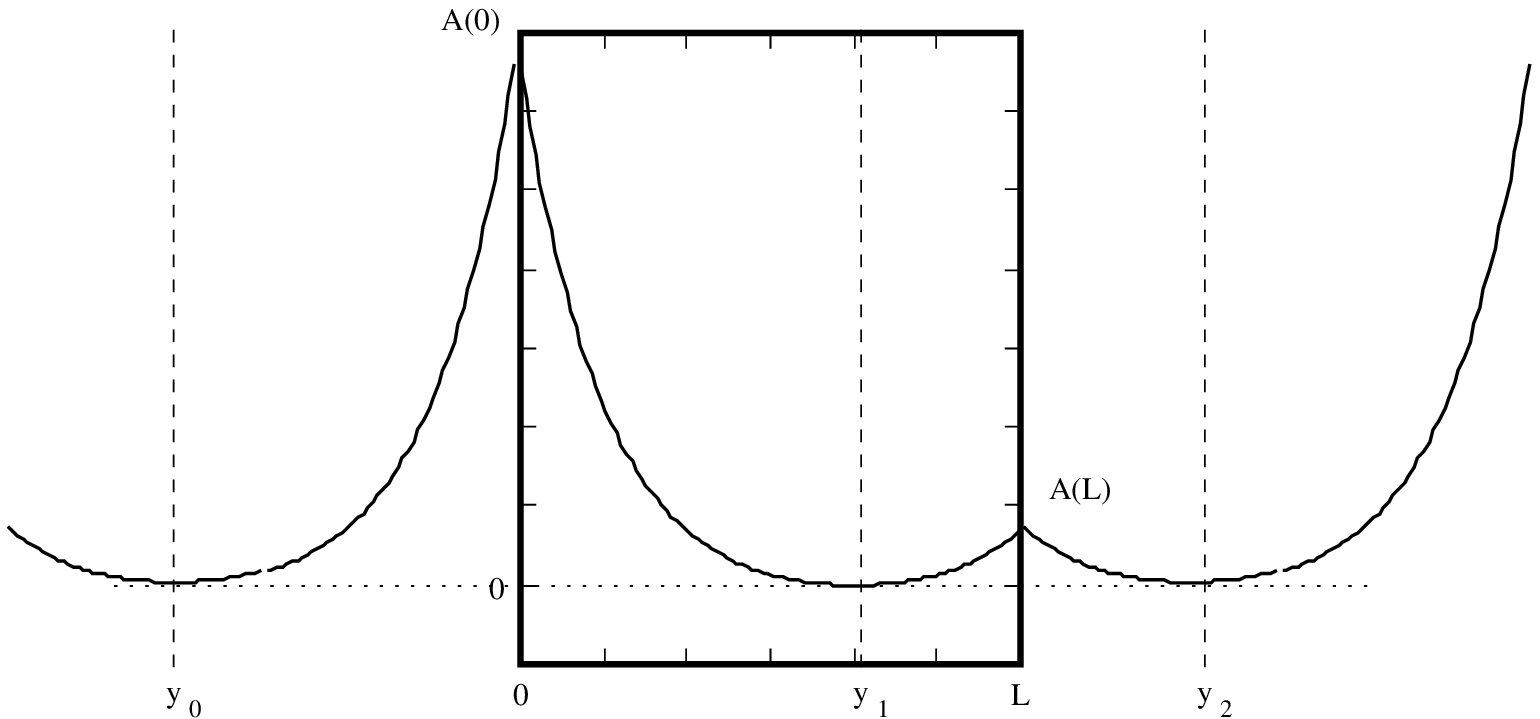}
  \parbox{12cm}{\vspace{1cm}
    FIG. \thefigcount \hspace{2pt} The two brane model in 
    the uncompactified case, with the $q_0$ brane at $y=0$ and the $q_1$ 
    brane at $y=L$. In the $S^1/\Bbb{Z}_2$ orbifold version (boxed), the two 
    branes sit at the two fixed points. The metric factor $A(y)$ is shown 
    schematically.\vspace{12pt}}
\end{center}

The jump conditions at the two branes are
\baray
{\kL\cosh(\kL\yL)\over\sinh(\kL\yL)}-{\kM\cosh(\kM\yM)\over\sinh(\kM\yM)}
&=&q_0\nonumber\\
{\kM\cosh[\kM(L-\yM)]\over\sinh[\kM(L-\yM)]}
-{\kR\cosh[\kR(L-\yR)]\over\sinh[\kR(L-\yR)]}&=&q_1.
\label{jump}
\earay

The expansion rate seen by observers
on the brane at $y=0$ is $H(0)=H/\sqrt{A(0)}$, where
\baray
{H^2\over A(0)}&=&{\kL^2\over\sinh^2(\kL\yL)}
={\kM^2\over\sinh^2(\kM\yM)}
\nonumber\\
&=&{[\kM^2-(\kL+q_0)^2][\kM^2-(\kL-q_0)^2]\over4q_0^2},
\label{HZ}
\earay
with $\kM^2-(\kL\pm q_0)^2>0$ or $<0$, in agreement with 
Ref\cite{horace}, which uses 
a slightly different approach. Similarly, the expansion rate seen 
by observers on the brane at $y=L$ is $H(L)=H/\sqrt{A(L)}$, 
where
\baray
{H^2\over A(L)}&=&{\kR^2\over\sinh^2[\kR(L-\yR)]}
={\kM^2\over\sinh^2[\kM(L-\yM)]}
\nonumber\\
&=&{[\kM^2-(\kR+q_1)^2][\kM^2-(\kR-q_1)^2]\over 4q_1^2},
\label{HL}
\earay
with
$\kM^2-(\kR\pm q_1)^2>0$ or $<0$. 
We can rescale $t$ so that 
$A(0)=1$, and the Hubble constants on the two branes are, respectively, 
$H(0)=H$ and $H(L)=H/\sqrt{A(L)}$.
Note that although Eqs.~(\ref{HZ}) and (\ref{HL}) appear to determine 
the expansion rates on the two branes completely in terms of local 
quantities (i.e., the local brane tensions, and bulk cosmological 
constants just outside each brane), the values of these quantities 
on/near the two branes are connected via $\kM$ and $\yM$.

The 4-D Newton's constant $G_N$ can be determined by
introducing a small matter density $\rho$ to the visible brane,
that is, $q_0 \to q_0 + \kappa^2 \rho/3 $, finding the Hubble
constant $H$, as in Eq.~(\ref{LeGN}) and Refs. \cite{oldcosmo,horace,cosmo1,cosmo},
and then requiring that $H^2=(8\pi G_N/3)(\Lambda_{eff}
+\rho)$; the result is
\be
\label{GNval}
4\pi G_Nq_0=\kappa^2\alpha_0\kL 
[1+2\kappa^2\Le(2\alpha_0+\kL)/3q_0\kL] 
\ee
where $\alpha_0 \equiv q_0 - k_0$. Although $L-$dependent,
the correction term is small if $G_N\Le/\kL^2\ll 1$, which is the case here,
so, to  a very good approximation, we have
\be
G_N = {\kappa^2 \alpha_0 \kL \over 4 \pi q_0}. 
\ee
Positivity of $G_N$ requires $\alpha_0>0$; 
to be specific, let us consider 
\be
0 \le \kM \le q_0 -\kL.
\label{condition}
\ee
Since the expansion rates $H$ and $H(L)$ given in Eqs.~(\ref{HZ}) and 
(\ref{HL}) depend on $\kM$, our goal is to express $H$ and $H(L)$ as 
functions of $L$ and the parameters $\kL$, $\kR$, $q_0$ and $q_1$. 
This requires an expression relating $L$ and $\kM$; from Eqs.~(\ref{match})
and (\ref{jump}) we find
\baray
\kM L =\sinh^{-1}\biggl({\kM\over H(0)}\biggr)
+\sinh^{-1}\biggl({\kM\over H(L)}\biggr) \nonumber \\
=\sinh^{-1}\biggl({2\kM q_0\over\sqrt{[\kM^2-(\kL+q_0)^2]
[\kM^2-(\kL-q_0)^2]}}\biggr)
\nonumber\\
+\sinh^{-1}\biggl({2\kM q_1\over
\sqrt{[\kM^2-(\kR+q_1)^2][\kM^2-(\kR-q_1)^2]}}\biggr).
\label{Lfind}
\earay
Here, Eq.~(\ref{Lfind}) is regarded as a relation
that determines $\kM$ in terms of $L$, $q_0$, $q_1$, $\kL$ and $\kR$. 
In the $5$-form field strength model, $\kM$, $\kL$ and $\kR$ will adjust
together as the background field strength is determined as a function of
$L$, $q_0$ and $q_1$.
In general, $H(0)\neq H(L)$. Because of the condition (\ref{condition}),
$H\to 0$ as $\kM \to \alpha_0=q_0-\kL$ from below. 
This means that the expansion rate as seen by observers on the visible 
brane becomes exponentially small for large $L$,
\be
H^2 \approx 4 \alpha_0^2 e^{-2 (\alpha_0 L - C)}
\label{Hgen}
\ee
where $\alpha_0>0$, and $\sinh(C) = \kM/H(L)$.
As $\kM \to \alpha_0$ from below, $H(L)$ approaches a 
$L$-independent constant given 
by Eq.~(\ref{HL}) and so does $C$. 
This implies that $\Le$ becomes exponentially small as $L$ increases.

Note that there is only one integration constant for the bulk 
cosmological constants. Consider the model of Eq.~(\ref{HTcouple}).
Here $k_0^2= -\kappa^2\phi/6$, $k_1^2= -\kappa^2(\phi+\mu_0)/6$ and
$k_2^2= -\kappa^2(\phi+\mu_0+\mu_1)/6$, where the integration constant
$\phi$ is negative and the brane charges $\mu_0$ and $\mu_1$ are 
essentially arbitrary constants.

In the symmetric case, where $\kL=\kR=k$ and $q_0=q_1=q$ 
(and let $\alpha=q-k=\alpha_0$), we have
$H(L)=H(0)=H$, and $\yM = -\yL = \yR = L/2$. The constant $C$ 
in Eq.~(\ref{Hgen}) becomes $C=\alpha L/2$, so, including a matter
density $\rho$ (which is treated as a perturbation),
\be
H^2 \approx 4 \alpha^2 e^{-\alpha L} + {2\kappa^2 \alpha k \over{3q_0}} \rho.
\ee
Thus, $\Le$ still decreases exponentially with $L$, but slower than in
the nonsymmetric case.

\subsection{The $S^1/\Bbb{Z}_2$ Orbifold Model}
\label{sec:twobraneorbi}

To this point, we have concentrated on spacetimes that are noncompact
in $y$, but similar results can be derived for the
compactified case. First, we may choose to identify $\kL=\kR=\kM$ and
derive $\kM$ and $H$ in terms of $L$. Next, we can
compactify the $y$ direction to a circle $S^1$ of length $2L$.
Placing the branes at $y=0$ and $y=L$, we may identify the two sides of $S^1$
to obtain a line segment.
That is, we perform a $\Bbb{Z}_2$ orbifold, 
with one brane sitting at each of the two fixed points ($y=0,L$).
This $S^1/\Bbb{Z}_2$ orbifold model is particularly simple, since there is 
only one bulk space between the branes sitting at the two end points.
This may be considered as an expanding (non-supersymmetric) version
of the Horava-Witten model \cite{phew}, and is discussed in Ref\cite{ira}.
The model has branes
with tension $\sigz=3q_0/\kappa^2$ at $y=0$ and
$\sigL=3q_1/\kappa^2$ at $y=L$,
separated by AdS space with bulk
cosmological constant $-6k^2/\kappa^2$, which is treated as an 
integration constant. The solution is
\be
A(y)={H^2\over k^2}\sinh^2[k(y-y_0)]
\ee
for $L>y>0$. Because of the orbifold symmetry of the model,
the jump conditions are
\baray
2k/q_0&=&\tanh ky_0
\nonumber\\
2k/q_1&=&\tanh[k(L-y_0)]
\earay
at $y=0$ and $L$, respectively. Combining the jump conditions implies
\be
{q_0\over 2k}={\tanh kL-q_1/2k\over
1-(q_1/2k)\tanh kL};
\label{qzdet}
\ee
if $q_1/2k=\pm 1$, then $q_0/2k=\mp 1$ irrespective of $kL$, but
for $q_1/2k\neq 1$, $q_0/2k\to 1$ as $kL\to\infty$.
According to our viewpoint, Eq.~(\ref{qzdet}) determines 
$k$ given $q_0,q_1$ and $L$. 

The expansion rates on the branes are $H(0)=H/\sqrt{A(0)}$
and $H(L)=H/\sqrt{A(L)}$, where $H^2(0)=k^2[(q_0/2k)^2-1]$
and $H^2(L)=k^2[(q_1/2k)^2-1]$. 
We see that $L$ is related to $k$ via
\be
L=\frac1{2k}\ln{\left[\frac{\left(1+2k/q_0\right)\left(1-2k/q_1\right)}
{\left(1-2k/q_0\right)\left(1+2k/q_1\right)}\right]}.
\ee
Eliminating $k$, we find that
\be
  H^2(0) = {q_0^2\over4}\left[
    {(q_0+q_1)^2-2q_0q_1t^2-2q_1^2t^2+(q_0+q_1)\Delta
      \over(q_0+q_1)^2-2q_0q_1t^2+(q_0+q_1)\Delta}\right].
\label{twoexact}
\ee
where $t\equiv\tanh kL$ and 
$\Delta\equiv\sqrt{(q_0+q_1)^2-4q_0q_1t^2}$.

\addtocounter{figcount}{1}
\setcounter{Hubblegraphportraiteps}{\value{figcount}}
\begin{center}
  \epsfbox{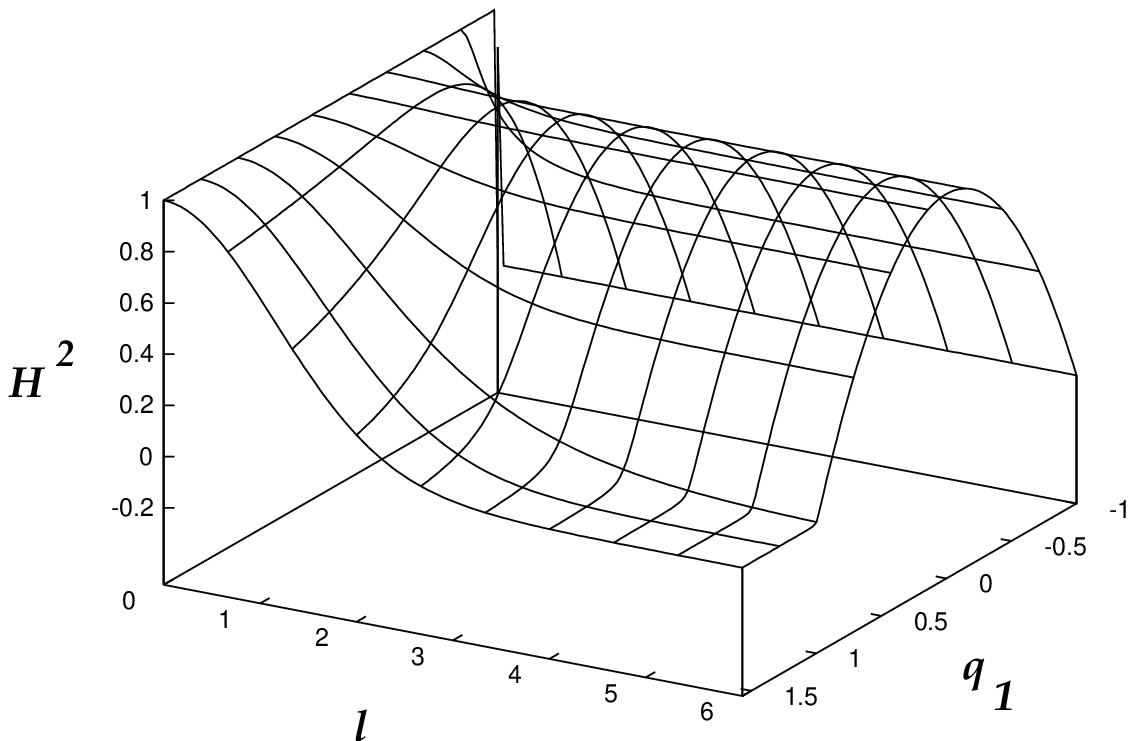}
  \parbox{12cm}{\vspace{1cm}
    FIG. \thefigcount \hspace{2pt} Hubble constant $H^2$ on the brane 
    at $L_0=0$ as a function of the dimensionless separation $l=kL$ and 
    the hidden brane tension $q_1$, in units such that the observable brane 
    tension $q_0=1$. Note that $H^2=0$ for $q_1=-1$, the RS scenario.
    \vspace{12pt}}
\end{center}

For large values of $q_0L$, this reduces to
\be
H^2(0)\approx {q_0^2(q_1+q_0)\over (q_1-q_0)}e^{-q_0L}
\label{Hz}
\ee
where $k \simeq q_0(1-e^{-q_0L})/2$.
Thus, for $q_0L\gg 1$ and $\vert q_1\vert>q_0>0$, the cosmological
constant on the $y=0$ brane is positive, and exponentially small.
Moreover, although Eq.~(\ref{Hz})
may appear singular as $q_1\to q_0$ in fact
\be\label{halfHz}
  H^2(0)\approx q_0^2e^{-q_0L/2}
\ee
in that case.

The behavior of Eq.~(\ref{twoexact}) is shown in 
Figure~\arabic{Hubblegraphportraiteps} where we have taken $q_0=1$ 
in some units.  For $q_1>q_0$ ({\em i.e.}, $q_1>1$), $H^2=H^2(0)$  
brane becomes exponentially 
small for large $l=kL$. As $q_1$ decreases toward $q_0=1$, $H^2$ 
still drops off exponentially, but at a slower rate. For still smaller $q_L$,
$H^2$ goes to a constant $\sim 1$ for large $l$ implying a large expansion
rate on the $L=0$ brane. However, $H^2_L=H^2(L)$ then becomes exponentially
small, as the roles of the branes at $y=0$ and $L$ reverse (and we may
then identify the brane at $y=L$ as the visible brane). 

\addtocounter{figcount}{1}
\setcounter{figRSeps}{\value{figcount}}
  \begin{center}
    \epsfbox{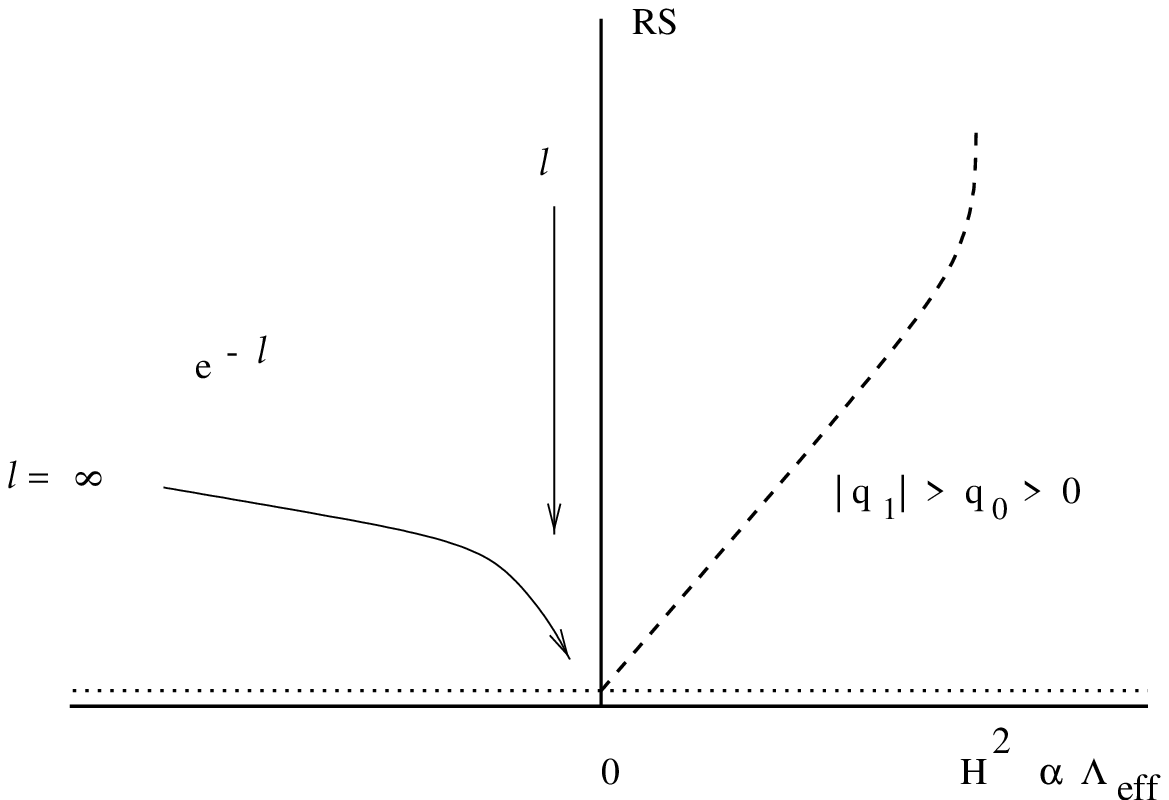}
    \parbox{12cm}{\vspace{1cm}
      FIG. \thefigcount \hspace{2pt} Hubble constant squared versus 
      separation.  
      More precisely, $e^{-l}=e^{-kL}$ vs $H^2$, which is proportional to the 
      4-D effective cosmological constant $\Le$. The dashed line corresponds 
      to our model. The solid vertical line corresponds to the Randall-Sundrum 
      model. The horizontal dotted line corresponds to the single brane model.
      $l=\infty$ at the intersection of the 3 lines.\vspace{12pt}}
  \end{center}

If $q_1=-q_0$, as considered in \cite{RS1},
then $y_0=\pm\infty$, so $H(0)=H(L)=0$ (see 
Figure~\arabic{Hubblegraphportraiteps}) and therefore
$\vert q_0\vert=2k$ for {\it any} finite non-zero $L$. In this limit,
\be
A(y) = \sinh^2[k(y-y_0)]/\sinh^2[ky_0] \to e^{-2k|y|},
\ee
the same as in the RS model \cite{RS1}. 
In general, our two brane model is qualitatively different from \cite{RS1}, 
since it involves two 
branes with tensions of different magnitudes and not necessarily
opposite signs, and interprets the bulk cosmological constant
as an integration constant derived from the brane tensions and separation,
but the RS model can be obtained as an appropriate limit of the more general
model presented here.

Figure~\arabic{figRSeps}, in which we plot $e^{-l}$ versus $H^2$,
illustrates the key features of Eq.~(\ref{twoexact}). 
The dashed line represents the two brane model described above, and as $l$
increases, the $\Lambda_{eff}$ follows the dashed line and approaches 
zero. For the RS model
$q_0+q_L=0=H^2$ for any separation $l$. (Of course, $q_0+q_L=0$ is
a fine-tuning.) Thus, the corresponding curve for the RS model is the 
solid vertical line at $H^2=0$. 
Suppose we start with only
one brane, which may be considered as the $l= \infty$ limit of the
two brane model. Then we find that the visible brane (at $y=0$) can have
any value of $\Lambda_{eff}$. This happens because the problem has
reduced to that of a single brane, {\em i.e.}, the jump
condition for the brane at $y=L=\infty$ is absent, so $k$
is no longer determined. It follows that $H^2$, and hence $\Lambda_{eff}$, 
can have any value; this is shown by the dotted
line along the $x$-axis. This illustrates that the difference is due to 
the interchange of limits, {\em i.e.,}, the difference between
solving the Einstein equation
and before taking $l\to\infty$, versus starting with $l= \infty$ before solving
the Einstein equation. For large, finite $l$, we see that one does not have 
to tune the brane tensions to obtain an exponentially small $\Lambda_{eff}$,
as long as the brane separation is relatively large.

For positive tension branes, there will always be a zero of $A(y)$ 
between the branes.  This ``particle horizon'', to be discussed below, 
is a barrier through which communication is impossible in finite time. 
In light of this, it is more useful to express the Hubble constant on the 
visible brane not in terms of the tension of the brane at $L$ and its 
location which can never be detected, but only with $q_0$, and $y_0$, 
the detectable quantities. Then Eqs.~(\ref{twoexact},\ref{Hz}) are
\be
\label{onebranecase}
H^2(0) = {q_0^2\over4}\left[1-\tanh^2(ky_0)\right]
\simeq {q_0^2\over2}e^{-q_0y_0}
\ee
where $y_0$ is a function of $q_0,q_L$ and $L$. We discuss the role of
horizons in determining the expansion rate on a given brane more completely
in \S\ref{sec:horizons}.

For positive tension branes, we shall see that the presence of the interbrane
horizons prevents us from explaining the mass hierarchy problem, although we
can obtain an exponentially small cosmological constant on one of the two
branes -- which we can identify as the visible brane -- in that case. 
When one of the branes has negative tension, there are no horizons, and
the warp factor will be small on that brane. Hence, we identify the negative
tension brane as the visible brane, and can explain the mass hierarchy problem
if the warp factor $A(y)$ is sufficiently small on that brane. In that case,
we also find that the expansion rate on the visible brane is small, but
not small enough to explain the cosmological constant problem: when scaled
appropriately, the expansion rate is of order the Higgs mass, which is
far smaller than the Planck mass, but far larger than the expansion rate
of our Universe. 
To find a solution to both problems, we need more than 
one separation distance and one bulk integration constant.
In general, this may be achieved in a multibrane world, with more than
two branes. However, we may also realize such a solution in the
two brane case, provided that the brane world is compactified but not
orbifolded. The properties of this model are discussed in detail
in \S\ref{sec:ccandhier}, and from a different perspective.

\section{The Cosmological Constant and The Hierarchy Problems}
\label{sec:ccandhier}

The above two brane orbifold model is able to provide an explanation for 
either the hierarchy\cite{RS1} or cosmological constant problem\cite{ira}, 
but not both.
However, it is possible to solve both problems simultaneously in the 
multibrane world generally. 
In this section, we present details of a relatively
simple model that solves both problems: a two brane model in which
the fifth dimension is compactified, but not orbifolded. 
To capture the key feature, which is generic in a multibrane world, 
let us first present a toy model.

\subsection{Probing A One-Brane Orbifold Model}

It is easy to construct a variation of the above two brane
$S^1/\Bbb{Z}_2$ orbifold model that foreshadows how the mass hierarchy
and cosmological constant problems might be solved simultaneously in a 
two-brane model. Cut out the line
segment $y_0\leq y\leq L$ and remove
the $q_1$ brane of the above $S^1/\Bbb{Z}_2$ model, so the
resulting orbifold model has length $y_0$ (with orbifold fixed points
at $y=0$ and $y=y_0$). The particle horizon is now at
the orbifold fixed point at $y=y_0$, and $A(y_0)=0=A^\prime(y_0)$
at that point. There is only one brane (the $q_0$ brane at the 
fixed point $y=0$) in this $S^1/\Bbb{Z}_2$ orbifold model.
$H^2$ is still given by Eq.~(\ref{onebranecase}) for $q_0y_0\gg 1$, 
which is exponentially small. 
Let us treat the $q_0$ brane as the Planck brane,
with $A(0)=1$, and introduce the visible brane as a probe brane
({\em i.e.}, with negligible brane tension, as in Ref\cite{lykken}) at 
$y=L_v$, somewhere between the two fixed points. The warp
(or hierarchy scale) factor on the visible brane is then
$A(L_v)\simeq e^{-q_0L_v}$, whereas
$\Le \propto  e^{-q_0y_0}$. Since $y_0>L_v$, the inequality
(\ref{important}) follows. In this simple two brane model, we see that,
if the hierarchy scale factor is exponentially small, 
the 4-D effective cosmological constant must be exponentially smaller.
We can of course choose to keep the $q_1$ brane for a three brane model.

This model is strictly correct only if the visible brane tension 
is exactly zero, which is a fine-tuning. 
A more careful treatment is necessary for an
arbitrary visible brane tension, even if it is very small.

\subsection{Two Brane Compactified Model}

Consider a two brane compactified model, which has 
a $S^1$ compactified $5$th dimension, with circumference $L_2-L_0$, 
as shown in the following figure.
The brane at $L_0$ has tension $\sigma_0$ and the brane at $L_1$ has 
tension $\sigma_1$, both of which are positive. 
Recall that $q_0=\kappa^2\sigma_0/3$ 
and $q_1=\kappa^2\sigma_1/3$. These two branes are separated by $L_1-L_0$
on one side and $L_2-L_1$ on the other side of the circle, 
where $L_0$ is identified with $L_2$. 
Without loss of generality, let 
$\sigma_0>\sigma_1>0$ ($q_0>q_1>0$).
Note that we require two integration constants, namely, the two bulk
cosmological constants $\Lambda_1$ 
and $\Lambda_2$, or equivalently, $k_1$ and $k_2$.

For a piecewise constant bulk cosmological constant, we can introduce two 
$5$-form field strengths, or alternatively, start with the model 
(\ref{mixedS}) with only one $5$-form field strength. Because the $5$th 
dimension is compactified, the  
brane charges, which are constant parameters, must add to zero: 
$e_0=-e_1$.  Here $k_1^2= -\kappa^2(\phi - (e_{(0)}+e_0)^2/2)/6$ and 
$k_2^2= -\kappa^2(\phi -e_{(0)}^2/2)/6$, where the integration constants are 
the constant background field strength $e_{(0)}>0$ and the constant $\phi<0$. 
This allows us to treat $k_1$ and $k_2$ as integration 
constants to be determined by the 5-D Einstein equation.

\addtocounter{figcount}{1}
\setcounter{bulkbehavioureps}{\value{figcount}}
\begin{center}
  \epsfbox{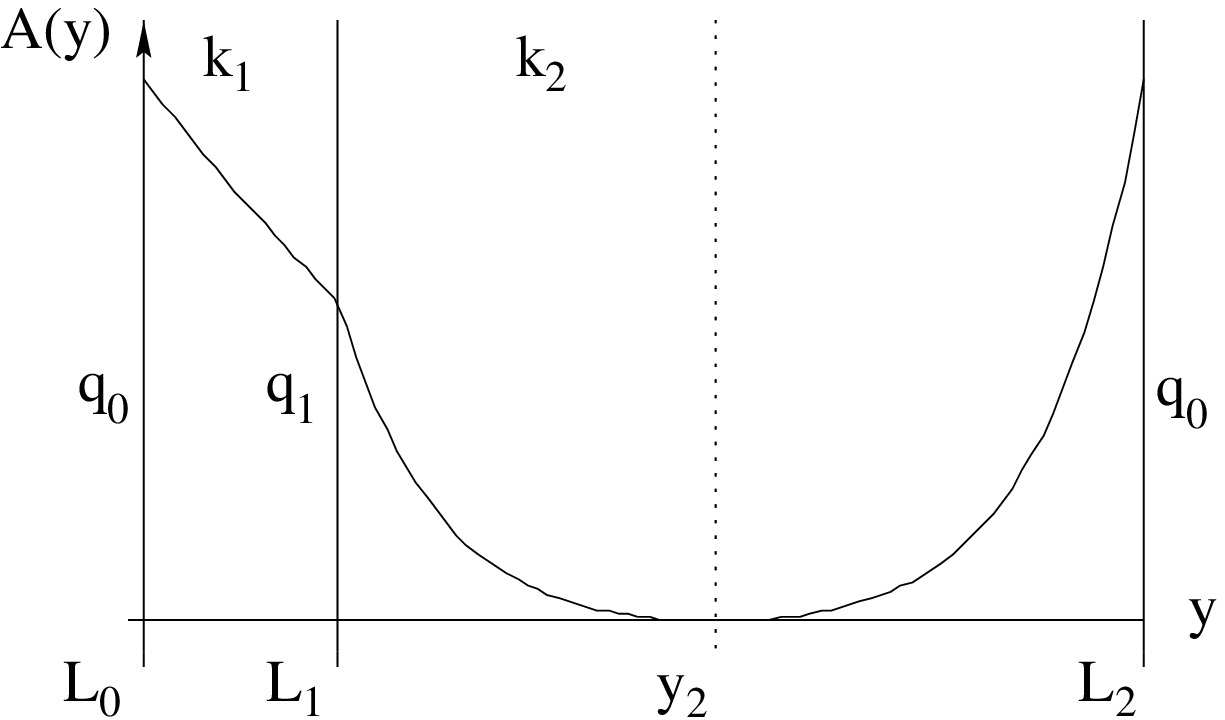}
  \parbox{12cm}{\vspace{1cm}
    FIG. \thefigcount \hspace{2pt} The two brane compactified model, in 
    which the hierarchy problem and the cosmological constant problem may be 
    simultaneously solved.  $L_0$ is identified with $L_2$, so the circle
    has circumference $L_2-L_0$. The brane at $L_0$ ($L_1$) is the Planck
(visible) brane. The metric factor $A(y)$ is shown schematically.\vspace{12pt}}
\end{center}

The solution $A(y)$ in the bulks is given by Eq.~(\ref{scale})
and $H(L_i)$ is given by Eq.~(\ref{hubbleconstdef}).
The continuity conditions of the metric $A(y)$ at the branes are
(\ref{boundary}) 
\baray
  {\sinh^2[k_1(L_1-y_1)]\over k_1^2}
    &=&{\sinh^2[k_2(L_1-y_2)]\over k_2^2}\\\nonumber
  {\sinh^2[k_2(L_2-y_2)]\over k_2^2}
    &=&{\sinh^2[k_1(L_0-y_1)]\over k_1^2}\
\earay
and the jump conditions at the branes are 
\baray
\label{twobccch}
  q_0 &=& k_2\coth[k_2(L_2-y_2)]-k_1\coth[k_1(L_0-y_1)]\\\nonumber
  q_1 &=& k_1\coth[k_1(L_1-y_1)]-k_2\coth[k_2(L_1-y_2)].
\earay
It is straightforward to obtain the large separation behavior.
However, let us first get an overall picture of the model.
The above conditions can be rewritten as
\baray
\label{cont_cond}
  k_1^2\left[\coth^2[k_1(L_1-y_1)]-1\right]
    &=& k_2^2{(\coth^2[k_2(L_2-y_2)]-1)(1-t_2^2)
    \over(1-t_2\coth[k_2(L_2-y_2)])^2}\\\nonumber
  k_2^2\left[\coth^2[k_2(L_2-y_2)]-1\right]
    &=& k_1^2{(\coth^2[k_1(L_1-y_1)]-1)(1-t_1^2)
    \over(1-t_1\coth[k_1(L_1-y_1)])^2}\\\nonumber
  q_0 = k_2\coth[k_2(L_2-y_2)]) &-& k_1{\coth[k_1(L_1-y_1)]-t_1
    \over1-t_1\coth[k_1(L_1-y_1)]}\\\label{hierarjump}
  q_1 = k_1\coth[k_1(L_1-y_1)]) &-& k_2{\coth[k_2(L_2-y_2)]-t_2
    \over1-t_2\coth[k_2(L_2-y_2)]},
\earay
where $t_{1,2}\equiv\tanh[k_{1,2}\Delta l_{1,2}]$, where
$\Delta l_1 \equiv k_1(L_1-L_0)$ and $\Delta l_2 \equiv k_2(L_2-L_1)$.
Let us consider the allowed regions in the $(k_1,k_2)$ plane.

\addtocounter{figcount}{1}
\setcounter{kphaseeps}{\value{figcount}}
\begin{center}
  \epsfbox{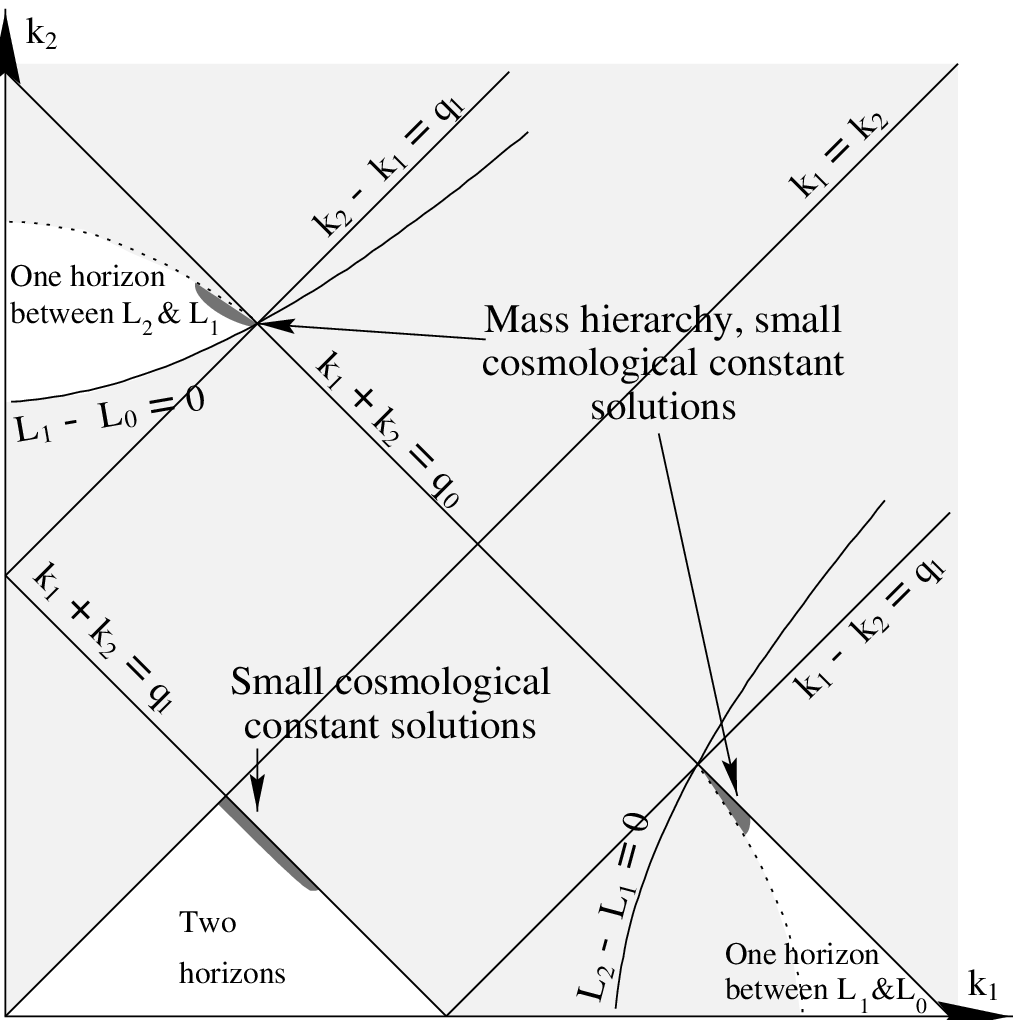}
  \parbox{12cm}{\vspace{1cm}
    FIG. \thefigcount \hspace{2pt} The allowed regions (white) of 
    ``$k$-space'' for which there are consistent solutions with 
    $H^2(L_{1,2})>0$ and $L_2-L_1>L_1-L_0$.  The regions with 
    small cosmological constant solutions are shaded dark.\vspace{12pt}}
\end{center}

The allowed regions in Figure~\arabic{kphaseeps} are those for which 
the cotanh terms in Eq.~(\ref{twobccch}) have the correct signs to 
correspond to the behavior of the scale factor $A(y)$ in the bulk 
between the branes.  
Some of the details appear in Appendix~\ref{app:twobranedetails}.  
Figure~\arabic{kphaseeps} shows the physically distinct allowed regions,
where one demands $L_2-L_1\ge L_1-L_0$ without loss of generality.
We see that there are two allowed regions:
\setcounter{Lcount}{0}
\begin{list}{(\arabic{Lcount})}
  {\usecounter{Lcount}}
  \item the ``one horizon'' region for $k_2>k_1$ is bounded by 
    the $k_1=0$ line, the 
    $q_0=k_1+k_2$ line and the $L_1-L_0=0$ curve. In this region, 
    \be
    k_2(L_2-L_1) > k_1(L_1-L_0) 
    \ee
    that is, the particle horizon is 
    between $L_1$ and $L_2$ segment of the circle, whose bulk 
    cosmological constant has a larger magnitude ($k_2>k_1$).
    (The other $k_1 \ge k_2 > 0$ region has an identical description, and 
    they are related by a simple interchange of the two segments of the 
    circle.)
    
  \item the ``two horizon'' region, where there is a particle horizon 
    in each of the two segments of the circle.
\end{list}
Let us consider the ``one horizon'' region where $k_2 \ge k_1 > 0$. 
This region is bounded by the $k_1=0$ line, the 
$q_0=k_1+k_2$ line and the $L_1-L_0=0$ curve. 
The $L_1-L_0=0$ curve is given by the equation $k_2^2=k_1^2+q_0q_1$, 
which can be derived from (\ref{hierarjump}). 
In this region, 
$k_2(L_2-L_1) > k_1(L_1-L_0)$, that is, the particle horizon is between
the $L_1$ and $L_2$ segment of the circle, that is, 
the segment with a larger magnitude
bulk cosmological constant ($k_2>k_1$).
This region is further divided into two subregions, where
$(L_2-L_1)<(L_1-L_0)$ (the upper part) and $(L_2-L_1) \ge (L_1-L_0)$
(the lower part).
Because of the relationship between $\Delta l_i$ and the behavior of the 
scale factors in the bulk, for
Figure~\arabic{bulkbehavioureps} we have
\baray\nonumber
  \Delta l_1 &\equiv& k_1(L_1-L_0)
    = \sinh^{-1}\left({k_1\over H(L_0)}\right)
    -\sinh^{-1}\left({k_1\over H(L_1)}\right)\\\nonumber
  \Delta l_2 &\equiv& k_2(L_2-L_1)
    = \sinh^{-1}\left({k_2 \over H(L_1)}\right)
    +\sinh^{-1}\left({k_2 \over H(L_2)}\right).
\earay

\subsection{Large Brane Separations}

It is easy to see that as $H(L_1)\to 0$, $\Delta l_2\to\infty$, but 
$\Delta l_1$ can remain positive and relatively small if 
$H(L_0)\equiv H(L_2)$ is smaller 
than $H(L_1)$.  The brane at $L_1$ will then be identified as the visible 
brane, on which $A(L_1)$ is large compared to $H(L_1)$ there, yet still 
exponentially small when compared to the fundamental (Planck) scale.

Following the remarks above, we expect $\Delta l_2$ to be large ($t_2\to1$).  
The boundary condition at $L_1$ can then be satisfied if $k_1\to 0$, or 
$\coth[k_1(L_1-y_1)]\to -1$.  The later solution will permit the two different 
scales required.  The second boundary condition will require 
$\coth[k_2(L_2-y_2)]\to 1$.  Writing
\baray\nonumber
  t_2 &\simeq& 1-2\exp[-2\Delta l_2]\\\nonumber
  \coth[k_1(L_1-y_1)] &\simeq& -1-\eta_1\\\nonumber
  \coth[k_2(L_2-y_2)] &\simeq& 1+\eta_2,
\earay
where $\eta_{1,2}$ are expected to be $O(\exp[-\Delta l_2])$.   
To zeroth order in $\eta_{1,2}$, the jump conditions are satisfied if
\be
\label{twocomk1}
  k_1 = {1\over2}(q_0-q_1),\quad k_2 = {1\over2}(q_0+q_1).
\ee
We imposed $q_0>q_1$ which give $k_i>0$ as required for consistency.  
The boundary conditions, to first order in $\eta_{1,2}$ lead to
\baray\nonumber
  \eta_1 &=& 2\left({q_0+q_1\over q_0-q_1}\right)^2
    \sqrt{\left({1+t_1\over1-t_1}\right)}e^{-\Delta l_2},\\\nonumber
  \eta_2 &=& 2\sqrt{\left({1-t_1\over1+t_1}\right)}e^{-\Delta l_2}.
\earay
We now demand that $\Delta l_1$ is large, but it must remain smaller than 
$\Delta l_2$ (see Appendix~\ref{app:twobranedetails}) so 
$t_1\simeq1-2\exp[-2\Delta l_1]$, which gives  
\baray
\label{twocomH}
  H^2(L_0) &=& H^2(L_2)\simeq (q_0+q_1)^2
    e^{-(\Delta l_2 + \Delta l_1)},\\\nonumber
  H^2(L_1) &\simeq &(q_0+q_1)^2
    e^{-(\Delta l_2 - \Delta l_1)}.
\earay
Figure~\arabic{bulkbehavioureps} shows schematically that the scale 
factor on the $L_1$ brane is exponentially smaller than on the 
$L_0$ brane:
\be
  {A(L_1)\over A(L_0)} = {H^2(L_0)\over H^2(L_1)}
  \simeq e^{-2\Delta l_1}
\label{twocomAorig}
\ee
to lowest order.  For a compactified $5$th dimension, with both 
$\Delta l_1 \simeq (q_0-q_1)(L_1-L_0)/2$ and 
$\Delta l_2 \simeq (q_0+q_1)(L_2-L_1)/2$ large, we find that  
$A(L_1)$ and $H^2(L_0)$ and $H^2(L_1)$ are all exponentially small.
The results are
\baray
\label{twocomA}
  A(L_1) &\simeq& e^{-(q_0-q_1)(L_1-L_0)}\nonumber\\
  H^2(L_0)&\simeq&   
          (q_0+q_1)^2
e^{-[(q_0+q_1)(L_2-L_1)+(q_0-q_1)(L_1-L_0)]/2}
\nonumber\\
  H^2(L_1) &\simeq& 
    (q_0+q_1)^2
e^{-[q_0+q_1)(L_2-L_1)-(q_0-q_1)(L_1-L_0)]/2}.
\earay
Since $L_1>L_0$ and $q_0>q_1$, the warp factor $A(L_1)$ is
exponentially small.

\subsection{$G_N$ and $\Le$}
 
Let us consider further the situation as shown in 
Figure~\arabic{bulkbehavioureps} in
terms of the 4-D effective action. The $y=L_0=0$ brane is referred 
to as the Planck (hidden) brane and the $y=L_1$ brane is referred to as the 
TeV (visible) brane. (Note that the visible brane tension may be of 
order Planck scale.) We adopt $A(0)\equiv 1$, so $A(L_1)=e^{-2 \Delta l_1}$,
where $\Delta l_1 \simeq (q_0-q_1)L_1/2$. 
We shall take $A(L_1) \simeq 10^{-32}$, so the mass 
$m_{H1}=\sqrt{A(L_1)}M_{H1}$ of the 
Higgs field $\hat \phi_1$ on the visible brane is around TeV,
while  the mass $m_{H0}=\sqrt{A(0)} M_{H0}$ of the Higgs field 
$\hat \phi_0=\phi_0$ on the hidden brane is comparable to the Planck mass.
In this model, the $y$ integration is over the entire $S^1$.
Up to exponentially small corrections, we have
\be
\int A(y) dy =
\int \frac{H^2}{k^2}\sinh^2{\left[k\left(y-y_0\right)
\right]}dy \simeq \frac1{2k_1} +\frac1{2k_2}
\ee
with $k=k_1, k_2$ for each side of the $y=L_0$ brane,
so, up to exponentially small corrections, this integral over $y$ 
yields the same $G_N$ as that in (\ref{cosmoGN01}).
In terms of brane tensions,
\be
\frac{1}{\kappa_N^2} \simeq \frac{2q_0}{(q_0^2-q_1^2)\kappa^2}.
\ee
Using the result Eq.~(\ref{twocomA}) for $H^2$, we find that the 
effective 4-D cosmological constant is
\be
\Le \simeq \frac{2\sigma_0(\sigma_0 + \sigma_1)}{(\sigma_0-\sigma_1)}
e^{-\Delta l_2 -\Delta l_1}
\ee 
where $\Delta l_1 \simeq (q_0-q_1)(L_1-L_0)/2$ and
$\Delta l_2 \simeq (q_0+q_1)(L_2-L_1)/2$.
In terms of $\hat \phi_i$ and $m_{Hi}$, $G_N$ and $\Le$ 
are the same for observers on  
both the visible ({\em i.e.}, the TeV) brane at $y=L_1$ and on the 
Planck brane at $y=L_0=0$. Since $H^2=\kappa_N^2 \Le/3$, we expect
the same Hubble constant for the two branes. However,
Eq.~(\ref{twocomH}) gives the ratio $H^2(L_0)/H^2(L_1)=A(L_1)$. This 
difference arises because Eq.~(\ref{twocomH}) is calculated in the
($\gamma_{\mu \nu}$, $\phi_i$, $M_{Hi}$) frame, 
while properties calculated from
the effective action ${\bf S}^{\left(4\right)}$ are in the 
($\hat \gamma_{\mu \nu}$, $\hat \phi_i$, $m_{Hi}$) frame. That is, in the 
($\hat \gamma_{\mu \nu}$, $\hat \phi_i$, $m_{Hi}$) frame, 
the Hubble constant seen by observers 
on the visible brane is simply $H^2=H^2(L_0)$. (More details on such 
$A(L_i)$ rescaling later.) 

As pointed out in Ref.\cite{RS1}, the Higgs mass $m_{Higgs}$ as 
seen by observers on the visible brane is given by 
$m_{Higgs}^2=A(L_1)M_H^2$. 
Using Eqs.~(\ref{twocomk1},\ref{twocomA},\ref{cosmoGNi}), we see 
that the observed Higgs mass is exponentially small compared to 
the Planck mass for large brane separation $L_1=L_1-L_0>0$,
\be
{m_{Higgs}^2\over m_{Planck}^2}\simeq 
\frac{2k_1 k_2 M_H^2 A(L_1)}{(k_1+k_2)M^3}
=\frac{4 \pi M_H^2 (\sigma_0^2-\sigma_1^2)}{3 M^6 \sigma_0}
e^{-8\pi (\sigma_0-\sigma_1)L_1/{3M^3}}
\ee
for $\sigma_0 > \sigma_1>0$. Thus, we obtain a Randall-Sundrum like
explanation of the hierarchy problem: an appropriate choice of the 
the warp factor can explain the hierarchy problem.
At the same time $\hat \phi_1^6$ and other higher-dimensional 
operators become more important in the effective theory as their 
couplings are exponentially enhanced, that is, 
$m_{c1}$ becomes TeV scale.

\subsection{An Inequality}
\label{sec:compactdisc}
 
It is clear from the above that the mass hierarchy problem can be
explained by this model if $A(L_1)\sim 10^{-32}$. What is not so obvious,
but shall be explained in \S\ref{sec:gmh}, is that the appropriate statement
of the cosmological constant problem is that $H^2(L_0)G_N\sim 10^{-122}$.
This may be understood most easily in terms of the effective theory
introduced in \S\ref{sec:effaction}, where we show that, with appropriate
choice of units, $G_N$ and $H$ may be regarded as the same on all branes,
but the Higgs mass on any particular brane is related to $G_N$ by
$G_Nm_{Higgs}^2\sim A(L_i)$. 
Assuming that $q_0,q_1\sim G_N^{-1/2}$, we see that Eq.~(\ref{twocomA})
(as well Eqs.~(\ref{twocomAorig}) and (\ref{twocomH}))
implies that if we can explain the mass hierarchy problem, then 
the observed expansion rate, in Planck units, is smaller than the
observed Higgs mass in Planck units by additional powers of
$A(L_i)\ll 1$, because $\Delta l_2>\Delta l_1$. 
Although the static models do not fix the size of
expansion rate precisely, even when the ratio of Higgs and Planck
masses is taken from observation, they guarantee that the value is
far smaller than the {\it Higgs} mass.
These considerations are based on the one-horizon model, and the inequality
$\Delta l_2>\Delta l_1$. 

In the above analysis, we show how two exponentially small factors can 
be generated, one for the hierarchy problem and the other for the 
cosmological constant problem. As an input, we choose a large radius 
$(L_2-L_0)/2\pi$ for the compactification. It is important to find a 
dynamical reason for such a large radius. Here let us address a much more 
modest question: why should the exponent for the effective cosmological 
constant, $m^4_{Planck}/\Le$, be bigger than the exponent for the hierarchy 
factor, $m^2_{Planck}/m^2_{Higgs}$. Since both exponential 
factors have similar origins, we actually have an inequality between these 
two exponents that is satisfied in nature.

The solution we have considered is the ``one horizon'' region in 
Figure~\arabic{kphaseeps}.  As we approach $q_0=k_1+k_2$ and 
$q_1=k_2-k_1$, both $L_1-L_0$ and $L_2-L_1$ become large. 
Since $\Delta l_2 > \Delta l_1$ (that is, 
$(q_0+q_1)(L_2-L_1)> (q_0-q_1)(L_1-L_0)$),
we have the inequality (\ref{important}), namely,
$\ln(m^4_{Planck}/\Le) > \ln(m^2_{Planck}/m^2_{Higgs})$, 
as is the case in nature.
However, we can actually make a stronger inequality, that is,
$L_2-L_1> L_1-L_0$ with $q_0+q_1> q_0-q_1$ ({\em i.e.}, the particle 
horizon is in the larger segment of the circle).
Let us assume that the positions 
of the branes are fixed, that is, the stationary branes satisfy their 
corresponding equations of motion. 
There are two distinct cases here:
\setcounter{Lcount}{0}
\begin{list}{(\alph{Lcount})}
  {\usecounter{Lcount}}
  \item $L_2-L_1>L_1-L_0>0$ and $k_2>k_1$; that is, the particle horizon 
    is in the larger segment of the circle between the branes.  This 
    case is schematically shown in Figure~\arabic{bulkbehavioureps}.
 \item The particle horizon is in the shorter segment of the circle.
    There are two equivalent descriptions of this case. We can either
    fix $L_2-L_1>L_1-L_0>0$, which implies that $\Delta l_2 < \Delta l_1$ 
    and $k_1>k_2$; or choose $L_1-L_0>L_2-L_1>0$ in the $k_2>k_1$ region
   with $\Delta l_2 > \Delta l_1$.  The Hubble constant in this second 
    case (with the particle horizon in the shorter segment of the circle) 
    is given by
   \be
       \tilde H^2(L_0) = (q_0+q_1)^2
         e^{-(\tilde \Delta l_2 + \tilde \Delta l_1)}
    \ee
    where $\tilde \Delta l_2 = (q_0-q_1)(L_2-L_1)/2$
   and $\tilde \Delta l_1= (q_0+q_1)(L_1-L_0)/2$.
\end{list}

Nature will pick the lower energy (or least action) solution, that is, 
the one with smaller $\Le$. 
Since the Newton's constant is essentially the same in both cases
more discussion on this later), we can simply compare
this Hubble constant to that in Eq.~(\ref{twocomH}):
\baray\nonumber
 (q_0+q_1)(L_2-L_1)+(q_0-q_1)(L_1-L_0) &>&
   (q_0-q_1)(L_2-L_1)+(q_0+q_1)(L_1-L_0)\\\label{inequale}
\Rightarrow L_2-L_1 &>& L_1-L_0.
\earay
This implies that the particle horizon should be in the larger segment
between the branes, as schematically shown in
Figure~\arabic{bulkbehavioureps}, where $L_2-L_1>L_1-L_0>0$ and
$k_2>k_1$.
Note that this inequality $L_2-L_1 > L_1-L_0$ together with $q_0>q_1>0$ 
is stronger than the inequality $(q_0+q_1)(L_2-L_1)> (q_0-q_1)(L_1-L_0)$.
Using this inequality (\ref{inequale}) and $q_0>q_1>0$, we can now compare
the exponent in Eq.~(\ref{twocomA}), where
\be
  \frac{m^2_{Higgs}}{m^2_{Planck}} \simeq A(L_1)
    \simeq e^{-\left(q_0-q_1\right)(L_1-L_0)} 
\ee
and the exponent in Eq.~(\ref{twocomH}), where we can write in terms of
$\Le$,
\be
  \Le \simeq \frac{2\sigma_0(q_0 + q_1)}{(q_0-q_1)}
    e^{[(q_0+q_1)(L_2-L_1)+(q_0-q_1)(L_1-L_0)]/2}
\ee
to obtain
\be
 \ln\biggl(\frac{m^4_{Planck}}{\Le}\biggr) > \ln\biggl(\frac{m^2_{Planck}}{m^2_{Higgs}}\biggr)
\ee
as is the case in nature, where $\ln({m^4_{Planck}}/{\Le}) \sim 2.3\times 122$
and $\ln({m^2_{Planck}}/{m^2_{Higgs}}) \sim  2.3 \times 32$.
To get a feeling for the magnitudes of the various quantities, we see that 
choosing $L_2-L_1 \sim 2 (L_1-L_0)$ and $\sigma_0 \sim 2 \sigma_1$ gives the 
correct order of magnitude. If we assume that the brane tensions are of 
Planck scale, the values of the cosmological constant and mass hierarchy 
observed in nature follow if $(L_1-L_0) \sim 10 m^{-1}_{Planck}$. 
Smaller brane tensions imply larger $(L_1-L_0)$.

In this model, the branes are stationary (that is, they satisfy
the brane equations of motion) and both $L_1$ and $L_2$
are stable. But we still need a dynamical explanation for why
$L_1$ (or $L_2$) is large. However, if we understand the hierarchy problem
via some means, then this brane world
model provides an explanation of the cosmological constant problem.
Now, the hierarchy problem may be understood from the renormalization 
group flow in 4-D quantum field theory.
It is well-known that the running of the couplings (marginal operators) 
are logarithmic versus the running of the mass terms (relevant operators).
A well-known example is the unification of the couplings in some
supersymmetric versions of the standard model.
If we accept some version of this picture as a resolution of the 
hierarchy problem, then the cosmological constant problem is explained 
by the inequality (\ref{important}) that appears in this two brane world 
model. In fact, the running of the couplings and masses in 4-D quantum 
field theory can be recast as the holographic renormalization group flow 
in the brane world scenario\cite{verlinde}. So, in a brane world model 
where the hierarchy problem is solved, the inequality (\ref{important})
tells us that we have an exponentially small 4-D cosmological constant.

\subsection{Two-Horizon Region}

There is another possible 
solution, namely, the ``two horizon'' region shown in 
Figure~\arabic{kphaseeps}.  In this case, the two branes are separated 
by particle horizons.  Since we are approaching $q_1=k_2+k_1$, the value of
$H^2/A(0)$ is finite, so the $q_1$ brane again has exponentially small 
cosmological constant. Within the 5-D classical Einstein theory, the two 
solutions are topologically distinct and so we cannot compare them. However, 
one may imagine a deviation from the pure AdS situation, or in quantum 
gravity or string/M theory, where they may be compared. 
Which solution does the two brane system pick? As is discussed in more detail 
in Appendix~\ref{app:twobranedetails}, the
$4D$ effective theory has a well-defined $\Le$, so we may compare the 
values of $\Le$ in the two cases.
For large brane separation $L_2-L_1$, we have seen that $\Le$ is 
exponentially small in both cases.
By comparing $\Le$ in the two cases, the ``one horizon'' case
(where the hierarchy problem may be solved) has a smaller $\Le$, 
due to smaller $k_2$ and maybe also $k_1$.
For finite brane separations,
and with appropriate choices of $q_i$, it is possible that
the second case is preferred. One can imagine that the brane world starts
with (two) particle horizons separating the two branes. As the brane 
separations (and the radius of $S^1$) increase, the bulk cosmological 
constants on two sides jump to the values for the hierarchy solution phase.
We do not expect this to happen in the 5-D theory. However, within the 
string/M theory framework, such a transition is akin to a topological change,
a possibility that deserves further study.

\section{Multibrane World}
\label{multibraneoverview}

It should be clear that a three brane or multibrane model can 
provide a similar solution to both fine 
tuning problems by the same mechanism; on the visible brane 
$H^2 \sim\exp(-2\Delta l_2)$, with a coefficient dependent on all 
three brane tensions, and $A(L_1)/A(L_0)\simeq\exp(-2\Delta l_1)$.
However, the 5-form field strength mechanism to provide piecewise 
constant bulk energy densities fails in an orbifolded space and it is 
not currently known how to achieve the different adjustable bulk cosmological 
constants.  A multibrane world with $5$-form field strengths providing 
the piecewise bulk cosmological constant can give a solution to both 
fine-tuning problems, in non-orbifolded spaces. Other variations of the 
models studied here should be explored further.

We can now give a general picture of the physics when the brane separations
are large.
In the stationary situation, we shall treat the brane positions $L_i$
as fixed, while the parameters $\ki$ between branes and $\yi$
are to be determined by the jump conditions, via Eqs.~(\ref{boundary}) 
and (\ref{jumpn}). 
By solving the jump conditions, the Hubble constant
seen by observers on the 
$i$th brane can be expressed in terms of the 5-D gravitational 
coupling $\kappa$, its brane tension $\sigma_i$ and the  
tensions $\sigma_{i-1}$ and $\sigma_{i+1}$ and positions 
$L_{i-1}$ and $L_{i+1}$ of its neighboring branes.
For large brane separations the relation (\ref{jumpn}) between the bulk 
parameters $k_i$ and $k_{i+1}$ and the brane tension becomes
\baray
\label{kepsilon}
&&  q_i = k_i\coth[k_i(L_i-y_i)]
    - k_{i+1}\coth[k_{i+1}(L_i-y_{i+1})],\nonumber \\
&&  \to s_i^- k_i-s_i^+ k_{i+1}
\earay
where $s_i^+$ is the sign of $A^\prime$ at $y\to L_i^+$, and the corrections 
are exponentially small for large separations. Therefore, 
for any of the four possible sign choices we see that $H^2(L_i)$ 
becomes exponentially small.
The implications of the various sign choices for the cosmological constant
and hierarchy problems can be summarized qualitatively in the following table,
where it is assumed that $k_{i+1}\ge k_i>0$. (The opposite inequality 
$k_{i+1}\le k_i$ does not yield any new possibilities, as they can be obtained
from the table below by exchanging some of the cases.)

\bigskip

\begin{center}
  \begin{tabular}{|c|c|l|c|c|c|c|c|}
    \hline
    &A(y)&Case&$q_i$&stability&$H^2(L_i)$&$G_N$&Hierarchy\\
    \hline\hline
    1&\epsfbox{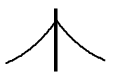}&$q_i\ge k_{i+1}+k_i$&+&\Y&+&+&\N\\
        \hline
    2&\epsfbox{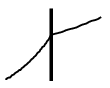}&$q_i\le k_{i+1}-k_i$&+&\Y&+&+&\Y\\
    \hline
    3&\epsfbox{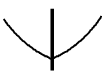}&$q_i\le -(k_{i+1}+k_i)$&-&\P&+&+&\Y\\
    \hline
    4&\epsfbox{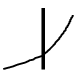}&$q_i\ge -(k_{i+1}-k_i)$&-&\N&+&+&\Y\\
       \hline
  \end{tabular}
\end{center}

We consider cases for which $H^2(L_i)>0$ only.
For case (1) the scale factor is locally peaked at the brane. 
The brane in case (3) can have no zeros between it and either neighboring 
brane, while the remaining cases may have a horizon on one side 
only. The $q_i$ column gives the sign of the brane tension; stability is
expected for a brane with a positive tension.  Generically, a negative 
tension brane is unstable, as in case (4), but, it is possible that 
some stabilization mechanism may be found in a more complicated model.
Stability is assumed 
for a negative tension brane if it sits at an orbifold fixed point, so 
that the fluctuation modes that will cause destabilization have presumably 
been projected out. In this case, the $\Bbb{Z}_2$ symmetry requires  
$k_{i+1}=k_i$. This is represented by ``~?~'' in case (3) in the 
table.

As we shall explain later, the Newton constant is always positive and is 
the same on all branes.  The value of the 4-D Newton's constant $G_N$ may 
be determined by the 
introduction of a small matter density in case (1), where $A(y)$ is peaked 
at the brane, as given in (\ref{LeGN}).
However, the same method cannot be applied to branes in the other cases,
where $A(y)$ is not peaked at the brane (see Appendix~\ref{app:GNandLE}).

In the above framework, and without fine-tuning, we see that there can be 
models that simultaneously
\setcounter{Lcount}{0}
\begin{list}{(\arabic{Lcount})}
  {\usecounter{Lcount}}
  \item have a stable positive tension brane,
  \item have an exponentially small 4-D positive cosmological constant, and
  \item yield a solution to the hierarchy problem.
\end{list}
Self-consistent models with all of these desirable properties 
require a combination of case (1) and case (2), so the 
simplest model must contain at least two branes.
Generically, a multibrane model can accommodate both an 
exponentially small cosmological constant and a solution to the 
hierarchy problem. 
The model of Ref\cite{ira} corresponds to the case (1) in the Table, 
in which the hierarchy problem is not solved. We shall show 
that the hierarchy problem in this model is simply the standard one,
so it may be solved by more conventional means.
The Randall-Sundrum solution to the hierarchy problem corresponds to 
the converging limit of case(3), where $k_{i+1}=k_i$, with a negative 
brane tension.  The hierarchy problem may be solved 
when the metric is non-vanishing before it reaches the next brane.  
This is indicated by a ``~\Y~'' in case(2) in the Table.
The simplest realization of such a solution to both the cosmological 
constant problem and the 
hierarchy problem is a two brane model in the compactified (but not 
orbifold) case. In this model, the hidden Planck brane corresponds
to case(1) and the visible brane corresponds to case(2).

In fact, the two brane compactified model, although simplest, is 
rather specific, and we can generalize the main result
to prove that $H^2/m_{Higgs}^2\ll 1$
in a multibrane model. Suppose that the visible brane is the
$i$th brane, with a warp factor $A(L_i)$ relative to the value
on the Planck brane, which is somewhere to its left, with
no particle horizons in the intervening region. Let $D_i=L_i-L_{i-1}$ be
the distance to brane $i-1$ just to the left of the $i$th brane, where the
warp factor is $A(L_{i-1})>A(L_i)$, and let $D_{i+1}=L_{i+1}-L_i$ be the
distance to the brane $i+1$ just to the right of the visible
brane, where the warp factor is $A(L_{i+1})<A(L_i)$. (This
presumes that there is no particle horizon between branes
$i$ and $i+1$; if there is, then take $A(L_{i+1})=0$ and let
$D_{i+1}$ be the distance from the $i$th brane to the horizon.)

Then instead of Eq.~(\ref{twocomA}), we find
\baray
H^2&\approx&4q_i^2A(L_i)\biggl[{A(L_i)\over A(L_{i-1})}\biggr]^{D_{i+1}/D_i}
e^{-2q_iD_{i+1}}\biggl[1+{\ln(A(L_{i-1})/A(L_i))\over 2q_iD_i}\biggr]^2
\nonumber\\& & \qquad\times
\biggl[1-{[A(L_{i+1})]^{1/2}[A(L_{i-1})]^{D_{i+1}/2D_i}\over[A(L_i)]^{1/2+
D_{i+1}/2D_i}}e^{q_iD_{i+1}}\biggr];
\label{Heqi}
\earay
once again, we see that $H^2/m_{Higgs}^2$ is exponentially small,
although the model is more complicated than before.
(Consistency requires non-negative $H^2$.)
The two brane, uncompactified version of Eq.~(\ref{Heqi}), which is
appropriate when horizons surround two branes, one of which is
the visible brane (with tension $q_V$ and warp factor $A_V$) 
and the other the
Planck brane (with tension $q_P$ and warp factor $A_P=1$) is
\be
H^2\approx 4q_V^2A_V^{1+D_0/D_P}e^{-2q_VD_0}
\biggl[1+{\ln(A_V^{-1})\over 2q_VD_P}\biggr]^2,
\label{Heq}
\ee
where $D_0$ is the distance from the visible brane to the 
horizon to its right, and $D_P$ is the distance from the
visible brane to the Planck brane to its left. Eq.~(\ref{twocomA})
can be obtained from the uncompactified model by assuming
a periodic sequence of identical two brane regions bounded
by horizons.

%%%%%%%%%%%%%%%%%%%%%%%%%%%%%%%%%%%%%%%%%%%%%%%%%%%%%%%%%%%%%%%%%%%%%%%%%%%%%%%%%%%%%%%%%%%%%%%%%%%%%%%%%%%%%%%%%%%%%%%%%%%%
\section{Global structure of the solutions}
\label{sec:globalstructure}

All of the solutions discussed in this paper consist of regions of 5-D 
anti-deSitter space joined across the $3$-branes, together with some 
additional identifications in some of the
solutions.  However, the form (\ref{metric}) of the 5-D anti-deSitter
metric used here is unconventional, and in addition in some cases possesses
coordinate singularities where the metric coefficient $A(y)$ vanishes.  
Coordinate singularities of this type are present in all solutions for
which all the branes have positive tension \footnote{
In the two brane case, the coordinate singularities can
be avoided (by choosing the free parameters appropriately)
when the two branes are not identical.  However, in the symmetric
two brane case, a coordinate singularity between the two branes is
unavoidable.}.
As discussed in the introduction, these zeros of $A(y)$
are particle horizons, in the sense that signals traveling from a brane
to the horizon are perceived by observers on that brane to take an
infinite time to reach the horizon (although the elapsed proper time 
or elapsed affine parameter for null signals is finite).  These
horizons are closely analogous to the horizons in Rindler space.
In this section, we show that the coordinate singularities are removable by
exhibiting the coordinate transformation between the form (\ref{metric})
of the metric and a more standard form of 5-D AdS space.  
We also derive the global structure of the various bulk
solutions we have been discussing.  
The key result that we shall obtain is that the solutions typically
split into different disconnected components across the particle
horizons.  The global structure of the RS model 
has been discussed in Ref.\cite{global}.

\subsection{Different coordinate systems on 5-D anti deSitter space}

We start by discussing the global structure of the spacetime
(\ref{metric}) when no branes are present, so that $L_i =
-\infty$ and $L_{i+1} = + \infty$. 
The metric (\ref{metric}) can be written 
\be
ds^2 = dy^2 - A(y) dt^2 + A(y) e^{2 H t} \left[ (dx^1)^2 + (dx^2)^2 +
(dx^3)^2 \right],
\label{metric1}
\ee
where the coefficient $A(y)$ is given by Eq.~(\ref{scale}),
\be
A(y) = {H^2 \over k_i^2} \sinh^2 \left[ k_i (y-y_i) \right].
\ee
We define new coordinate $T$, $Y$ and $X^l$ for $1 \le l \le 3$ by
\baray
\label{eqn:tf1}
T &=& {1 \over 2} \tanh \left[ k_i (y-y_i)/2 \right] \left\{ ( 1 + H^2
{\bf x}^2 ) e^{H t} - e^{-H t} \right\} \\
Y &=& {1 \over 2} \tanh \left[ k_i (y-y_i)/2 \right] \left\{ ( 1 - H^2
{\bf x}^2 ) e^{H t} + e^{-H t} \right\} \\
X^l &=& H \tanh \left[ k_i (y-y_i)/2 \right] e^{H t} x^l,
\earay
where ${\bf x}^2 = (x^1)^2 + (x^2)^2 + (x^3)^2$.
The inverse of this transformation is given by
\baray
\label{eqn:tf2}
\tanh^2 \left[ k_i(y-y_i)/2 \right] &=& {\bf X}^2 + Y^2 - T^2 
 \\
\exp \left[ 2 H t \right] &=& { (T + Y)^2 \over {\bf X}^2 + Y^2 - T^2 }
\\
H x^l &=& {X^l \over T + Y },
\earay
where the sign of $y-y_i$ is taken to be the same as the sign of $T + Y$,
and where ${\bf X}^2 = (X^1)^2 + (X^2)^2 + (X^3)^2$.  In the new
coordinates $(T,Y,X^1,X^2,X^3)$ the metric takes the standard,
conformally-flat AdS form 
\be
 ds^2 = {4 / k_i^2 \over (1 + T^2 - {\bf X}^2 - Y^2)^2} \left[ -dT^2 + dY^2 +
(dX^1)^2 + (dX^2)^2 + (dX^3)^2 \right].
\label{metric2}
\ee
See Appendix~\ref{app:coords} for details \footnote{
We note that another conformally flat coordinate system is obtained
from the transformation 
$$
{\bar T} = { e^{- H t} \over H \, \tanh \left[ k_i (y-y_i) \right] },
\ \ \ \ \ \ \ \ \ \ 
%$$
%$$
{\bar Y} = { e^{- H t} \over H \, \sinh \left[ k_i (y-y_i) \right] }.
$$
In the coordinates $({\bar T},{\bar Y},x^1,x^2,x^3)$ the metric takes
the form
\be
ds^2 = {1 \over k_i^2 {\bar Y}^2} \left[ -d{\bar T}^2 + d {\bar Y}^2 +
(dx^1)^2 + (dx^2)^2 + (dx^3)^2 \right].
\label{conf_flat_nonzero_H}
\ee
The domain of the original coordinate system $(t,y,x^1,x^2,x^3)$ is
the region $|{\bar T}| > |{\bar Y}|$.}.

We now discuss some properties of the coordinate transformation.  From
Eq.\ (\ref{eqn:tf2}), the original coordinates $(t,y,x^1,x^2,x^3)$ cover only 
the region 
$Y^2 + {\bf X}^2 > T^2$ of the full spacetime.  If we define $R =
\sqrt{{\bf X}^2 + Y^2}$, the metric can also be written as
\be
ds^2 = {4 / k_i^2 \over (1 + T^2 - R^2)^2} \left[ -dT^2 + dR^2 + R^2
d\Omega_3^2 \right],
\label{metric2a}
\ee
where $d \Omega_3^2$ is the metric on the unit three sphere.
The
conformal factor diverges at the hypersurface $R = \sqrt{1+T^2}$, which is the 
timelike boundary at infinity of  the spacetime. The Penrose diagram is 
Fig.\ 20 (ii) of Ref.\ \cite{HE}. 

\addtocounter{figcount}{1}
\setcounter{hyperboloideps}{\value{figcount}}
\begin{center}
  \epsfbox{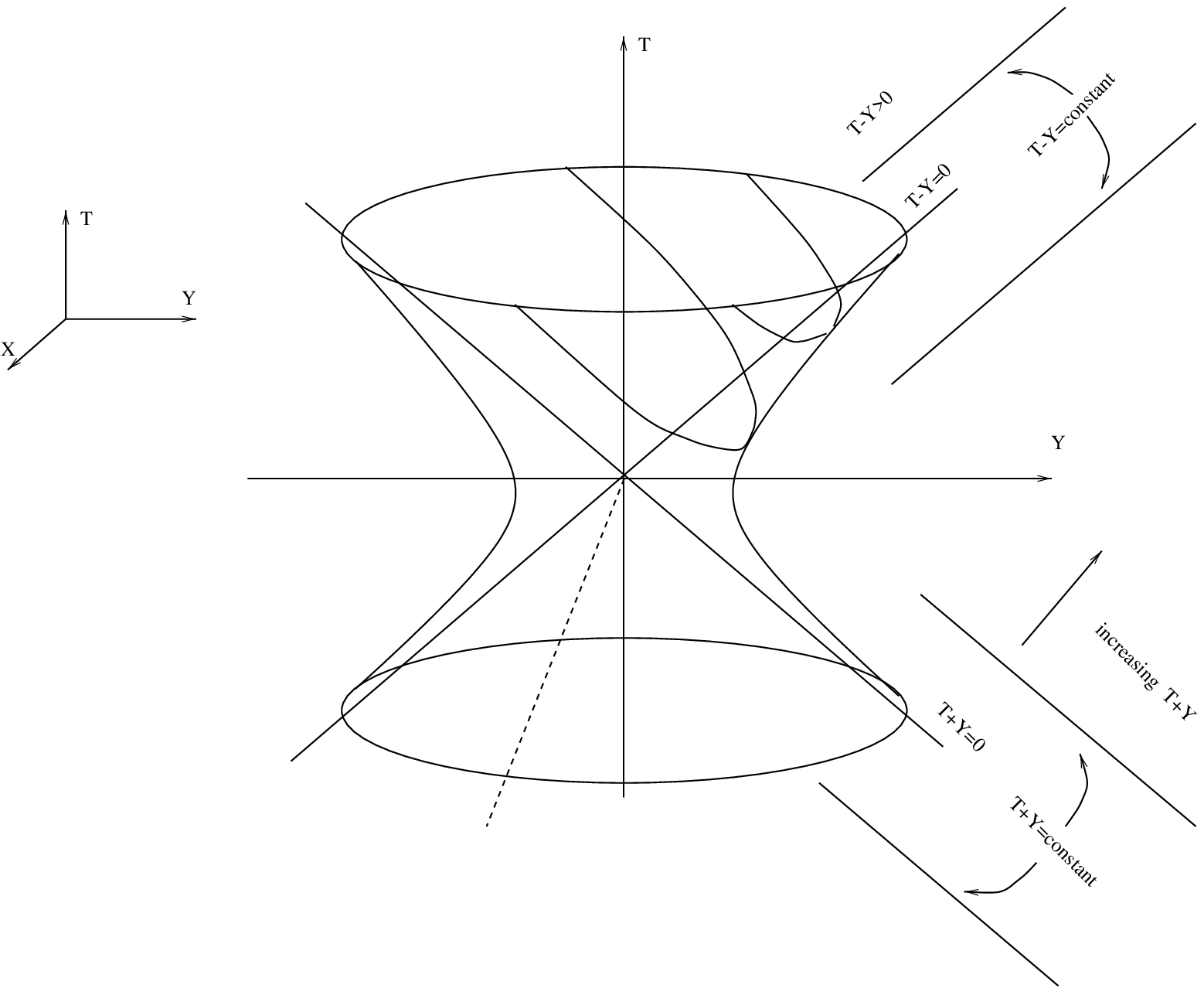}
  \parbox{12cm}{\vspace{1cm}
    FIG. \thefigcount\hspace{2pt} The shape of the hypersurface
  $\alpha=\alpha_0$ in the new coordinates $(T,Y,X^1,X^2,X^3)$. Two of the
  coordinates, $X^2$ and $X^3$, are suppressed.\vspace{12pt}}
\end{center}

Another useful coordinate system is
given by $R = \alpha \cosh \psi$, $T = \alpha \sinh \psi$, in which
the metric takes the form
\be
ds^2 = {4 / k_i^2 \over (1  - \alpha^2)^2} \left[ - \alpha^2 d\psi^2 +
d\alpha^2 + \alpha^2 \cosh^2 \psi \, d\Omega_3^2 \right].
\label{metric2b}
\ee
The range of these coordinates is $0 < \alpha < \infty$ and $-\infty <
\psi < \infty$. 
This coordinate system covers only the region $R<|T|$ of the full
spacetime.  From Eq.\ (\ref{eqn:tf2}), the coordinate $\alpha$ is
related to the original coordinates $t,y,x^i$ by
\be
\alpha = \left| \, \tanh \left[ k_i(y - y_i)/2 \right] \, \right|.
\label{alpharel}
\ee

Consider now the hypersurface $\alpha = \alpha_0$, where $\alpha_0 >
0$ is a constant.  In the coordinates $(T,Y,X^1,X^2,X^3)$, this
hypersurface is the hyperboloid 
\be
Y^2 + {\bf X}^2 - T^2 = \alpha_0^2.
\label{fullsurf}
\ee
From  Eq.\ (\ref{metric2b}), the induced metric on this hypersurface is
\be
{}^{(4)} ds^2 = {4 \alpha_0^2  \over k_i^2 (1  - \alpha_0^2)^2} \left[
- d\psi^2 +  \cosh^2 \psi \, d\Omega_3^2 \right],
\label{indm}
\ee
which is the spatially compact ($k=+1$), geodesically complete version
of 4-D deSitter space.  Using Eq.\ (\ref{alpharel}), we see that in
the original 
coordinates $(t,y,x^1,x^2,x^3)$, the hypersurface (\ref{fullsurf})
consists of the union of the two surfaces 
\be
y = y_i + 2 \tanh^{-1}(\alpha_0)/k_i
\label{surfright}
\ee
and
\be
y = y_i - 2 \tanh^{-1}(\alpha_0)/k_i,
\label{surfleft}
\ee
on the right and left hand sides of the coordinate singularity at $y =
y_i$.  Thus, the two hypersurfaces (\ref{surfright}) and
(\ref{surfleft}) are in fact connected, even though they appear
disconnected in the original coordinates $(t,y,x^1,x^2,x^3)$.  
The hypersurfaces (\ref{surfright}) and (\ref{surfleft}) are mapped
onto the regions $Y+T > 0$ and $Y+T < 0$ respectively of
of the hypersurface (\ref{fullsurf}).  The hyperboloid $\alpha =
\alpha_0$ is illustrated in Fig.\ \arabic{hyperboloideps}.
Note that the induced metric on each of the hypersurfaces
(\ref{surfright}) and (\ref{surfleft}) is, from Eq.\ (\ref{metric}), the
spatially non-compact ($k=0$), geodesically incomplete version of 4-D
deSitter metric \cite{HE}.
Note also that it follows from
Eq.\ (\ref{eqn:tf1}) that if the vector
$\partial / \partial t$ is taken to be future directed for $y>y_i$,
then it is past directed for $y<y_i$, just as in Rindler spacetime.

\subsection{The class of geodesically complete multibrane solutions}
\label{sec:fullmultibrane}

The various multibrane solutions discussed in the earlier sections of
this paper are geodesically incomplete.  In particular, all the
``branes'' discussed in the earlier sections are surfaces of constant
$y$, and it is clear from the discussion above that a surface of
constant $y$ comprises in reality just half of a brane.  Our goal in
this section is to find geodesically complete extensions of those
multibrane solutions, in which the ``other halves'' of all the branes
are present, and also to elucidate the global structure of those
extensions.

In attempting to find the geodesically complete extensions, it is
convenient not to start with the multibrane solutions discussed
earlier in the paper, but instead to start by directly constructing a class of
geodesically complete multibrane solutions via cutting and pasting
regions of 5-D AdS space in the form (\ref{metric2a}).  The metric
is 
\be
ds^2 = {4 \over k^2 \, (1  + T^2 - R^2)^2} \left[ -  dT^2 +
dR^2 + R^2 \, d\Omega_3^2 \right],
\label{metric2c}
\ee
where we have written $k$ for $k_i$.  Consider now a submanifold of
this manifold given by
\be
\alpha_1^2 \le R^2 - T^2 \le \alpha_2^2.
\ee
We shall make use of three types of such submanifolds:
\begin{itemize}
\item Type I, where $\alpha_1^2 = -\infty$ and $0 < \alpha_2^2 < 1$.
Such regions contain the ``particle horizon'' $R = |T|$ and are
spatially compact.  They have a single 4-D boundary at $\alpha = \alpha_2$.
\item Type II, where $0 < \alpha_1^2 < \alpha_2^2 < 1$, which are
spatially compact.  They have two 4-D boundaries at $\alpha=\alpha_1$ and
at $\alpha=\alpha_2$.
\item Type III, where $0 < \alpha_1^2 < 1$ and $\alpha_2^2 = 1$, which are
spatially non-compact.  They have a single 4-D boundary at $\alpha =
\alpha_1$.  
\end{itemize}
Each such submanifold is characterized by three numbers: the values of
$\alpha_1^2, \alpha_2^2$ and of $k$.

Consider now the class of solutions to Einsteins equations that one
can obtain by gluing together submanifolds of the above type at their
boundaries.  It is clear from the number of boundaries that each type
of submanifold possesses that such solutions, if connected, must be in
one of the following two classes:

\begin{itemize}
\item Class A:  The solution starts at one ``end''
with either at type I region or a type III region, then has one or
more type II regions (with either orientation, i.e, $\alpha$ can be
increasing or decreasing), and then finishes at the other ``end'' again
with either a type I region or a type III region.  
\item Class B:  The solution has no ``ends'' and instead consists
of one or more type II regions joined together in a circular fashion.  
\end{itemize}
Additional solutions can of course be generated from these classes of
solutions by 
performing additional identifications, and also by considering
solutions with more than one connected component.

The junction conditions in any of these solutions are
the same as those 
obtained earlier in the paper, except that they are now expressed in
terms of different quantities.  Continuity of the induced metric (\ref{indm})
across any boundary or brane requires that the quantity
\be
{ \alpha^2 \over k^2 (1 - \alpha^2)^2}
\ee
take the same value on both sides of the brane.  The Israel
junction condition is the requirement that the rescaled brane tension
$q$ at any boundary is the sum of two terms, one from each submanifold
on either side of the brane.  The absolute value of each of these
terms is
\be
{ k ( 1 + \alpha^2 ) \over 2 \alpha}.
\ee
The sign of each of these terms is positive for the boundary of a
type I region and for the second boundary $\alpha = \alpha_2 >
\alpha_1$ of a type II region.  It is negative for the boundary of a
type III region, and for the first boundary $\alpha = \alpha_1 <
\alpha_2$ of a type II region.

If one requires that all brane tensions be positive, then class A solutions
with type III regions on both ends are disallowed, as are all class B
solutions.  Thus, there are two different types of solutions for
positive tension branes: spatially compact class A solutions with type
I regions on both ends, and spatially non-compact class A solutions
with a type I region on one end and a type III region on the other end.

\subsection{Relation to earlier solutions}
\label{sec:horizons}

Consider now how the above geodesically complete multibrane solutions
are related to the multibrane solutions
discussed earlier in the paper, which were characterized by parameters
$k_i$, $L_i$ and $y_i$.

First consider the case of solutions
with no particle horizons (which requires negative tension branes
if the bulk regions are all AdS as in this paper).  For these
solutions, there is a simple and one-to-one correspondence 
between the solutions of Sec.\ \ref{sec:multibrane} and those of Sec.\
\ref{sec:fullmultibrane} above.  For a given solution $(k_i,y_i,L_i)$,
one can obtain the geodesically complete extension as follows.  For
each interbrane region $L_{i-1} < y < L_{i}$ where $L_{i-1}$ and $L_i$
are both finite, one constructs a region of type II with the
corresponding parameters obtained from Eq.\ (\ref{alpharel}), namely
$\alpha_1 = \min( {\hat \alpha}_1, {\hat \alpha}_2)$ and  
$\alpha_2 = \max( {\hat \alpha}_1, {\hat \alpha}_2)$, where
\be
{\hat \alpha}_1^2 = \tanh^2 \left[ k_i(L_{i-1} - y_i) /2 \right]
\ee
and
\be
{\hat \alpha}_2^2 = \tanh^2 \left[ k_i(L_{i} - y_i) /2 \right].
\ee
For any regions $L_{i-1} < y <L_{i}$ where
one of the boundaries is at $|y| = \infty$, one similarly constructs a
region of type III.  Then, one glues the regions of type II and III together to
get a geodesically complete solution.  The junction conditions will be
automatically satisfied.  Each interbrane region in the original
solution maps onto the subset $Y + T>0$ of a type II or type III
submanifold of the full solution.  Each $y=$ constant half-brane
in the original solution acquires an additional half-brane in the full
solution to make it a geodesically complete brane.

Now consider solutions with particle horizons.  First, for interbrane
regions not containing particle horizons one follows the same
procedure as above to construct the corresponding portions of the
maximally extended solution.   
Second, for each interbrane region $L_{i-1} < y < L_i$ containing a particle
horizon $y=y_i$, there are two possibilities for the global structure.

\begin{itemize}

\item One can map the entire interbrane region, including both sides
of the particle horizon, into a single region of type I.  Because the
type I region has a single boundary $\alpha = \alpha_2$, this
structure requires that the surfaces $y=L_{i-1}$ and $y=L_{i}$ both
map onto the boundary $\alpha = \alpha_2$, and 
therefore both surfaces must be equidistant from the particle horizon at
$y=y_i$.  More generally, the entire multibrane solution must be reflection
symmetric under the mapping $y-y_i \to -(y-y_i)$, since the regions
$y>y_i$ and $y<y_i$ are the $Y+T>0$ and $Y+T<0$ portions of the same
bulk region.

\item If the above property of symmetry under reflection through the
particle horizon is not satisfied by the multibrane solution
$(k_i,L_i,y_i)$, then the only possible global structure is as
follows.  One
identifies the region $y>y_i$ to the right of the particle horizon with the
$Y+T>0$ portion of a type I submanifold, and one identifies the region
$y<y_i$ to the left of the particle horizon with the $Y+T <0$ portion
of a {\it different} type I submanifold.  Therefore the maximally
extended multibrane solution can have several disconnected components.
\end{itemize}

The general structure of a multibrane solution $(k_i,L_i,y_i)$ without
any additional identifications and without any special symmetries is
therefore as follows.  The spacetime can be split up into cells
bounded by particle horizons, but with no particle horizons within any
given cell.  Then, the maximally extended solution consists of a
number of different connected components, one for each cell.
Since the physics of each connected component is independent from all
the others, it is natural to restrict attention to solutions with only
one connected component or cell.

This global structure justifies our prescription of Sec.\ \ref{sec:effaction}
for obtaining an effective 4-D theory of restricting the 
integral over the fifth dimension to between pairs of particle horizons.
The global structure also explains the fact that the Hubble constant
$H(L_i)$ on any brane can be expressed entirely in terms of the
tensions and locations of the branes in the same connected component,
without reference to the detailed arrangement of external branes.
For example, consider two branes
in an {\it uncompactified} spacetime. Let the two branes
be sandwiched between two horizons, and assume there is no
horizon between the branes. Let the two branes be at $y=0$
and $L$, and let the bounding horizons be at
$y=-y_0$ and $y=L+y_2$. Assume that $A(0)=1>A(L)$. The expansion
rates on the two branes turn out to be
\baray
H^2(0)&\approx& 4q_L^2\biggl[1+{\ln(1/A(L))\over 2q_LL}\biggr]^2
[A(L)]^{1+y_2/L}e^{-2q_Ly_2}\nonumber\\
H^2(L)&\approx& 4q_L^2\biggl[1+{\ln(1/A(L))\over 2q_LL}\biggr]^2
[A(L)]^{y_2/L}e^{-2q_Ly_2},
\label{newHeq}
\earay
where $q_L$ is the tension on the brane at $L$, and we have
written the solution in terms of $A(L)$, which may be regarded
as an observable quantity, the ratio 
$(m_{Higgs}/m_{Planck})^2$ found on the visible brane.

For solutions with additional identifications, the existence of
particle horizons can be compatible with the maximally extended
solution having only one connected component.  For example, consider
the the two brane compactified model of Sec.\ \ref{sec:ccandhier}.
That model has branes at $y=L_0$, $y=L_1$ and $y=L_2$, but $L_0$ and
$L_2$ are identified.  When there is only one horizon, then the
maximally extended solution is connected and starts with a type I
region at one end, then has a type II region, then has another type I
region at the other end.  When there are two horizons the maximally
extended solution has two connected components.

The global structure analysis also shows that the multibrane solutions
of the earlier sections are a subset of the full set of multibrane
solutions, for the following reason.  In the earlier sections, it was
always assumed that the value of the bulk cosmological constant was
the same on both sides of any particle horizon in a given interbrane
region.  However, since the two sides of a particle horizon typically
correspond to two {\it different} type I regions in the maximally extended
solution, there is no reason to make this restriction.  The most
general class of multibrane solutions, when translated into the 
old $(t,y,x^1,x^2,x^3)$ coordinates, allows interbrane regions with
particle horizons where the value of $k$ appears to change
discontinuously as one crosses the particle horizon.
The discontinuity is only apparent as the two sides of the particle
horizon are not physically connected.

\section{4-D Newton's Constant and Particle Horizons}
\label{sec:gmh}

There are some issues that we like to resolve in this section:
the determination of the 4-D Newton's constant $G_N$ and when a large 
warp factor helps to solve the hierarchy problems.
We have already assumed the some of the results of this section in 
earlier discussions. Here we like to clarify and justify them.

\subsection{The Issue}

Naive determinations of $G_N$ sometimes yield different answers. 
In the multibrane model, we face the following issue. 
We know of 3 approaches to determine the 4-D Newton's constant $G_N$:
\setcounter{Lcount}{0}
\begin{list}{(\roman{Lcount})}
  {\usecounter{Lcount}}
  \item integrate over the $5$th dimension in the 5-D action to obtain 
    the low-energy effective 4-D action\cite{RS1}, as discussed earlier;
  \item solve for the trapped gravity mode and determine its probability
    (wavefunction squared) on the brane\cite{RS2};
  \item calculate the Hubble constant and determine the coefficient of 
    the 4-D cosmological constant or the  matter density contribution 
    term\cite{oldcosmo}.
\end{list}
  
So far in this paper, we have used the approaches (i) and (iii). In approach 
(iii), note that the 
value of $G_N$ depends only on the local ({\em i.e.}, neighboring) 
properties of the brane on which we measure the value of $G_N$:
\be 
\label{cosmoGNi}
G_N =  G_5\frac{2k_ik_{i+1}}{k_i + k_{i+1}}
\ee
for the $i$th brane, where $8\pi G_5=\kappa^2$. Recall that $k_i$ 
($k_{i+1}$) is a function of $q_i$ as well as $q_{i-1}$ and $L_i-L_{i-1}$ 
($q_{i+1}$ and $L_{i+1}-L_i$) (where the explicit dependence is 
model-dependent), so $G_N$ depends on neighboring properties only. 
Specifically, it does not depend on the total number of branes in the model.
On the other hand, if we use the first approach, and naively integrate 
over the whole $5$th dimension ({\em i.e.}, from $y=-\infty$ to $y=\infty$ 
in the uncompactified case), we will get a $G_N$ that depends on properties 
of all the branes in the model, specifically, on $N$, the number of branes. 
This is also the case in the second approach. If we 
normalize the gravity trapped mode over all $y$, we will typically 
get a $G_N$ that is roughly $N$ times smaller than for
the single brane model. That is, naively, the answer from (i) and (ii) 
will be different than that from (iii). 

The resolution to this puzzle is already evident in the global structure
analysis just discussed, where a zero of the metric corresponds to a 
particle horizon. There, we see that the metric used is geodesically 
incomplete, so we should not integrate beyond the particle horizon.
In calculating $G_N$ for the $i$th brane using 
the first approach, we should integrate over the $5$th dimension only 
to the nearest particle horizon ({\em i.e.}, the zero of the metric) 
on each of the brane. This also applies to the normalization of the 
trapped gravity mode (or any other 
mode) in the second approach. This way, the three approaches 
produce the same answer for $G_N$, as should be the case. 
Furthermore, $G_N$ and the effective cosmological constant is the same 
for all observers within the the same particle horizons,
as pointed out earlier.

The reasoning presented below may be summarized as follows. 
It takes infinite time for a signal emitted from a brane to reach 
the particle horizon, as seen by observers on the brane. 
The trapped gravity mode on the brane is treated 
as a perturbation of the metric in the ``almost'' conformally flat background 
metric\cite{RS2}. The nearest particle horizons in the $y$ coordinate 
map to $z= \pm \infty$ in the ``conformal metric'' coordinate. 
Thus, the normalization of the gravity mode over the entire $z$ coordinate 
only covers the region between particle horizons in the $y$ coordinate,  
so the properties of branes separated by particle
horizons are treated independently (wavefunctions and so $G_N$ are
independently normalized). 

For the negative tension brane in the Randall-Sundrum model, and for
the visible brane in the compactified model discussed above, a naive
determination using approach (iii) yields a negative $G_N$, 
contradicting the above result. In Appendix~\ref{app:GNandLE},
we discuss the ambiguity and the problem arising in this approach.

This above result also means that we cannot compare the 
mass scales between branes that are separated by particle horizons.
In this case, the mass hierarchy problem in this 
brane world scenario is no worse than the standard hierarchy problem.
We can still compare mass scales between branes that are not separated by  
particle horizons, and between mass scales measured by observers on the 
same (visible) brane. 
This implies that solutions to the mass hierarchy problem can be 
addressed only between branes that are not separated by particle horizons. 
Our discussion below follows closely the original discussion of 
Ref\cite{RS1,RS2}.

\subsection{Green's Function and Newton's Constant}
\label{sec:gfg}

Instead of bringing the metric into the conformally flat
form (\ref{conf_flat_nonzero_H}), we can bring the metric (\ref{metric}) into
an almost ``conformally flat'' form, where the $y$ coordinate becomes the $z$ 
coordinate given by:
\be
dz=\pm\frac{dy}{\frac{  H}{k}\sinh{\left[k\left(y-y_0\right)\right]}}=
\pm\frac{d\left[k\left(y-y_0\right)\right]}{H\sinh{\left[k\left(y-y_0
\right)\right]}}.
\ee
Here we have in mind that $H$ is very small, and all measurements 
involving $G_N$ are at distances much smaller than $1/H$.
This above equation can be readily integrated, and the result is:
\be
  H\left(z+z_0\right)=-\ln{\tanh{\frac{k}2\left|y-y_0\right|}}
\ee
where $z_0$ is defined so that when $y=0$ we also have $z=0$. 
We choose the ``$-$'' sign so that 
when $y\rightarrow y_0$ we will have $z\rightarrow \infty$. 
We observe that the space between the horizons $\left(-y_0,+y_0\right)$ 
is mapped into the entire real line in the new coordinate system. 
If the model contains many branes separated by horizons,
each interval will be mapped into the entire real line, so we end up
with a collection of
disconnected spaces, each space containing one single brane.
Using the above relationship between $y$ and $z$ we obtain the 
conformal factor of the metric:
\be
\frac{ H^2}{k^2}\sinh^2{\left[k\left(y-y_0\right)\right]}=
\frac{ H^2}{k^2\sinh^2{\left[  H\left(z+z_0\right)\right]}}.
\ee
According to the result from \cite{RS1,CS_JE_TJH} we obtain the 
following wavefunction for the trapped graviton mode:
\be
\psi_0\left(z\right)=N\left(\frac{  H}{k\sinh{\left[  H\left(z
+z_0\right)\right]}}\right)^{\frac32}
\ee
where $N$ is a constant that is determined by the 
normalization condition
\be
\int_{-\infty}^{+\infty}\vert\psi_0\left(z\right)\vert^2dz=1.
\ee
Calculating the integral we obtain
\be
-\frac{2N^2}{k}\frac{ H^2}{k^2}\left\{\frac14\ln{\frac{\sqrt{1+
\frac{H^2}{k^2}}+1}
{\sqrt{1+\frac{H^2}{k^2}}-1}}-\frac12\frac{\sqrt{1+\frac{H^2}{k^2}}}
{\frac{H^2}{k^2}}\right\}=1.
\ee
Expanding for small $H/k$ we obtain:
\be
N=\sqrt{k}\left\{1-\frac{H^2}{4k^2}+
\frac{H^2}{4k^2}\ln\left(\frac{8k^2}{H^2}+1\right)+
O\left(\frac{H^4}{k^4}\right)\right\}.
\ee
In the limit $H\rightarrow 0$ we obtain $N=\sqrt{k}$.

In order to obtain the relationship between the 5-D and
4-D Newton constants, we first note that
the relation between the physical perturbation and the trapped gravity
wavefunction is \cite{CS_JE_TJH}
\be
 \left. h_{\mu\nu}\right|_{physical}(x,z)
 \propto\hat h_{\mu\nu}(x)A^{\left(2-d\right)/4}A(z)\psi_0\left(z\right).
\ee
It turns out that $\psi_0(z)\propto A^{(d-2)/4}$ for this mode,
so that the physical perturbation is simply $\hat h_{\mu\nu}(x)A(z)$.
To get the right scaling of $G_{eff}$ on the brane, 
one also has to consider the source term carefully.
For a point particle confined to the $i$th brane 
and at rest at ${\bf x}={\bf x_0}$, only the $00$ component of the
energy momentum tensor is nonzero:
\be
T_{00}={M\over A(z_i)}\delta^{(3)}({\bf x}-{\bf x_0})\delta(z-z_i),
\ee
where the brane is at $z=z_i$, $A(z)=A(z_i)$ and $M$ is
the particle mass measured on the brane relative to the {\it
physical} metric, $A(y_i)\hat h_{\mu\nu}$. The corresponding
metric perturbation is
(apart from possible exponentially small corrections \cite{GT,TM})
\baray
 \hat h_{00}({\bf x})&\simeq&-{2G_5\vert\psi_0(z_i)\vert^2M\over A(z_i)
 \vert{\bf x}-{\bf x}_0\vert}
 \nonumber\\
 &=&-{2G_N m\over\vert{\bf x}-{\bf x_0}\vert},
\earay
where $m=M\sqrt{A(z_i)}$ is the mass of the particle in the {\it
effective} 4-D theory. (A similar analysis on this issue appeared while
this paper was being completed\cite{grinstein}.)

To see what the effective Newton constant is for brane-bound observers
using the physical metric, $h_{\mu\nu}=A(z_i)\hat h_{\mu\nu}$, let us first
change coordinates to a system in which the unperturbed 4-D metric is Minkowski
on the brane. Then it is clear that the mass produces a perturbation
which is identical to $\hat h_{00}$, so all we have to do is re-express
the solution found above in the new coordinate system. Since a coordinate
separation $r$ corresponds to a distance $R=r(L_i)=r\sqrt{A(z_i)}$, and 
$m=M\sqrt{A(z_i)}$, we find that the perturbation is 
$\hat h_{00}=-2G_NA(z_i)M/R$ {\em i.e.} it is identical to Newton's law, but
with a constant of gravitation $G_{eff}=G_NA(z_i)$.

We would also like to know the corrections to Newton's force law that
would be deduced on different branes. We find that, for two masses
$M_1$ and $M_2$ separated by a distance $r(L_i)$ on the brane at $L_i$,
\be
 F=-\frac{G_N A(L_i)M_1M_2}{r^2(L_i)}[1 + \frac{2C^2}{r^2(L_i)}]
\label{forceq}
\ee
where the masses $M_1$ and $M_2$ and the distance $r(L_i)$ are 
measured with the induced metric $\gamma_{\mu \nu}$. Here we also 
include the correction term due to gravity KK modes, where
$C^2(k_i,k_{i+1})$ is a function of the bulk cosmological constants
on the two sides of the Planck brane. We shall determine this function in 
the next section. When the the bulk cosmological constants on the 
two sides of the brane are equal, $C^2=k^{-2}$. 
In terms of the rescaled metric $\hat \gamma_{\mu \nu}$, 
$m_i=\sqrt{A(L_i)}M_i$ and $r(L_i)= r\sqrt{A(L_i)}$,  
and relative to the induced metric there is a force
\be
\hat F=A(L_i)F=-\frac{G_N m_1m_2}{r^2}[1 + \frac{2C^2}{A(L_i)r^2}].
\ee
This reproduces the familiar Newton's gravitational law as given by 
the 4-D effective theory (\ref{eff4da}).
We see that the correction to Newton's force law on the visible brane 
becomes important at $TeV$ scale distances.
The Newton's force law in different metrics is discussed in terms of 
Feynman diagram in Appendix~\ref{app:newtonsforcelaw}.

Consider the scenario in Figure~\arabic{2braneNewtonconsteps}.
The above observation shows that the Newton's constant as seen 
by observers on each brane in Figure~\arabic{2braneNewtonconsteps} 
yields the same relation (\ref{cosmoGNi}), even though the ratio 
$A(0)/A(L)$ can be exponentially large. The metric blow-ups beyond the 
particle horizons are unobservable. In fact, $G_N$ as seen by observers 
on the two branes in the $S^1/\Bbb{Z}_2$ orbifold model will have the 
same value.By now, the reason is clear.  In changing to the almost 
``conformally flat'' metric, we have to determine $G_N$ separately for 
each brane, with each region between the horizons being mapped into real 
line, $-\infty < z < +\infty$. Each region will contain a trapped 
graviton, and they do not ``communicate'' with each other across the 
particle horizon at $y=y_0$ since they belong to separate spaces.  For 
an observer on the brane at $y=0$, it takes infinite time for a 
light-like signal to travel from the brane to $y_0$:
\be
\Delta t=\frac1H\left\{\ln{\left[\tanh{k\left(y_0-y_{initial}\right)}
\right]}-
\ln{\left[\tanh{k\left(y_0-y_{final}\right)}\right]}\right\}
\ee
where $y_{initial}=0$ and $\Delta t$ is clearly divergent
when $y_{final}\rightarrow y_0$.
In terms of the 5-D parameters $\kappa^2$ and $q_i$, we see that
$A(0)/A(L)$ can be exponentially large. Since this huge factor is
not measurable by observers on either brane, it has nothing to do with
the hierarchy problem, that is, it neither solves nor
worsens the standard hierarchy problem.

\section{Corrections to Newton's Gravitational Law}
\label{sec:corrnewton}

In the two brane model where both the hierarchy and the cosmological 
constant problems are solved, we note that the bulk cosmological constants 
on the two sides of the Planck brane are generically different.
Here we consider this asymmetric situation, in particular the correction 
to Newton's gravitational law and the leading post-Newtonian effects. Let 
us summarize the key observation. 
The results are similar to that in the symmetric case,
that is, the leading order post-Newtonian effects are the same as that 
from Einstein theory, and the correction to Newton's gravitational law
is still $r^{-2}$ as in the symmetric case\cite{RS2}, with a different 
coefficient. We shall calculate this coefficient below. 
Since we consider the case where the effective 4-D cosmological constant 
is exponentially small, it is safe to ignore it ({\em i.e.}, set it to zero)
in most applications. This simplifies the analysis. The analysis 
below follows closely that of Randall and Sundrum\cite{RS2} and that of 
Garriga and Tanaka\cite{GT,GT2,zura}. 
Once we have the Green's function, we can also determine the 
correction to Newton's law for masses on the visible brane.
Although the $5$th dimension is compactified in the model,
the gravity KK spectrum is still expected to be continuous
because the model has a particle horizon.
 
The Planck brane at $y=0$ is a surface layer separating two AdS spaces with 
different cosmological constants. 
We want to change the coordinates to Gaussian normal coordinates so that we 
can apply Israel's junction conditions. We expect the required  change to be
of the order of the perturbation induced by the mass located on the brane. 
Let us 
denote the new coordinates by:
\be
\bar x^a_{\pm}=x^a_{\pm}+\xi^a_{\pm}\left(x^a\right).
\ee
In general we will have to make different changes in coordinates for the two sides of 
the brane, so from now on we will suppress the $\pm$. In the new coordinates the metric will be:
\be
ds^2=\bar g_{\mu\nu}d\bar x^{\mu}d\bar x^{\nu}+2\bar g_{\mu5}d\bar x^{\mu}d\bar y+\bar g_{55}d\bar y^2.
\ee
The metric transforms as:
\be
g_{ab}=\bar g_{pq}\frac{\partial\bar x^p}{\partial x^a}
\frac{\partial\bar x^q}{\partial x^b}=\bar g_{ab}+
\bar g_{aq}\xi_{,b}^{q}+\bar g_{pb}\xi_{,a}^{p}+O\left(\left(\xi^c\right)^2\right).
\ee
We include now fluctuations around this background.
We want to choose the functions $\xi^a (x^b)$ so that the perturbations $h_{ab}$ satisfy the RS
gauge condition while the coordinates $\bar x^a$ are Gaussian normal. Consequently we impose the conditions:
\begin{displaymath}
\begin{array}[h]{lllllllll}
\bar h_{55} & = & 0\;,\hspace{12pt}\bar h_{\mu5} & = & 0\;,\hspace{12pt}\bar g_{\mu5} & = & 0\;,\hspace{12pt}
\bar g_{55} & = & 1 
\nonumber \\
h_{55} & = & 0\;,\hspace{12pt}h_{\mu5} & = & 0\;,\hspace{12pt}h_{\mu}^{\hspace{8pt}\mu} & = & 0\;,\hspace{12pt}
h_{\mu\nu}^{\hspace{12pt},\nu} & = & 0.
\end{array}
\end{displaymath}
We transform now from the Gaussian normal coordinates to the Randall-Sundrum coordinates: 
\baray
ds^2=\left[\bar g_{\mu\nu}+\bar g_{\left(\mu\right|\lambda}\xi^{\lambda}_{,\left|\nu\right)}+\bar h_{\mu\nu}
\right]dx^{\mu}dx^{\nu}+2\left[\bar g_{\mu5}+\bar g_{\mu\lambda}\xi^{\lambda}_{,5}+\bar g_{55}\xi^5_{,\mu}
+\bar h_{\mu5}\right]dx^{\mu}dy+ \nonumber \\
\left[\bar g_{55}+2\bar g_{55}\xi^5_{,5}+\bar h_{55}\right]dy^2=
\left[g_{\mu\nu}+h_{\mu\nu}\right]dx^{\mu}dx^{\nu}+2\left[g_{\mu5}+h_{\mu5}\right]dx^{\mu}dy+
\left[g_{55}+h_{55}\right]dy^2.
\earay
The condition $h_{55}=0$ imposes $\xi^5_{,5}=0$ so this function can depend only on the coordinates parallel to the 
brane, $x^{\rho}$. The condition $h_{\mu5}=0$ imposes the condition:
\be
\bar g_{\mu\lambda}\xi^{\lambda}_{,5}+\xi^5_{,\mu}=0 \Longrightarrow 
\gamma_{\mu\lambda}\xi^{\lambda}_{,5}+\xi^5_{,\mu}=0
\ee
since we keep only first order terms. This fixes the $y$-dependence of the functions $\xi^{\mu}$; we have 
to impose continuity of these functions at $y=0$, $\xi^{\mu}_+\left(0\right)=\xi^{\mu}_-\left(0\right)$ :
\be
\xi^{\mu}=-\frac1{2k}\gamma^{\mu\lambda}\xi^5_{\lambda}+\hat\xi^{\mu}\left(x^{\rho}\right).
\ee
The functions $\hat\xi^{\mu}\left(x^{\rho}\right)$ will be fixed by choosing the harmonic gauge on the brane,
while the functions $\xi^5\left(x^{\rho}\right)$ will be determined by the tracelessness of the RS perturbations.
In the $\bar x^a$ coordinates the background metric is given by:
\be
\bar g_{\mu\nu}=\eta_{\mu\nu}e^{-2k\left|y+\xi^5\right|}\simeq\gamma_{\mu\nu}\left(1-2k\xi^5+\cdots\right)
\ee
so we can obtain the relationship between the perturbations on the two coordinate systems:
\be
h_{\mu\nu}=\bar h_{\mu\nu}-2k\gamma_{\mu\nu}\xi^5-\frac1{2k}\gamma_{\left(\mu\right|\lambda}\gamma^{\lambda\rho}
\xi^5_{,\rho\left|\nu\right)}+\gamma_{\left(\mu\right|\lambda}\xi^{\lambda}_{,\left|\nu\right)}=
\bar h_{\mu\nu}-2k\gamma_{\mu\nu}\xi^5-\frac1k\xi^5_{,\mu\nu}+\hat\xi_{\left(\mu,\nu\right)}.
\ee
 
Now the metric must be continuous at the brane. Expressing the metric in the Randall-Sundrum
coordinates, the background metric is continuous at $y=0$, so the fluctuations must also be 
continuous at $y=0$. Using the fact that the functions $\xi^{\mu}$ must be continuous at $y=0$, 
and that $\xi^5$ is independent of $y$, we obtain the following condition:
\be
\label{warp_factor}
k_+\xi^5_{+}\left(x^\rho\right)=k_-\xi^5_{-}\left(x^\rho\right).
\ee
We can understand this condition in terms of the ``brane-bending'' effect:
the bending of the brane should be the same as an observer in the bulk
approaches the brane from either side, $y>0$ or $y<0$, 
so we can ``glue'' the two AdS spaces together along the brane.  
The relation (\ref{warp_factor}) can also be understood in the following way.
There are 5 physical polarizations for the 5-D graviton. For the massive 
KK modes, this is precisely what one needs for a spin-2 particle.
For the massless mode, the scalar component, if present, will contribute 
a Brans-Dicke-like interaction that is clearly ruled out. Fortunately, 
this scalar mode is gauged away by $\xi^5(x^\rho)$. Since there is only one 
scalar mode to be gauged away, $\xi^5_{\pm}$ must be related.

The jump condition at the brane, $\bar y=0$, can now be written as:
\baray
\partial_{\bar y}\left[\gamma_{\mu\nu}+\bar h_{\mu\nu}\right]\vert_{0+}-
\partial_{\bar y}\left[\gamma_{\mu\nu}+\bar h_{\mu\nu}\right]\vert_{0-}=
\nonumber \\
-\frac{2\kappa^2}{3}\left[\sigma\left(\gamma_{\mu\nu}+\bar h_{\mu\nu}\right)+
3T_{\mu\nu}-T\gamma_{\mu\nu} \right]
\earay
where $T_{\mu \nu}$ is the energy-momentum tensor and $T$ its trace.
Using the explicit form of the background metric and the fact that the bulk
cosmological constants are determined to give an (almost) flat brane,
the above equation can be simplified to the following form:
\be
\frac12\left[\partial_{\bar y}\bar h_{\mu\nu}|_{0+}-
\partial_{\bar y}\bar h_{\mu\nu}|_{0-}\right]+
\left(k_++k_-\right)\bar h_{\mu\nu}=
-\kappa\left[T_{\mu\nu}-\frac13\gamma_{\mu\nu}T\right].
\ee
To obtain this equation we used the Csaki-Shirman solution for the brane 
tension\cite{CSh,horace} :
\be
\kappa^2\sigma^2=\frac32\left(\sqrt{\Lambda_+}+\sqrt{\Lambda_-}\right)^2.
\ee
The following steps are the same as in the paper \cite{GT}: Combining the 
equation of motion for the metric perturbations and the jump condition, we 
obtain the Green function:
\be
\left\{a^{-2}\Box^{\left(4\right)}+\partial_y^2-
4\left(k^2_+\theta\left(y\right)+k^2_-\theta\left(-y\right)\right)+
2\left(k_++k_-\right)\delta\left(y\right)\right\}
G_R\left(x,x^{\prime}\right)=\delta^5\left(x-x^{\prime}\right).
\ee
In the non-orbifolded case the Green function will have contributions 
from both ``symmetric'' and ``antisymmetric'' KK modes. 
In the orbifolded case the ``antisymmetric'' modes are projected out. 
The tracelessness of the Randall-Sundrum metric fluctuations will result 
in the following equation for the 
$\hat\xi^5_{\pm}\left(x^{\rho}\right)$ functions:
\be
\Box^{\left(4\right)}\left[\hat\xi^5_{+}+\hat\xi^5_{-}\right]=\frac13\kappa^2T 
\Longrightarrow \Box^{\left(4\right)}\left[k_+\hat\xi^5_++k_-\hat\xi^5_-\right]=
2\kappa^2\frac{k_+k_-}{k_++k_-}\frac{T}3 .
\ee
We will use this result later when we calculate the corrections 
to Newton's law. The other four functions, 
$\hat\xi^{\mu}\left(x^{\rho}\right)$ provide the gauge freedom needed to 
restore the 4-D linearized Einstein equations. We will choose the usual 
harmonic gauge \cite{Wald}, 
\be
\partial^{\mu}\left(\bar h_{\mu\nu}-
\frac12\eta_{\mu\nu}\bar h_{\lambda}^{\lambda}\right)=0
\ee
so that the linearized Einstein equations for different components of the 
metric will  decouple. Notice that the harmonic gauge condition is satisfied 
``up to KK corrections''.  The contribution from 
$\hat\xi^5_{\pm}\left(x^{\rho}\right)$ will
change the numeric coefficient of $T_{\lambda}^{\lambda}$ will change from 
$\frac13$ to $\frac12$ so that the linearized Einstein equations become:
\be
\Box^{\left(4\right)}\left[\bar h_{\mu\nu}-
\frac12\eta_{\mu\nu}\bar h_{\lambda}^{\lambda}\right]=
-2\kappa^2\left[T_{\mu\nu}-\frac12\eta_{\mu\nu}T\right].
\ee

The set-up consists of a single positive tension brane placed at $z=0$  and the bulk is made 
of two slices of $AdS_5$ with different cosmological constants, $\Lambda _{i}$.
The background metric can be expressed as:
\be
ds^2=e^{-B\left(z\right)}\left[\eta_{ab}dx^adx^b+dz^2\right]=
\frac1{\left(k_iz+1\right)^2}\left[\eta_{ab}dx^adx^b+dz^2\right].
\ee
On each side of the brane, the RS graviton satisfies the usual Schroedinger-like equation,
\be
\label{RS_Schr1}
-\partial^2_z\psi\left(z\right)+V\left(z\right)\psi\left(z\right)
=m^2\psi \left(z\right)
\ee
where:
\be
V\left(z\right)=\frac9{16}B^{\prime}\left(z\right)^2-
\frac34B^{\prime\prime}\left(z\right)
\ee
\be
B\left(z\right)=2\ln{\left(k_S\vert z\vert+k_Az+1\right)}
\ee
where $k_S=\left(k_-+k_+\right)/2$ and $k_A=\left(k_+-k_-\right)/2$, 
so that the equation for the wave function becomes:
\be
\label{RS_Schr2}
\left[-\partial^2_z\psi\left(z\right)+\frac{15\left(k_A+k_Ssgn\left(z\right)\right)}
{4\left(k_S\vert z\vert+k_Az+1\right)^2}-3k_S\delta\left(z\right)\right]\psi 
\left(z\right)=m^2\psi \left(z\right).
\ee
In order to solve the above equation, we first make the change of variable, 
$u=k_i\vert z \vert+1$ and then the change of function 
$\psi\left(u\right)=\sqrt{u}\chi\left(u\right)$ so that the equation becomes:
\be
u^2\partial_u^2\chi+u\partial_u\chi+\left(\frac{m^2}{k^2}u^2-4\right)\chi=0 .
\ee
The $m=0$ mode has the following solution:
\baray
\psi\left(z\right)=\left\{\begin{array}{lcr}
&\frac{N}{\left(k_-\vert z\vert+1\right)^{\frac32}},\; &z<0 \\
&\frac{N}{\left(k_+\vert z\vert+1\right)^{\frac32}},\; &z>0
\end{array}\right.
\earay
The normalization constant $N$ gives the 4-D Newton constant, 
\be
G_N=Ge^{-B(0)/4}\vert\psi\left(0\right)\vert^2=GN^2=G\frac{2k_+k_-}
{k_++k_-} .
\ee
In the limit $k_+=k_-=k$ we obtain the RS 
relationship between $G$ and $G_N$: $G_N=kG$

For the continuum mode, $m\not=0$, the solutions of Eq.~(\ref{RS_Schr2}) are linear combinations
of the Bessel  functions $J_2\left(\frac{m}{k_i}u\right)$ and 
$Y_2\left(\frac{m}{k_i}u\right)$, so the most general solution of 
Eq.~(\ref{RS_Schr2}) is:
\baray
\label{psiabcd}
\psi\left(z\right)=\left\{\begin{array}{lcr}
&af_-\left(z\right)+bg_-\left(z\right),\; &z<0 \\
&cf_+\left(z\right)+dg_+\left(z\right),\; &z>0
\end{array}\right.
\earay
where:
\be
f_{\pm}\left(z\right)=
\sqrt{\vert z \vert+\frac1{k_{\pm}}}J_2\left(m\left(\vert z\vert+
\frac1{k_{\pm}}\right)\right)\;,\;
g_{\pm}\left(z\right)=
\sqrt{\vert z \vert+\frac1{k_{\pm}}}Y_2\left(m\left(\vert z\vert+
\frac1{k_{\pm}}\right)\right).
\ee
The coefficients $a$, $b$, $c$ and $d$ are determined by the continuity 
condition at $z=0$ and the wavefunction normalizations. The calculation 
is straightforward but a little tedious.
The result allows us to determine the correction to 
the Newton's gravitational law for the Planck brane:
\be
V\left(r\right)\sim -\frac{G_Nm_1m_2}{r}-\int_0^{\infty}\frac{G_N}{k}
\frac{m}{2k}\frac{e^{-mr}}{r}dm=
-\frac{G_Nm_1m_2}{r}\left(1+\frac{C^2}{2r^2}+\dots\right)
\ee
where 
\be
C^2=\frac1{4k_+^2}\frac1{1-\rho+\rho^2}\left[\frac{10\rho\left(1-\rho+\rho^2\right)+
3\left(1+\rho^4\right)}{3\rho\left(1+\rho^2\right)+2\rho^2}\right]^2
\ee
where $\rho=k_+/k_-$. In the symmetric case ($\rho=1$), $C^2=k^{-2}$.
For observers on the visible brane at $L$, 
\be
V\left(r\right)\sim 
-\frac{G_Nm_1m_2}{r}\left(1+\frac{C^2}{2A(L)r^2}+\dots\right)
\ee
where $A(L)\sim 10^{-30}$ is the warp factor.
The physical fluctuations of the metric will be given by:
$\bar h_{\mu\nu}=h_{\mu\nu}+\gamma_{\mu\nu}\left[k_+\xi^5_++k_-\xi^5_-\right]$
where $h_{\mu\nu}$ is determined by $V\left(r\right)$.
Using the relationship $G_N=2k_+k_-G/\left(k_++k_-\right)$  we obtain the metric perturbations:
\be
\bar h_{00}=\frac{2G_NM}r\left(1+\frac{2C^2}{3A(L)r^2}\right)\;,\;
\bar h_{ij}=\frac{2G_NM}r\left(1+\frac{C^2}{3A(L)r^2}\right)\delta_{ij}
\ee
where $L$ can be $0$ or $L_i$ 
These solutions differ from the ones obtained by expanding the Schwazschild 
solution to linear order by a coordinate transformation.

\section{Discussions}
\label{sec:discussion}

The viewpoint developed in this paper leads to a connection between
the cosmological constant and mass hierarchy problems, resulting,
rather generally, to relationships of 
the form $(H/m_{Planck})^2\sim(m_{Higgs}/m_{Planck})^{2p}$, with $p>1$,
when the expansion of the Universe is dominated by cosmological
constant. Such a relationship is reminiscent of the so-called
``large numbers'' of cosmology, which have been variously
either dismissed as numerology or regarded
as physically significant clues to some fundamental connection
between the microphysical world and cosmology.
Most explanations of how these numbers could arise from
physics have tended to invoke variations on the theory of 4-D
gravitation, such as advocated by Dirac \cite{dirac} or
Brans and Dicke \cite{bdt}. However, Zeldovich \cite{zeld}
proposed a different idea, that it is the cosmological constant,
$\Le$, not $H$ that should appear in the large numbers, and
that they therefore tell us something fundamental about
the relationship of $\Le$ to the rest of physics. The 
viewpoint developed here is intermediate, since it is based
on multibrane solutions in 5-D gravitation, 
but only involves static spacetimes. In the context of
such solutions, $\ln(m^2_{Planck}/m^2_{Higgs})$ depends on 
some combination of dimensionless separations between branes,
and $\ln(m^4_{Planck}/\Le)$ depends on a different combination of
the separations,  and is always larger than 
$\ln(m^2_{Planck}/m^2_{Higgs})$. The values of these dimensionless
numbers depend on the particular arrangements of branes, which
are not determined in these static models, and we do not yet
understand what determines these arrangements fundamentally.
Moreover, even if we regard $\ln(m^2_{Planck}/m^2_{Higgs})$
as observable, so one combination of brane displacements is
knowable, the other combination, $\ln(m^4_{Planck}/\Le)$, depends
on a different combination, and is not fixed. It is clear
that for {\it dynamical} spacetimes, we
should expect some movement of the branes, resulting in
variability of both $\ln(m^4_{Planck}/\Le)$ and
$\ln(m^2_{Planck}/m^2_{Higgs})$. For some reason, which we
do not address here, the branes must have been relatively
stationary since {\it before} the epoch of cosmological nucleosynthesis,
avoiding, in particular, changes in $\ln(m^2_{Planck}/m^2_{Higgs})$ 
that could alter the predicted light element abundances.

In a 4-D spacetime theory, the cosmological constant includes
contributions from microphysics, which has many scales.
For example, among other contributions, supersymmetry breaking
around the electroweak scale
will introduce a cosmological constant of the order of $TeV^4$,
which is many orders of magnitude bigger than the value observed, around
$\Le \sim (10^{-12} eV)^4$. This is the cosmological constant problem.
This problem will generically persist if we consider a higher-dimensional
theory and compactify it to 4-D, because the resulting
effective 4-D theory will face the same fine-tuning problem.
This problem may be avoided by considering higher dimensional theory
with non-factorizable geometry, and the Randall-Sundrum model is the
simplest example. In fact, the RS model has an (exponential) warp
factor, which provides a solution to the well-known hierarchy problem.
In Ref\cite{ira}, the RS model is extended to solve the
cosmological constant problem.
A key feature of the model is that the effective 4-D cosmological constant
seen by an observer on the brane is not the vacuum energy density on the
brane, but a non-linear combination of the brane tension and the bulk
cosmological constants on the two sides of the brane.
If other branes are relatively far from the visible brane, that
non-linear combination becomes exponentially small, and the observer
thinks he/she is seeing an exponentially small cosmological constant.
For observers on the brane, it is natural to find an effective 4-D
description. Then one may ask what happens to the argument for the
need of fine-tuning in 4-D theory?
The issue turns out to be quite interesting.
Let us consider the single brane case first.
If one insists on a 4-D description for
observers on the brane, one ends up with a novel description via
the AdS/CFT correspondence\cite{AdS,gubser}.
Recall that, besides the trapped graviton, there is a continuous
spectrum of Kaluza-Klein modes from 5-D gravity.
Thus, in addition to the standard model fields (and whatever other fields 
that may exist) on the visible brane in an effective 4-D theory, an 
appropriate conformal field theory (CFT) must be included on the $3$-brane,
to account for the effects of the gravity KK modes. 
That is, the CFT is equivalent/dual to the 5-D
gravitational effects on the brane. This CFT interacts strongly and couples
to other brane modes only via gravity.
In fact, the classical correction to Newton's gravitational law in the 
5-D picture becomes a quantum effect ({\em i.e.}, the CFT contribution 
to the vacuum polarization of the graviton) in the 4-D picture\cite{DuffLiu}.
Let us take the strongly interacting $SU(N)$ ${\bf N}=4$ super 
Yang-Mills theory as the CFT. 
In the symmetric case, the bulk cosmological constant is
related to the c-number of CFT \cite{ctheorem,gubser}, 
$2 \pi ^4 \kappa ^2 \Lambda= 27(N^2-1)$.
Note that $N$ is discrete but not $\Lambda$.
In this paper, we have explored the idea of adjusting the bulk cosmological
constant in the context of static solutions of Einstein's equations, but
it is reasonable to expect that similar adjustments occur in a dynamical
spacetime. When branes move slowly, we would expect the bulk $\Lambda$
to vary slowly and continuously, perhaps just making transitions along a
sequence of static models like the ones constructed here.
As $\Lambda$ varies continuously, we have to consider another
CFT when $2 \pi ^4 \kappa ^2 \Lambda= 27(N^2-1)$
cannot be satisfied for integer $N$.
In the uncompactified multibrane model, we expect the AdS/CFT
correspondence to continue to be valid. Thus, as branes move, the
theory explores the space of strongly interacting CFT theories,
not just the parameter space of a given quantum field theory. 
That is, the effective 4-D theory changes as the
brane separation changes. One should no longer apply here
the argument for the need of fine-tuning inside a given 4-D theory
to obtain a very small 4-D cosmological constant.

\section{Summary and Remarks}
\label{sec:summary}

In this paper we argue that, for parallel $3$-branes that are relatively
far apart, an exponentially small cosmological constant is 
quite generic. We present a two brane model which solves the hierarchy 
problem \`a la Randall-Sundrum and has an exponentially small cosmological 
constant for observers on the visible brane. Moreover, the cosmological
constant, in Planck units, turns out to be much smaller than the Higgs
mass in Planck units, naturally and generically. To be explicit,
we write down the 4-D low energy effective action for the model.
We consider this model as another motivation for the brane world.

We see that the inequality (\ref{important}) is robust. This 
follows from the fact that both the visible brane and the Planck brane 
must be inside the same particle horizons. The cosmological constant
is exponentially small roughly as a function of the distance of the 
Planck brane from the particle horizon while the hierarchy factor is
exponentially small roughly as a function of the distance of the
Planck brane from the visible brane. Since the visible brane must be 
between the Planck brane and the particle horizon, the inequality
(\ref{important}) follows.
 
We discuss the physical implications of the particle horizons that
may appear in the AdS bulk. In the process, we clarify some 
confusing issues in the literature. We also discuss a possible
``phase transition'' in the two brane model, which deserves 
more attention. We have concentrated exclusively on bulks that are
purely AdS. Similar, simultaneous explanations for the cosmological 
constant and
mass hierarchy problems may also be possible when the bulks are not
pure AdS; this question also needs further study.

There are interesting questions that remain to be addressed.
Are the solutions stable under quantum effects ? We believe so. Both 
the hierarchy problem and the cosmological
constant problems are solved by involving factors that are
exponential functions of brane separations, so they become exponentially 
small as brane separation becomes large. This behavior resembles
the Yukawa force, which is robust under quantum corrections. This leads us 
to believe that quantum corrections will not 
destroy the exponential behaviors in the brane world.

It will be interesting to find a reason why the $3$-branes like to 
stay parallel. The dynamics of these brane world models will be 
important. String/M theory realization is another important question.

\section*{Acknowledgements}

We thank Philip Argyres, Gia Dvali, Zurab Kakushadze, Juan Maldacena, 
Vatche Sahakian and Adrian Salzmann for discussions.
This research is partially supported by NSF (S.-H.H.T. and E.E.F.) and NASA
(I.W.).

\appendix

\section{The Static Solution to the Brane Equation of Motion}
\label{app:eqnofmotion}

We can now check that the metric (\ref{scale}) satisfies all 
the equations of motion derived from the action (\ref{Action}). 
The Einstein equations, the equation 
of motion of the field $A_{a_1\cdots a_{d-1}}$, and the equation 
for the induced metric have already been checked. Here we check 
the equation for the embedding coordinates, {\em i.e.}, the brane 
equation of motion Eq.~(\ref{eq:wscalars}). In this appendix, we shall
express the brane equation of motion in terms of the metric 
(\ref{metric}). There are three types of embedding coordinates,
the spatial coordinates tangential to the brane, the time coordinate 
and the coordinate normal to the brane. 
We consider them separately.
\begin{itemize}
\item
The $a=j=1,2,3$ case. \\
This case is trivial, since all the terms in Eq.~(\ref{eq:wscalars}) are zero.  The last term
contains the field-strength tensor with two indices repeated:
$F^i_{\mu_1\cdots\mu_4}\epsilon^{\mu_1\cdots\mu_4}$ and is zero due to the antisymmetry of the tensor.
The first term can also be proven to be zero.  The embedding coordinates are scalars with respect to the
intrinsic coordinates of the brane, so we obtain:

\be
\nabla^{\mu}\nabla_{\mu}\xi^j_n=\frac1{\sqrt{\vert\gamma_n\vert}}\partial^{\mu}
\left[\sqrt{\vert\gamma_n\vert}\partial_{\mu}\xi_n^j\right]=\frac1{\sqrt{\vert\gamma_n\vert}}\partial^j
\left[\sqrt{\vert\gamma_n\vert}\right]=0
\ee
since $\vert\gamma_n\vert$ depends only on $\xi^0$ and $Y_n$. 
We now prove that the second term of the equation is also zero:
\baray
&& \Gamma^j_{ab}\partial_{\alpha}X^a\partial_{\beta}X^b\gamma_n^{\alpha\beta}=
\Gamma^j_{ab}\delta^a_{\alpha}\delta^b_{\beta}\gamma_n^{\alpha\beta}=\Gamma^j_{\alpha\beta}g^{\alpha\beta}=
\frac12g^{ja}\left[\frac{\partial g_{a\beta}}{\partial\xi^{\alpha}}+\frac{\partial g_{\alpha a}}{\partial\xi^{\beta}}-
\frac{\partial g_{\alpha\beta}}{\partial x^a}\right]g^{\alpha\beta}=\nonumber \\
&& \frac12g^{jj}\left[\frac{\partial g_{j\beta}}{\partial\xi^{\alpha}}+
\frac{\partial g_{\alpha j}}{\partial\xi^{\beta}}\right]g^{\alpha\beta}=
g^{jj}\sum_{\alpha}\frac{\partial g_{j\alpha}}{\partial\xi^{\alpha}}=g^{jj}\frac{\partial g_{jj}}{\partial \xi^j}=0
\earay
since the metric is diagonal and does not depend on $\xi^j$.
\item
The case $a=0$.
In this case the term containing the field strength tensor is also zero due to repeated indices and antisymmetry,
but the first two terms are no longer zero. However, it will turn out that the two terms will cancel each other.
The first term can be easily calculated:
\baray
&& \frac1{\sqrt{\vert\gamma_n\vert}}\partial^0\left[\sqrt{\vert\gamma_n\vert}\partial_0X^0\right]=
\frac{g^{00}}{\sqrt{\vert\gamma_n\vert}}\partial_0\left[\sqrt{\vert\gamma_n\vert}\right]=
\frac{g^{00}}{\sqrt{\vert g\left(L_n\right)\vert}}\partial_0\left[\sqrt{\vert g\left(L_n\right)\vert}\right].
\earay
In terms of the metric(\ref{metric}), this is equal to 
$-\frac{4H}{A\left(L_n\right)}$.
The second term can be calculated as in the previous case:
\baray
&& \Gamma^0_{\alpha\beta}g^{\alpha\beta}=\frac12g^{0a}\left[\frac{\partial g_{a\beta}}{\partial\xi^{\alpha}}
+\frac{\partial g_{\alpha a}}{\partial\xi^{\beta}}-\frac{\partial g_{\alpha\beta}}{\partial\xi^a}\right]g^{\alpha\beta}=
\frac12g^{00}\left[\frac{\partial g_{0\beta}}{\partial\xi^{\alpha}}+\frac{\partial g_{\alpha0}}{\partial\xi^{\beta}}\right]
g^{\alpha\beta}-\nonumber \\
&& \frac12g^{00}\frac{\partial g_{\alpha\beta}}{\partial\xi^0}g^{\alpha\beta}=
\frac12g^{00}\left[2\frac{\partial g_{00}}{\partial\xi^0}\right]-\frac12g^{00}\frac1{g}\frac{\partial g}{\partial\xi^0}=
-g^{00}\frac1{\sqrt{\vert g\vert}}\frac{\partial\sqrt{\vert g\vert}}{\partial\xi^0}
\earay
which equals $\frac{4H}{A\left(L_n\right)}$ in the metric(\ref{metric}).
The two terms cancel each other, so this equation is identically satisfied.

\item
The case $a=5$.
If $a=5$ the term containing the field strength tensor is no longer zero, so we calculate this term first:
\baray
&& F^5_{a_1\cdots a_4}\partial_{\mu_1}X^{a_1}\cdots\partial_{\mu_4}X^{a_4}\epsilon^{\mu_1\cdots\mu_4}=\nonumber \\
&& F^5_{\mu_1\cdots_mu_4}\epsilon^{\mu_1\cdots\mu_4}=g^{5a}F_{a\mu1\cdots\mu_4}\epsilon^{\mu_1\cdots\mu_4}=
4!F_{50123}=4!F_{01235}.
\earay
The equation of motion for the field $A$ has already been solved, and the solution is:
\be
F^{a_1\cdots a_5}=\frac{e}{\sqrt{\left(\vert g\vert\right)}}\epsilon^{a_1\cdots a_5}.
\ee
We can use this result to calculate the value of $F_{01235}$:
\baray
&& F_{01235}=F^{a_1\cdots a_5}g_{1a_1}\cdots g_{5a_5}=F^{01235}g_{11}\cdots g_{55}=\nonumber \\
&& gF^{01235}=-\sqrt{\vert g\vert}e\epsilon^{01235}=-e\sqrt{\vert g\vert}.
\earay
The second term is:
\baray
&& \Gamma^5_{ab}\partial_{\alpha}X^a\partial_{\beta}X^b\gamma^{\alpha\beta}=\Gamma^5_{\alpha\beta}g^{\alpha\beta}=
\frac12g^{5a}\left[\frac{\partial g_{a\beta}}{\partial\xi^{\alpha}}+\frac{\partial g_{\alpha a}}{\partial\xi^{\beta}}
-\frac{\partial g_{\alpha\beta}}{\partial\xi^a}\right]g^{\alpha\beta}=\nonumber \\
&& -\frac12g^{\alpha\beta}\frac{\partial g{\alpha\beta}}{\partial y}=
-\frac12\frac1g\frac{\partial g}{\partial y}=-\frac1{\sqrt{\vert g\vert}}\frac{\partial\sqrt{\vert g\vert}}{\partial y}.
\earay
Finally using the fact that $\sqrt{\vert\gamma_n\vert}=\sqrt{\vert g\left(L_n\right)\vert}$, the equation becomes:
\baray\label{tensionvscharge}
\sigma_n\frac{\partial\sqrt{\vert g\vert}}{\partial y}+e_ne\sqrt{\vert g\vert}=0.
\earay
In the case where the derivative of $g_{\mu \nu}$ has a jump across the brane,
we should take the average of the values on the two sides.
This yields Eq.(\ref{branemot}).
Naively, we expect this equation to impose a non-trivial constraint 
between the brane tension and the brane charge, but in the static case we are 
interested in, this equation will be identically satisfied if the Einstein 
equation is satisfied.
\end{itemize}

A similar result can be obtained for the brane equation of motion in 
the other models discussed in \S\ref{sec:integconst}. 

\section{Details of the two brane model}
\label{app:twobranedetails}

Using the continuity and junction conditions
(\ref{twobccch},\ref{cont_cond}),
we can solve for each of the $\coth$ functions in (\ref{hierarjump}) in
terms 
of $k_1,k_2,q_0,q_1$, giving
\baray\nonumber
\coth\left[k_1\left(L_0-y_1\right)\right]=\frac{k_2^2-k_1^2-q_0^2}{2q_0k_1}
\;,
\;\coth\left[k_1\left(L_1-y_1\right)\right]=\frac{k_1^2-k_2^2+q_1^2}{2q_1k_1}\\
\coth\left[k_2\left(L_1-y_2\right)\right]=\frac{k_1^2-k_2^2-q_1^2}{2q_1k_2}
\;,
\;\coth\left[k_2\left(L_2-y_2\right)\right]=\frac{k_2^2-k_1^2+q_0^2}{2q_0k_2}.
\label{cothk}
\earay
Following from Figure~\arabic{bulktypeseps} and the subsequent discussion, the
possible 
values of the various $\coth$ functions can be deduced.  Using
(\ref{cothk}), these 
inequalities are constraints on the allowed values of $k_1$ and $k_2$,
given $q_0$ 
and $q_1$. Here $\coth\left[k_1\left(L_0-y_1\right)\right]\le-1$ and 
$\coth\left[k_2\left(L_2-y_2\right)\right]\ge1$ always for positive
tension branes, 
which imply $k_2^2\le(k_1-q_0)^2$ and $k_1^2\le(k_2-q_0)^2$ respectively.
The other 
inequalities depend on the behaviour of the scale factor in the bulks
between the 
branes:
\begin{itemize}
  \item if there is a horizon between $L_1$ and $L_0$ only,
    \baray\nonumber
      \coth\left[k_1\left(L_1-y_1\right)\right]\ge1\quad&\Rightarrow
      k_2^2\le(k_1-q_1)^2,\\\label{app:tb:H1}
      \coth\left[k_2\left(L_1-y_2\right)\right]\ge1\quad&\Rightarrow
      k_1^2\ge(k_2+q_1)^2;
    \earay
  \item if there is a horizon between $L_2$ and $L_1$ only,
    \baray\nonumber
      \coth\left[k_1\left(L_1-y_1\right)\right]\le-1\quad&\Rightarrow
      k_2^2\ge(k_1+q_1)^2,\\\label{app:tb:H2}
      \coth\left[k_2\left(L_1-y_2\right)\right]\le-1\quad&\Rightarrow
      k_1^2\le(k_2-q_1)^2;
    \earay
  \item if there is a horizon in both bulks,
    \baray\nonumber
      \coth\left[k_1\left(L_1-y_1\right)\right]\ge1\quad&\Rightarrow
      k_2^2\le(k_1-q_1)^2,\\\label{app:tb:2H}
      \coth\left[k_2\left(L_1-y_2\right)\right]\le-1\quad&\Rightarrow
      k_1^2\le(k_2-q_1)^2.
    \earay
\end{itemize}
The allowed regions of the ``k-plane'' can be further constrained by
solving 
(\ref{cothk}) for the brane separations in terms of $k_1$ and $k_2$:
\baray\nonumber
L_1-L_0=\frac1{2k_1}\ln\left[\frac{\left(k_1+k_2+q_1\right)\left(k_1-k_2+q_1\right)\left(k_1+k_2+q_0\right)\left(k_1-k_2+q_0\right)}
{\left(k_1+k_2-q_0\right)\left(k_1-k_2-q_0\right)\left(k_1+k_2-q_1\right)\left(k_1-k_2-q_1\right)}\right]\\\label{ls}
L_2-L_1=\frac1{2k_2}\ln\left[\frac{\left(k_2+k_1+q_1\right)\left(k_2-k_1+q_1\right)\left(k_2+k_1+q_0\right)\left(k_2-k_1+q_0\right)}
{\left(k_2+k_1-q_1\right)\left(k_2-k_1-q_1\right)\left(k_2+k_1-q_0\right)\left(k_2-k_1-q_0\right)}\right].
\earay
When there is no horizon between $L_0$ and $L_1$, $L_2-L_1$ is positive is
the 
region allowed by (\ref{app:tb:H2}), but the condition $L_1-L_0\ge0$
requires that 
$k_2^2\ge k_1^2+q_0q_1$.  There is a reflection of this constraint about
the line 
$k_1=k_2$ in order for $L_2-L_1\ge 0$ when there is no horizon in the
second bulk.  
Finally, we have imposed $L_2-L_1\ge L_1-L_0$ to distinguish the
physically distinct 
configurations of the branes and horizons between them.  The function 
$(L_2-L_1)-(L_1-L_0)$, obtained from (\ref{ls}), can be evaluated
numerically, 
revealing the regions in which this last condition is satisfied.
Combining all 
constraints gives Figure~\arabic{kphaseeps}.  

We find in Section~\ref{sec:ccandhier} that when there is a horizon in the
second 
bulk, in the limit of large $\Delta l_{1,2}$, the hierarchy scale factor
is 
$m_{Higgs}^2/m_{Planck}^2\sim\exp[-2\Delta l_1]$, and the cosmological
constant 
scale factor is $\Le/m_{Planck}^4\sim\exp[-(\Delta l_1 + \Delta l_2)]$
(with a horizon 
in the first bulk only, the roles of $\Delta l_1$ and $\Delta l_2$ will be
reversed).  
We now prove that in our model, the hierarchy scale factor is always
greater than that 
for the cosmological constant.  The curve along which $\Delta l_2=\Delta
l_1$ can be 
obtained from (\ref{ls}) most simply by defining $s=k_2+k_1$ and
$t=k_2-k_1$.  The 
condition $\Delta l_2=\Delta l_1$ becomes
\[
  {(t+q_1)(t+q_0)\over(t-q_1)(t-q_0)} =
{(t-q_1)(t-q_0)\over(t+q_1)(t+q_0)}
\]
which has $t=0$ or $k_2=k_1$ as its solution for positive brane tensions.
Above 
this line $\Delta l_2>\Delta l_1$.  Then for $k_2>k_1$, the only allowed
region in 
Figure~\arabic{kphaseeps} is that for the single horizon in the second 
bulk; there $\Delta l_2>\Delta l_1$, making the hierarchy scale larger.
For 
$k_2<k_1$ there are two allowed regions, one of which contains a horizon
in both 
bulks, and hence does not provide a solution to the hierarchy problem; the
other 
region has a horizon in the first bulk only.  In that region $\Delta l_2$
determines 
the hierarchy scale and $\Delta l_1$ determines the cosmological 
constant scale, and since $\Delta l_1>\Delta l_2$, the hierarchy scale
factor is 
again the larger.

A calculation similar to that which led to (\ref{twocomH}) reveals that,
when there is a particle horizon in each of the two bulks,  
\be
H^2(L_1)\sim\exp[-(\Delta l_1+\Delta l_2)]
\ee
with a complicated coefficient of proportionality.  
The Hubble constant in such a case will be greater than that 
with a single horizon 
because $k_2$ is less, giving a smaller $\Delta l_2$; this is evident from 
Figure~\arabic{kphaseeps}. Hence the single horizon model is an
energetically favorable configuration, if such a comparison is meaningful.

\section{Details of coordinate transformation}
\label{app:coords}

In this appendix we describe the derivation of the coordinate
transformation (\ref{eqn:tf1})--(\ref{eqn:tf2}).  We start by scaling
out all the dimensional constants by defining ${\hat t} = H t$,
${\hat y} = k_i(y-y_i)$, and ${\hat x}^l = H x^l$.  The metric
(\ref{metric1}) can now be written as
\be
k_i^2 ds^2 = d{\hat y}^2 - \sinh^2 {\hat y} \, d{\hat t}^2 + \sinh^2
{\hat y} \, e^{2 {\hat t}} \, d {\bf {\hat x}}^2.
\ee
If we focus attention on the first two terms in this line element, we
see that the coordinate singularity at ${\hat y}=0$ is analogous to
the coordinate singularity of the two dimensional Rindler metric $-
y^2 dt^2 + dy^2$, and can be eliminated by the coordinate
transformation
\baray
{\bar y} &=& \tanh({\hat y}/2) \, \cosh {\hat t} \nonumber \\
{\bar t} &=& \tanh({\hat y}/2) \, \sinh {\hat t}.
\earay
The metric now takes the form $k_i^2 ds^2 = \Omega^2 d {\bar s}^2$,
where 
\be
\Omega^2 = {4 \over (1 - {\bar y}^2 + {\bar t}^2)^2 }
\ee
and
\be
\label{metric3}
d{\bar s}^2 = - d {\bar
t}^2 + d {\bar y}^2 + ({\bar t} + {\bar y})^2 d {\bf {\hat x}}^2.
\ee
The coordinate singularity at ${\hat y}=0$ has now been eliminated in
the first two terms of the metric, but persists as the coordinate
singularity ${\bar t} + {\bar y}=0$ of the last term in (\ref{metric3}).
However, the metric (\ref{metric3}) is just 5-D Minkowski spacetime
\footnote{
We note that for the metric 
$$
k_i^2ds^2=d{\hat y}^2-A(\hat y)d{\hat t}^2+A(\hat y)e^{2{\hat
t}}d{\bf\hat x}^2,
$$  
if we define coordinates $\overline{y}=f(\hat y)\cosh{\hat t}$ and
$\overline{t}=f(\hat y)\sinh{\hat t}$ and choose $(f^\prime/f)^2
=1/A(\hat y)$ (where prime denotes differentiation with respect to
$\hat y$), then $k_i^2ds^2=(f^\prime)^{-2}d\overline{s}^2$,
with $d\overline{s}^2$ given by Eq.\ (\ref{metric3}).}.
In double null coordinates ${\bar u} = {\bar t} + {\bar y}$, ${\bar v}
= {\bar t} - {\bar y}$ it can be written as
\be
d {\bar s}^2 = - d{\bar u} d{\bar v} + {\bar u}^2 d {\bf {\hat x}}^2.
\ee
The transformation to Lorentzian
coordinates for this flat metric can be obtained by solving for the
exponential map from the tangent space at
the point ${\bar u}=1$, ${\bar v}=0$, ${\hat x}^l=0$ to the full
spacetime.  This yields the coordinate transformation 
\baray
U &=& {\bar u} \nonumber \\
V &=& {\bar v} + {\bar u} {\hat {\bf x}}^2 \nonumber \\
X^l &=& {\bar u} {\hat x}^l,
\earay
and the metric (\ref{metric3}) now takes the simple form $- dU dV + d
{\bf X}^2$.  If we now define coordinates $T$ and $Y$ by $U = T + Y$
and $V = T - Y$, and combine together all the successive coordinate
transformations of this appendix, we obtain the transformations
(\ref{eqn:tf1})--(\ref{eqn:tf2}) and the metric (\ref{metric2}) given
in the body of the paper.

\section{$G_N$ and $\Le$}
\label{app:GNandLE}

Here we explain why the determination of $G_N$ from
cosmology via the Hubble constant is sometimes misleading.
Recall that $H^2=\kappa^2_N \Le/3$, where 
the 4-D gravitational coupling is given by Eq.~(\ref{GNdefine}),
while the 4-D effective cosmological constant $\Le$ is given by
Eq.~(\ref{dcosmoconst}),
\be
\label{lambdakL}
  \Le = \sum  A^2(L_i) \sigma_i - F(k_i, L_i),
\ee
so both $\kappa^2_N$ and $\Le$ are dependent on the positions $L_i$ of 
the branes within particle horizons and the piecewise constant $\Lambda$, 
or equivalently, the $k_i$.
Since $\sigma_i$, $\kappa^2_N$, $L_i$ and $k_i$ are related by Einstein's 
equation, a perturbation on one of the brane tensions, $\sigma_i \to
\sigma_i +\delta \sigma_i$, requires a 
corresponding change in some of the other quantities. 
This back-reaction must be taken into account.
Let us see when this back-reaction is expected to be small and 
when it may be sizable.
Let us first consider the single positive tension brane model, where
the brane is not charged. Here, 
$H^2=(q^2-4k^2)/4$, $\kappa^2 \Le/3=q-2k$ and $\kappa^2_N=2k\kappa^2$. 
We can vary the brane tension,  
or equivalently $q$ by a small positive amount, $q \to q+\delta q$. Keeping
$k$ constant, we find that $dH^2/dq=q/2>0$ and $\kappa^2_N$ does not change.
This implies that the $d\Le/dq>0$. So we see that 
$\kappa^2_N= 3\delta H^2/{\delta \Le}$. As we shall see, a similar 
result will be obtained in a 
multibrane model if the metric is peaked at $q$ brane.

Now, let us consider a model where the hierarchy problem can be solved. 
Besides the visible brane, the model 
must involve at least another brane, where the metric factor $A(y)$ is 
(exponentially) bigger. 
To be specific, let us consider a two brane model, where $\sigma_1$ is the 
visible brane tension and $\sigma_0$ is the Planck brane tension.
\be
        \Le  = \sigma_0 + A^2(L_1) \sigma_1 - F(k_1,k_2,L_1,L_2)
\ee
where $A(L_1)$ is exponentially small. So the direct visible brane tension 
contribution to $\Le$ is exponentially small, and a small change in 
$\sigma_1$ is totally negligible. 
However, as $\sigma_1 \to \sigma_1 + \delta \sigma_1$, either $\sigma_0$,
$A(L_1)$ or $F(k_1,k_2,L_1,L_2)$ will adjust accordingly,
depending on the details of dynamics of the model. 
Since the influence of $\sigma_1$ on the other quantities are not 
necessarily exponentially suppressed, the induced change in $\Le$ can be
large.
It is easy to see scenarios where increasing $\sigma_1$ 
actually causes $\Le$ to decrease, or where
$\Le$ still increases, but $G_N$ decreases in a way such that $H^2$ decreases.

\section{Newton's Force Law}
\label{app:newtonsforcelaw}

Here we consider the Newton's force law
between two point masses on the visible brane explicitly. 
An easy way is to compare it to the electric force law, since the 
electromagnetic field $A_{\mu}$ which is confined on the brane
does not undergo a rescaling. 

Let us consider the electromagnetic field and two charged scalar fields 
confined on the visible brane at $y=L$:
\be
\label{appD1}
 {\bf S}^{\left(4\right)}=\int d^4x \sqrt{-\hat\gamma}
 [R^{\left(4\right)}/{2\kappa_N^2} 
- \frac14 F_{\mu \nu}F^{\mu \nu} +
 \sum_j \frac{A(L)}{2}D_{\mu}\phi^{+}_j D^{\mu}\phi_j 
- A(L)^2 M_j^2\phi_j^2/2].
\ee
In solving the hierarchy problem, we expect 
$A(L)\sim 10^{-30}$. The electric charge $e$ in 
$D_{\mu}=\partial_{\mu} -ieA_{\mu}$ is of 
order unity. As before\cite{RS1}, we can rescale the scalar fields
$\hat \phi_j = \sqrt{A(L)}\phi_j$ and the masses $m_j= \sqrt{A(L)}M_j$ 
($j=1,2$) to obtain
\be
\label{appD2}
 {\bf S}^{\left(4\right)}=\int d^4x \sqrt{-\hat\gamma}
 [R^{\left(4\right)}/{2\kappa_N^2}
- \frac14 F_{\mu \nu}F^{\mu \nu} + 
 \sum_j \frac12 D_{\mu} \hat\phi^{+}_j D^{\mu}\hat\phi_j 
-\frac12m_j^2\hat\phi_j^2]
\ee
which is in the standard familiar form. 
Note that there is no rescaling for gauge field $A_{\mu}$. 
As pointed out before, the effective action is independent of the way 
we define the warp factor, $A(0)=1$, and
$A(L)=\left(m_{Higgs}/m_{Planck}\right)^2$, or
$A(0)=\left(m_{Planck}/m_{Higgs}\right)^2$ and $A(L)=1$.
The effective action is also invariant to any further changes in the
$\left\{x^{\mu}\right\}$ coordinates.
The difference in the coordinates parallel to the brane,
$\sum_{\mu}\left(x_a^{\mu}-x_b^{\mu}\right)^2$, will
represent the physical distance measured by an observer on the brane
only if $A(L)=1$, so we may use this convention.

To warm up, let us first use the action (\ref{appD2}) to calculate the 
Newton's force law by considering the
one graviton exchange between the two scalar fields in the low energy
approximation with linearized gravity. In this large distance case, 
the energy-momentum tensor is 
$T_{\mu \nu} \simeq \eta _{\mu \nu} \sum_j m_j^2 \hat \phi_j^2+ ...$.
The result is well-known:
\be
     -G_N \frac{m^2_1m^2_2}{r^2}, \hspace{22pt}     e^2  \frac{m_1m_2}{r^2}
\ee
where we also give the one-photon exchange case. Comparing to the 
usual electric force law $e^2/r^2$, we notice that we must factor out the 
$m_1 m_2$ factor in the one-graviton exchange term to
obtain the usual Newton's force law $-G_N m_1 m_2/r^2$.
This is the result in the $(\hat \gamma_{\mu \nu},\hat\phi_j,m_j)$ frame.
Suppose we use the action (\ref{appD1}) instead, that is, in 
the $(\gamma_{\mu \nu},\phi_j,M_j)$ frame. The one-graviton exchange now gives
\be
         -G_N \frac{(A(L)^2M^2_1)(A(L)^2M^2_2)}{r^2} .
\ee
In the one-photon exchange case, we note that the coupling term
$ieA(L)A^{\mu}\phi^{+}\partial_{\mu}\phi$ implies the electric charge in 
this case to be $eA(L)$. Using the Fourier transform of 
$A(L)\partial_{\mu}\partial^{\mu}\phi - A(L)^2 M^2\phi \simeq 0$, we see that
\be
        (eA(L))^2\frac{(\sqrt{A(L)}M_1)(\sqrt{A(L)}M_2)}{r^2} .
\ee
To obtain the electric force law $e^2/r^2$, we have to factor out a 
$M_1 M_2 A(L)^3$ factor, so the Newton's force law is given by
\be
-G_N \frac{A(L)M_1 M_2}{r^2}=-G_N \frac{m_1m_2}{r^2}
\ee
so we obtain the Newton's force law of the effective theory, as expected.
Since both forces are proportional to $r^{-2}$, a rescaling of $r$ does 
not change the result.

We can also compare the relative strengths of the electric and gravitational 
forces between two particles:
\be
\frac{F_G}{F_{el}}=-G_NA\left(L\right)\frac{M_1M_2}{e^2} .
\ee
Although the gravitational force is so much weaker than the 
electric force in the visible brane, the ratio is increased by a factor of 
about $\left(m_{Planck}/m_{Higgs}\right)^2$ in the Planck brane. 
The variation of the relative strength 
of the forces as a function of the brane separation is analogous to the 
running of the couplings in ordinary field theory.

\end{document}